\documentclass[]{aa}
\usepackage{graphicx}
\usepackage[authoryear]{natbib}
\usepackage{txfonts}

\newcommand{\tcross}{t_\mathrm{cross}}

\newcommand{\mydotfill}{\leaders\hbox to 2pt{\hss.\hss}\hfill\phantom{.}}

\begin{document}

   \title{Algorithmic comparisons of decaying, isothermal, \\ supersonic turbulence}

   \subtitle{}

   \author{S.~Kitsionas\inst{1}\fnmsep\thanks{Current address: Hellenic-American Educational Foundation, Psychiko College, Stefanou Delta 15, GR-15452 P. Psychiko, Greece, e-mail: skitsionas@googlemail.com}
           \and C.~Federrath\inst{2,3}
           \and R.~S.~Klessen\inst{1,2}
           \and W. Schmidt\inst{4}
           \and D.~J.~Price\inst{5,6}
           \and L.~J.~Dursi\inst{7}
           \and M.~Gritschneder\inst{8}
           \and \\ S.~Walch\inst{8,9}
           \and R.~Piontek\inst{1}
           \and Jongsoo Kim\inst{10}
           \and A.-K.~Jappsen\inst{1,7,9}
           \and P.~Ciecielag\inst{8,11}
           \and M.-M.~Mac~Low\inst{12}
          }

   \institute{Astrophysikalisches Institut Potsdam, An der Sternwarte 16, D-14482 Potsdam, Germany
         \and
              Institut f\"{u}r Theoretische Astrophysik, Universit\"{a}t Heidelberg, Albert-Ueberle-Str. 2, D-69120 Heidelberg, Germany
         \and
              Max-Planck-Institut f\"{u}r Astronomie, K\"{o}nigstuhl 17, D-69117 Heidelberg, Germany
         \and
              Lehrstuhl f\"{u}r Astronomie, Institut f\"{u}r Theoretische Physik und Astrophysik, Universit\"{a}t W\"{u}rzburg, Am Hubland, D-97074 W\"{u}rzburg, Germany
         \and
              School of Physics, University of Exeter, Stocker Road, EX4 4QL Exeter, U.K.
         \and
              CSPA, School of Mathematical Sciences, Monash University, Clayton Vic 3168, Australia
         \and
              Canadian Institute for Theoretical Astrophysics, University of Toronto, 60 St. George Street, Toronto, ON, M5S 3H8, Canada
         \and
              Universit\"{a}ts-Sternwarte M\"{u}nchen, Scheinerstr. 1, D-81679 M\"{u}nchen, Germany
         \and
              School of Physics \& Astronomy, Cardiff University, 5 The Parade, CF24 3AA Cardiff, U.K.
         \and
              Korea Astronomy and Space Science Institute, 61-1 Hwaam-dong, Yuseong-gu, Daejeon, 305-348, Republic of Korea
         \and
              N. Copernicus Astronomical Center, Bartycka 18, 00-716 Warsaw, Poland
         \and
              American Museum of Natural History, Department of Astrophysics, Central Park West at 79th Street, New York, NY, 10024-5192, U.S.A.
             }

\date{}

\abstract
{Simulations of astrophysical turbulence have reached such a level of sophistication that quantitative results are now starting to emerge. However, contradicting results have been reported in the literature with respect to the performance of the numerical techniques employed for its study and their relevance to the physical systems modelled.}
{We aim at characterising the performance of a variety of hydrodynamics codes including different particle-based and grid-based techniques on the modelling of decaying supersonic turbulence. This is the first such large-scale comparison ever conducted.}
{We modelled driven, compressible, supersonic, isothermal turbulence with an RMS Mach number of $M_\mathrm{rms}\!\sim\!4$, and then let it decay in the absence of gravity, using runs performed with four different grid codes (\texttt{ENZO}, \texttt{FLASH}, \texttt{TVD}, \texttt{ZEUS}) and three different SPH codes (\texttt{GADGET}, \texttt{PHANTOM}, \texttt{VINE}). We additionally analysed two calculations denoted as \texttt{PHANTOM~A} and \texttt{PHANTOM~B} using two different implementations of artificial viscosity in \texttt{PHANTOM}. We analysed the results of our numerical experiments using volume-averaged quantities like the RMS Mach number, volume- and density-weighted velocity Fourier spectrum functions, and probability distribution functions of density, velocity, and velocity derivatives.}
{Our analysis indicates that grid codes tend to be less dissipative than SPH codes, though details of the techniques used can make large differences in both cases. For example, the Morris \& Monaghan viscosity implementation for SPH results in less dissipation (\texttt{PHANTOM~B} and \texttt{VINE} versus \texttt{GADGET} and \texttt{PHANTOM~A}). For grid codes, using a smaller diffusion parameter leads to less dissipation, but results in a larger bottleneck effect (our \texttt{ENZO} versus \texttt{FLASH} runs). As a general result, we find that by using a similar number of resolution elements $N$ for each spatial direction means that all codes (both grid-based and particle-based) show encouraging similarity of all statistical quantities for isotropic supersonic turbulence on spatial scales $k\lesssim N/32$ (all scales resolved by more than 32 grid cells), while scales smaller than that are significantly affected by the specific implementation of the algorithm for solving the equations of hydrodynamics. At comparable numerical resolution ($N_\mathrm{particles}\approx N_\mathrm{cells}$), the SPH runs were on average about ten times more computationally intensive than the grid runs, although with variations of up to a factor of ten between the different SPH runs and between the different grid runs.}
{At the resolutions employed here, the ability to model supersonic to transonic flows is comparable across the various codes used in this study.}

\keywords{Hydrodynamics -- shock waves -- methods: numerical -- methods: statistical -- turbulence}

\maketitle

\section{Introduction}
Laboratory and terrestrial fluid dynamics are often described as incompressible flow \citep[e.g.,][]{Lesieur1997}; 
however, astrophysical fluids are usually characterised by highly compressible supersonic turbulent 
motions \citep[see e.g.][]{ElmegreenScalo2004,ScaloElmegreen2004}. For 
example, the large observed line widths in Galactic and extragalactic 
molecular clouds and star-forming regions show direct evidence of chaotic velocity fields with magnitudes in excess of the sound 
speed. This random motion carries enough kinetic energy to counterbalance 
and sometimes overcompensate for the effects of self-gravity in these clouds 
\citep[e.g.][]{BallesterosEtAl2007,BlitzEtAl2007}. The 
intricate interplay between supersonic turbulence and self-gravity 
determines the overall dynamical evolution of these clouds and their 
observable features, such as their density structure, the star formation 
rate within them, and their lifetimes \citep[see e.g.][]{MacLowKlessen2004,McKeeOstriker2007}.

Despite turbulence being a universal phenomenon,
it is also one of the least understood natural phenomena. Turbulence arises 
as a result of the nonlinear terms in the Navier-Stokes equation that governs 
the dynamical behaviour of gases and fluids \citep{Frisch1995,Lesieur1997}. A 
self-consistent mathematical formulation does not exist. Thus, analytical research mostly focuses on finding appropriate closure equations 
that capture the bulk behaviour of the system.

As a first approach, turbulence is characterised by two spatial scales that 
are connected by a self-similar cascade of kinetic energy that occurs over the so-called 
inertial range. Energy is injected into the system on some large scale $L$
and dissipated on small scales $\ell$ that are comparable to the 
viscous length $\ell_{\rm visc}$. For incompressible turbulence, \citet{Kolmogorov1941c} described a 
simple heuristic model based on dimensional analysis that captures the basic behaviour of the flow surprisingly well. He assumed that 
turbulence driven on a large scale $L$ forms eddies on that scale. These 
eddies interact to form slightly smaller eddies, transferring some of their 
energy to the smaller scale. The smaller eddies in turn form even smaller 
ones, and so on, until energy has cascaded all the way down to the 
dissipation scale $\ell_{\rm visc}$.

In order to maintain a steady state, equal amounts of energy must be 
transferred from each scale in the cascade to the next, and eventually 
dissipated, at a rate $\dot{E} = \eta v_{\rm L}^3/L$, where $v_{\rm L}$ is 
the typical velocity on scale $L$ and $\eta$ is a constant determined empirically. 
\citet{Kolmogorov1941c} assumes this rate is constant throughout the scales, 
leading to $v_{\rm L} \propto L^{1/3}$, or equivalently $v_k \propto k^{-1/3}$ 
for wavenumbers $k \propto 1/L$. The kinetic energy in the wavenumber 
interval $[k,k+dk]$ is $E_\mathrm{kin}(k) \propto v^2_{\rm L} \propto L^{2/3} 
\propto k^{-2/3}$ and consequently the energy spectrum function $E(k) = 
dE_\mathrm{kin}/dk \propto k^{-5/3}$. This 
describes the self-similar cascade of turbulent kinetic energy. Most of this 
energy resides at the top of this cascade near the driving scale, and the 
spectrum drops off steeply below $\ell_{\rm visc}$. Because of the apparently 
local nature of the cascade in wavenumber space, the viscosity only 
determines the behaviour of the energy distribution at the bottom of the cascade below 
$\ell_{\rm visc}$, while the driving only determines the behaviour near the 
top of the cascade on and above $L$.

Supersonic flows in highly compressible gas create strong density 
perturbations. Early attempts to understand turbulence in the interstellar medium 
\citep{Weizsaecker1943,Weizsaecker1951,Chandrasekhar1949} were based on insights 
drawn from incompressible turbulence. An attempt to analytically derive the 
density spectrum and resulting gravitational collapse criterion was first made 
by \citet{Chandrasekhar1951a,Chandrasekhar1951b}. This work was followed up by several authors, 
culminating in the work by \citet{Sasao1973} on density fluctuations in 
self-gravitating media. Larson (1981) qualitatively applied the basic idea of 
density fluctuations driven by supersonic turbulence to the problem of star 
formation. \citet{BonazzolaEtAl1992} used a re-normalization group 
technique to examine how the slope of the turbulent velocity spectrum could influence 
gravitational collapse. This approach was combined with low-resolution 
numerical models to derive an effective adiabatic index for subsonic 
compressible turbulence by \citet{PanisPerault1998}. 

In supersonic turbulence, shock waves offer additional possibilities for 
dissipation. They can transfer energy between widely separated scales, 
removing the local nature of the turbulent cascade typical of incompressible 
turbulence. The spectrum may shift only slightly, however, as the power spectrum (Fourier spectrum) of a step function representative of a perfect shock wave is $k^{-2}$. \citet{Boldyrev2002} has 
proposed a theory of velocity structure function scaling based on the work of 
\citet{SheLeveque1994} using the assumption that dissipation in supersonic 
turbulence primarily occurs in sheet-like shocks, rather than linear filaments 
at the centres of vortex tubes \citep[see also][]{BoldyrevNordlundPadoan2002,BoldyrevNordlundPadoan2002b}. 
Transport properties of supersonic turbulent flows in the 
astrophysical context have been discussed by \citet{AvillezMacLow2002} and 
\citet{KlessenLin2003}, and the fractal dimension of turbulent media by \citet{FederrathKlessenSchmidt2009}. 

As satisfying analytic models are rare, especially when dealing with 
compressible and supersonic turbulent flows, special attention is drawn to 
numerical approaches. A wide range of methods are used to study 
turbulence, ranging from simulating statistical processes such as random walks 
\citep[e.g.][]{MetzlerKlafter2000}, remapping models, or certain Hamiltonian 
systems \citep{Isichenko1992}, to hydrodynamic large-eddy 
simulations (LES). In LES only the largest spatial scales are resolved 
directly using a hydrodynamic integrator. For the 
turbulent dynamics on smaller scales, a so-called subgrid scale (SGS) model is 
utilised. Among astrophysicists, the most often used SGS model is numerical 
dissipation, i.e.~performing Implicit LES. It is not possible to 
optimise the use of closure models for astrophysical turbulence through 
comparisons with laboratory experiments. Therefore, the representation of 
the SGS behaviour provided by numerical dissipation must be sufficient, and indeed provides a reasonably good approximation \citep{BenziEtAl2008}.

In the current study we focus on comparing different Implicit LES schemes. Our goal 
is to assess the applicability of different numerical schemes to the modelling 
of supersonic turbulent flows, and to compare their validity and accuracy in 
the astrophysical context. To keep this comparison simple, we focus on 
purely hydrodynamic turbulence in isothermal gaseous media in regions with periodic boundaries, and study the decay of fully developed 
turbulence. We follow the dissipation of kinetic energy due to the 
numerical viscosity intrinsic to any numerical scheme, and characterise 
the turbulent velocity field using volume- and density-weighted velocity power spectra\footnote{Volume-weighted velocity power spectra are often refered to as kinetic energy spectra for incompressible turbulence \citep[e.g.,][]{Frisch1995}. However, for compressible turbulence the kinetic energy is proportional to the density-weighted velocity power spectrum.}, and probability
distribution functions of density, velocity and velocity derivatives. We remind the reader that in supersonic turbulence 
energy is not only dissipated below $\ell_{\rm visc}^{\rm art}$ by the action of 
artificial viscosity on the smallest scale eddies, but also in shocks. In most 
of the codes employed here artificial viscosity is necessary also for the 
modelling of shocks. The main aim of our comparisons is to characterise the 
role of artificial viscosity in dissipating energy below 
$\ell_{\rm visc}^{\rm art}$ rather than the use of artificial viscosity in the modelling of 
shocks. A discussion on the shock capturing ability of 
the codes used in this study is provided in \S~\ref{codes} \citep[for a 
comprehensive such comparison, see also][]{TaskerEtAl2008}. One of the fundamental questions we want to 
address is at which numerical resolution are different numerical schemes 
capable of modelling supersonic turbulence adequately.

This is the first such comparative study; there has been no coherent 
comparison of the various hydrodynamic codes used in the literature 
for the study of supersonic turbulence\footnote{A limited comparison of two codes was presented for self-gravitating turbulence in \citet{KlessenHeitschMacLow2000} and \citet{HeitschMacLowKlessen2001}.}. In spite of the fact that results from 
different codes appear to contradict each other and lead to different 
interpretations of the role of turbulence in astrophysics -- e.g. the 
hydrodynamic simulations of \citet{BallesterosEtAl2006} and \citet{PadoanEtAl2007} in which different power-law slopes are obtained from the 
velocity power spectra leading to different interpretations of the role of 
turbulence on cloud fragmentation and the resulting core mass function -- it 
has not been properly checked whether at least some of these differences are 
due to differences in the numerical schemes employed. We perform and analyse here a 
first set of low-resolution calculations aiming at extending our 
investigations in the future to higher resolution simulations, achieved either 
directly, or by using adaptive resolution techniques like Adaptive Mesh Refinement \citep{KritsukNormanPadoan2006,SchmidtEtAl2009}, or Particle Splitting \citep{KitsionasWhitworth2002}, or by using Subgrid Scale Models \citep{SchmidtNiemeyerHillebrandt2006a,SchmidtEtAl2006b}, and/or combinations of the above. It should be emphasised that the typical number of resolution elements used here for the SPH calculations is quite large (number of particles $N=215^3$) compared to the typical number of particles used in existing studies of supersonic turbulence and cloud fragmentation in the literature \citep[e.g.][]{BallesterosEtAl2006}. On the other hand, the number of resolution elements used for the grid codes presented here is rather small (number of grid cells $N=256^3$) compared to what is the current state-of-the-art resolution for such studies \citep[e.g.][]{KritsukEtAl2007,PadoanEtAl2007,FederrathDuvalKlessenSchmidtMacLow2009,LemasterStone2009,SchmidtEtAl2009}.

The structure of our current study is as follows: in \S~\ref{codes}, we 
describe the setup of the experiments conducted as well 
as list the most important features of the codes used. In \S~\ref{statistics}, we 
review the statistical measures used in this paper for the analysis 
of supersonic turbulence. In \S~\ref{driving}, we present the initial 
conditions employed for the decay simulations, while in \S~\ref{results} we 
discuss the results of the decay experiments and the comparison of the 
performance of the various codes. We summarise the computational efficiency of the codes and runs in \S~\ref{sec:efficiency}, and present our conclusions in 
\S~\ref{sec:conclusions}.

\section{Experimental setup and numerical codes} \label{codes}
Our aim is to study the decay of supersonic hydrodynamic turbulence using different grid- and particle-based codes\footnote{We use the term particle code as the generic antonym of grid code. In general, a particle code is a numerical scheme that uses sampling points that are not fixed in space but rather move, resembling in this respect the properties of fluid particles. In particular, all particle codes used here are different implementations of the Smoothed Particle Hydrodynamics (SPH) technique first introduced by \citet{GingoldMonaghan1977} and \citet{Lucy1977}.} and compare the performance of the codes in this experiment. Therefore, we need a turbulent gas distribution that will serve as an initial condition for all codes. For simplicity, the turbulent initial conditions are produced with one of the particle/SPH codes, and then the particle code data is interpolated onto a grid. This grid, in turn, provides the initial conditions for the grid-based codes.

The turbulent gas distribution is created with \texttt{GADGET} \citep{SpringelYoshidaWhite2001}. We start with a box of side $L\,=\,0.29\,\mathrm{pc}$. Inside this box, $215^3\,=\,9,938,375$ particles were distributed homogeneously representing a static, uniform, isothermal gas with temperature $T\,=\,11.4\,\mathrm{K}$ (corresponding to a sound speed $c_{\rm s}\,=\,0.2\,\mathrm{km\,s^{-1}}$), and mass $M\,=\,120\,\mathrm{M}_{\odot}$. We impose a turbulent velocity field within our box using the driving scheme of \citet{MacLow1999}, as it has been implemented for \texttt{GADGET} by \citet{JappsenEtAl2005}. Turbulence is driven on large scales (at wavenumbers between $k\,=\,1$ and $k\,=\,2$), aiming at an RMS Mach number equal to $M_\mathrm{rms}\!\sim\!3.9$. The RMS velocity has then reached $V_\mathrm{rms}\,=\,M_\mathrm{rms}\,c_\mathrm{s}\,\sim\,0.78\,\mathrm{km\,s^{-1}}$.

We drive turbulence for a few code unit times. The code unit for time is arbitrarily chosen to be equal to the free-fall time $t_{\rm ff}\,=\,\sqrt{3\pi/(32 G\rho_0)}$, of the initial homogeneous and static gas distribution. Using $\rho_0\,=\,M/L^3$, we obtain $t_{\rm ff}\!\sim\!0.115\,\mathrm{Myr}$. This does not imply that the gas in our box is or will become self-gravitating. A useful time unit for the comparison of non-self-gravitating turbulent flows is the turbulent crossing time $\tcross\,=\,L/(2V_\mathrm{rms})\,=\,L/(2M_\mathrm{rms}c_\mathrm{s})\,\sim\,0.182\,\mathrm{Myr}$, which is defined as the time it takes for a typical turbulent fluctuation to cross half of the computational domain \citep[defined equivalently to][]{KritsukEtAl2007,FederrathKlessenSchmidt2009}.

In \S~\ref{driving_time}, we discuss for how long the turbulence needs to be driven to reach a statistical steady state. After driving has finished, the density and the velocity of the \texttt{GADGET}-particle distributions are interpolated onto a grid with $256^3$ cells. We use this grid as the initial condition for following the decay with the grid codes. The particle distribution at the end of the driving phase provides the initial conditions to the SPH codes.

We follow the decay of turbulence for about six turbulent crossing times $\tcross$. The self-gravity of the gas is kept switched off at all times and the influence of magnetic fields is neglected, i.e.~we follow the hydrodynamic decay only. An isothermal equation of state is used. Periodic boundary conditions are adopted.

Snapshots were taken at 0.0, 0.06, 0.31, 0.62, 3.1, and $6.2\;\tcross$. For the SPH codes, the particle distribution is interpolated onto a grid at each snapshot. The grid codes naturally provide their density and velocity fields on grids. From these grids, we then calculate spatially averaged quantities like the RMS Mach number, velocity power spectra, and probability distribution functions of several quantities including the density, the velocity, its derivatives, and combinations of density and velocity.

In this paper, we include turbulence decay experiments using the SPH codes \texttt{GADGET}, \texttt{PHANTOM} (runs: \texttt{A}, \texttt{B}), and \texttt{VINE}, as well as the grid codes \texttt{ENZO}, \texttt{FLASH}, \texttt{TVD}, and \texttt{ZEUS}. In the following paragraphs, the general features of the codes used in this work are listed briefly. For more details on each of the codes, please refer to the references given below. All codes were run in parallel. The parallelisation architectures that the codes were run on, the total number of CPUs used, and the number of CPU hours consumed are listed in \S~\ref{sec:conclusions} for each of the runs studied here.

Codes that are used to perform simulations of supersonic turbulence are chosen for their performance in the highly compressible regime typical of astrophysical flows. Thus, codes with the ability to capture accurately the sharp discontinuities and high density ratios that result in supersonic turbulence are preferred for this study.

Grid-based codes for supersonic flows are often based on finite volume Riemann-solvers using the Godunov scheme \citep[see e.g.][]{Toro1997}. By their conservative nature, these codes maintain correct shock speeds, and, because of the Riemann-solver approach in calculating fluxes, they maintain very sharp discontinuities across a shock (typically within a few zones). These methods are often implemented in a dimensionally split way \citep{Strang1968}.

A very common such method used for astrophysical flows is the Piecewise Parabolic Method (PPM), described in \citet{ColellaWoodward1984}, which is formally third-order accurate in space for smooth flows but switches to linear order to maintain step-function-like shocks or contact discontinuities near such features. Although not typically crucial, a small amount of artificial viscosity is implemented to ensure no or little oscillations behind shocks. Both \texttt{ENZO} \citep{NormanBryan1999,OSheaEtAl2004} and \texttt{FLASH} \citep{FryxellEtAl2000,DubeyEtAl2008}, used in this paper, are of this type. \texttt{ENZO} was used here in version 1.0.1, and \texttt{FLASH} was used in version 2.5. \texttt{FLASH} in particular has been extensively tested against laboratory experiments \citep{CalderEtAl2002} and other codes \citep{DimonteEtAl2004,HeitmannEtAl2005}. For the \texttt{FLASH} run of this work, a PPM diffusion parameter of $K=0.1$ \citep{ColellaWoodward1984} was used, whereas for the \texttt{ENZO} run the PPM diffusion parameter was set to $K=0.0$. This provides an additional test of the effects of artificial shock diffusion on the results of grid codes using the PPM. The effects of varying the PPM diffusion parameter has been investigated before: setting $K=0.0$ provides less dissipation, but produces stronger post-shock oscillations and a more pronounced bottleneck-effect \citep{KritsukEtAl2007,FederrathDuvalKlessenSchmidtMacLow2009}.

Other methods that attempt to maintain flatness of the solution behind shocks include Total Variation Diminishing (TVD) methods, which impose restrictions on the reconstruction of flow variables to ensure that the total variation of variables strictly diminishes over time. The \texttt{TVD} code used here employs an approximate (second-order-accurate), Roe-type (upwind) Riemann solver and TVD interpolation to maintain sharp shocks and smooth flow behind the shocks. Recipes for building the isothermal MHD code based on the \texttt{TVD} scheme are presented in \citet{KimEtAl1999}. For the turbulence comparisons of this project, we have used an isothermal hydrodynamic version of the code.

The \texttt{ZEUS} code \citep{StoneNorman1992a,StoneNorman1992b} is a second-order accurate code using the \citet{VanLeer1977} monotonic advection algorithm. It resolves shocks using artificial viscosity, and does not include explicit techniques for keeping shocks sharp. A staggered mesh approach is adopted: the velocity is stored at cell interfaces, and density and energy at cell centres. This version of \texttt{ZEUS} is however different from the official version of \texttt{ZEUS-MP}, which was employed for instance by \citet{VernaleoReynolds2006} and \citet{HayesEtAl2006}, in that the version used here was parallelised by R.~Piontek.

An entirely different method is employed in Smoothed Particle Hydrodynamics (SPH) codes \citep[e.g.][]{Lucy1977,GingoldMonaghan1977,Monaghan1988,Monaghan1992,Monaghan2005}. SPH codes do not involve a grid at all, but track (fixed)-mass packets of fluid in a Lagrangian sense. While this approach to fluid dynamics requires the use of an artificial viscosity to maintain shock structure \citep[e.g.][]{MonaghanGingold1983}, it has the advantage for highly compressible flows that resolution automatically increases in high-density regions, as the Lagrangian fluid packets follow the mass flow. Three SPH codes, namely \texttt{GADGET}, \texttt{PHANTOM} (runs: \texttt{A}, \texttt{B}), and \texttt{VINE}, are used in this work. More details of these codes are given below.

\texttt{GADGET} is an MPI parallel tree-SPH and $N$-body code designed by \citet{SpringelYoshidaWhite2001}. We here used \texttt{GADGET} version 1. The code uses individual and adaptive timesteps for all particles, and it combines this with a scheme for dynamic tree update. As a time integrator, it uses a variant of the leapfrog integrator, involving an explicit predictor step. The smoothing lengths are derived according to the commonly used M4 kernel \citep{MonaghanLattanzio1985,SpringelYoshidaWhite2001}. In the \texttt{GADGET} run performed here, we set the number of neighbours of each SPH particle to $40\pm5$ neighbours. The influence of changing the number of SPH neighbours $N_\mathrm{neigh}$ has been investigated by \citet{AttwoodGoodwinWhitworth2007}. They argued that using a fixed number of neighbours prohibiting any variation in $N_\mathrm{neigh}$ may reduce numerical dissipation. \citet{CommerconEtAl2008} also studied the influence of changing the number of SPH neighbours in simulations of gravitational fragmentation. They find a weak dependency of their results on the number of neighbours indicating that increasing $N_\mathrm{neigh}$ speeds up gravitational fragmentation slightly. All SPH codes used here employed roughly the same number of SPH neighbours (see below). A number of about $N_\mathrm{neigh}\!\sim\!50\pm5$ SPH neighbours is the typical setup for most SPH calculations reported in the literature. Thus, our comparison of SPH runs with each other should not be systematically affected by our choice of SPH neighbours. However, varying the number of neighbours should be investigated in a detailed systematic study of supersonic turbulence in the future.

\texttt{PHANTOM} is a low-memory, highly efficient SPH code written especially for studying non-self-gravitating problems. The code is made very efficient by using a simple neighbour finding scheme based on a fixed grid and linked lists of particles. In particular, it uses an $\eta\,=\,1.2$ term in calculating the smoothing length $h$ through $h\,=\,\eta\,(m/\rho)^{1/3}$. This corresponds to about 58 SPH neighbours in a uniform density distribution, though it is the $h-\rho$ relation (to a tolerance of $10^{-4}$) that provides the deciding criterion, not the neighbour number. The code implements the full "grad-h" SPH formulation developed by \citet{PriceMonaghan2004} and \citet{PriceMonaghan2007}, whereby the smoothing length and density are mutually dependent and iterated self-consistently, resulting in exact conservation of momentum, energy and entropy in the SPH equations. Shocks are treated using the \citet{Monaghan1997} formulation of artificial viscosity in SPH as described in \citet{Price2004}, though modified slightly in \texttt{PHANTOM} to allow a more efficient calculation. Timestepping is performed using a Kick-Drift-Kick leapfrog integrator. The standard \texttt{PHANTOM} run is labeled as \texttt{PHANTOM~A}. We conduct a second run with \texttt{PHANTOM}, labeled \texttt{PHANTOM~B}, in which dissipation is reduced away from shocks using the viscosity switch proposed by \citet{MorrisMonaghan1997}.

\texttt{VINE} \citep{WetzsteinEtAl2008,NelsonWetzsteinNaab2008,GritschnederEtAl2009} is an OpenMP parallel, tree-SPH and $N$-body code. The code scales linearly up to a high number of processors, and is designed for a combined usage of generic CPUs and the special purpose hardware GRAPE. Time integration is performed here with a Drift-Kick-Drift leapfrog integrator, which allows for individual particle timesteps. The smoothing lengths are derived according to the M4 kernel \citep{MonaghanLattanzio1985}. In the \texttt{VINE} simulations, we set the number of neighbours of each SPH particle to $50\pm5$. Shocks are treated with the time dependent artificial viscosity prescription introduced by \citet{MorrisMonaghan1997} (i.e.~similar to the \texttt{PHANTOM~B} run).

\section{Methods for the analysis of statistical measures} \label{statistics}

\subsection{SPH interpolation onto a grid}
For interpolating the density and the velocity distribution of the SPH particles onto a grid, we use the generic SPH interpolation formula \citep[e.g.][]{Monaghan1992}
\begin{equation} \label{eq:gridinterpolation}
A({\bf r}) = \sum_i \frac{A_i}{\rho_i} \frac{m_i}{h_i^3} W\left(\frac{|{\bf r} - {\bf r}_i|}{h_i}\right)\;,
\end{equation}
where $\rho_i$ is the density, $m_i$ is the mass, and $h_i$ is the smoothing length of the $i$-th particle. The vector ${\bf r}$ is the position vector to the centre of each grid cell, $A_i$ is the particle quantity to be interpolated to the grid (density and velocity for our purposes), and $W(s)$ is the kernel function used for smoothing the particle mass in space to derive an SPH density. The M4 kernel, which is based on spline functions (Monaghan \& Lattanzio 1985),
\begin{equation}
W(s) = \frac{1}{\pi}\left\{ \begin{array}{ll}
1 - 3 s^2 / 2 + 3 s^3 / 4, & \quad 0 \leq s \leq 1, \\
(2 - s)^3 / 4,             & \quad 1 \leq s \leq 2, \\
0,                         & \quad s \geq 2,
\end{array} \right.
\end{equation}
is used for the interpolation. For each grid point, the above summation is over a limited number of neighbouring SPH particles due to the compact support of the kernel function, which vanishes for $|{\bf r} - {\bf r}_i| \geq 2 h_i$.


\subsection{Volume-weighted velocity power spectrum} \label{sec:method_spectra_vw}
The velocity power spectrum is calculated as follows. For each velocity component we take the Fourier transform of the velocity field ${\bf v}\,=\,(v_1, v_2, v_3)$. We denote these Fourier transforms as ${\bf v}'\,=\,(v'_1, v'_2, v'_3)$. Using these definitions the volume-weighted velocity power spectrum is defined as
\begin{equation} \label{eq:spect_vol}
E(k) = \frac{1}{2}\,\left({\bf v}'\cdot\overline{{\bf v}'}\right)\;,
\end{equation}
where $\overline{{\bf v}'} = (\overline{v'_1}, \overline{v'_2}, \overline{v'_3} )$ is the complex conjugate of the transformed velocity. A wavenumber mapping, ${\bf k} = (k_1, k_2, k_3)$, is applied on the cells of the $E(k)$-cube, with each $k_1$, $k_2$, and $k_3$ ranging from $-N/2$ to $(N/2)-1$. The compressible (longitudinal) part of the velocity power spectrum is calculated as
\begin{equation}
E_{\rm com}(k) = E(k)\,\frac{ ({{\bf v}'} \cdot {\bf k})\,(\overline{{\bf v}'} \cdot {\bf k}) }{ ({\bf k} \cdot {\bf k})\,({{\bf v}'} \cdot \overline{{\bf v}'}) }\;,
\end{equation}
and the solenoidal (transverse) part as
\begin{equation}
E_{\rm sol}(k) = E(k) - E_{\rm com}(k)\;.
\end{equation}
For each wavenumber $k\,=\,\left|{\bf k}\right|$, we collect from the $E(k)$-cube all cells lying at distances in the $[k,\,k\!+\!1]$ interval\footnote{Distances are 
measured with respect to the $(0, 0, 0)$ cell, i.e.~the central cell.}. The mean of the $E(k)$-values of these cells is normalised by the area of the sphere element with radius $k+0.5$. This gives the volume-weighted velocity power spectrum $E(k)$. The above process is repeated for $E_{\rm sol}(k)$ and $E_{\rm com}(k)$.

To calculate the volume-weighted dissipation rate, we estimate at each snapshot the integral over the volume-weighted velocity power spectrum. This is formally given as 
\begin{equation} \label{eq:dissipationrate}
\frac{1}{2}\,V_\mathrm{rms}^2 = \int_0^\infty E(k)\,\mathrm{d}k\;,
\end{equation}
and is estimated numerically. The dissipation rate is then given as the rate of change of this integral as the decay proceeds. 

The sonic scale, $l_{\rm s}\,=\,2 \pi\,k_{\rm s}$, separates supersonic turbulent flows on large scales (small $k$) from subsonic turbulent flows on smaller scales (high $k$). We estimate the sonic scale by solving the following equation,
\begin{equation} \label{eq:sonic_scale}
\frac{1}{2}\,c_{\rm s}^2 \simeq \int_{k_{\rm s} = 2 \pi /l_{\rm s}}^\infty E(k)\,\mathrm{d}k\;,
\end{equation}
implicitly for the sonic wavenumber $k_{\rm s}$.

\subsection{Density-weighted velocity power spectrum} \label{sec:method_spectra_mw}
For density-weighted velocity power spectra, we substitute
\begin{equation}
{\bf v}_\mathrm{mw} = \left(\rho/\rho_0\right)^{1/2}\,{\bf v}\;,
\end{equation}
where $\rho_0$ is the mean density in the cube. Then, the above process for the calculation of velocity power spectra is repeated. The density-weighted velocity power spectrum is defined as
\begin{equation} \label{eq:spect_rho}
E_{\rm mw}(k) = \frac{1}{2}\,\left({\bf v}_\mathrm{mw}'\cdot\overline{{\bf v}_\mathrm{mw}'}\right)\,,
\end{equation}
in analogy to equation~(\ref{eq:spect_vol}). The density-weighted dissipation rate is computed in analogy to equation~(\ref{eq:dissipationrate}), from the density-weighted velocity power spectrum $E_{\rm mw}(k)$.

\subsection{Probability distribution functions} \label{sec:method_PDFs}
Using all the cells in our grid, we obtain the cumulative distributions, $F$, of the following quantities: logarithm of the density, the three velocity components $v_i$ with $i\,=\,1$, 2, 3, the logarithm of the trace free rate of strain, $|S^*|$, and the logarithm of vorticity, $\omega = \left|{\bf \nabla} \times {\bf v}\right|$. To obtain these distributions, the corresponding quantities are binned linearly. From the cumulative distribution, $F$, we derive the probability distribution function (PDF) of quantity $A$ at each bin $n$ by computing
\begin{equation}
\mathrm{PDF}_n(A) = \frac{F_n(A)-F_{n-1}(A)}{A_n-A_{n-1}}\;.
\end{equation}

For the trace free rate of strain, the spatial derivatives of the velocity components are first calculated as
\begin{equation}
v_{i,j} = \frac{\partial v_i}{\partial x_j}\;.
\end{equation}
The (symmetric) strain tensor has components, $S_{ij} = 0.5 (v_{i,j} + v_{j,i})$. The rate of strain is then
\begin{equation}
|S|^2 = 2 \sum_{i} \sum_{j} S_{ij} S_{ij}\;.
\end{equation}
The trace of the strain tensor is $d = \sum_{i} S_{ii}$, so that the trace free strain tensor has components $S^*_{ij} = S_{ij} - \delta_{ij}\,d/3$, where $\delta_{ij}$ is the Kronecker unit function. The trace free rate of strain then becomes
\begin{equation} \label{eq:strain}
|S^*|^2 = 2 \sum_{i} \sum_{j} S^*_{ij} S^*_{ij}\;.
\end{equation}
$|S^*|$ gives the rate at which a fluid element is deformed without changing its volume, e.g. by the act of a shear flow.

\section{Driving phase} \label{driving}

\subsection{Driving time} \label{driving_time}
\begin{figure*}
\centerline{\includegraphics[width=0.38\linewidth]{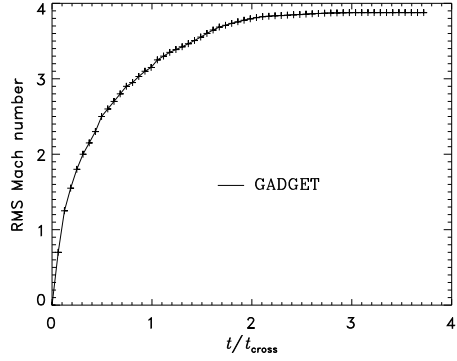}}
\caption{Time evolution of the RMS Mach number during the driving phase. After driving for about 2 turbulent crossing times $\tcross$ (see text), the RMS Mach number has reached a statistical steady state. The initial conditions for the decaying turbulence code comparison were chosen randomly at $t=2.5\tcross$ in the regime of fully developed supersonic turbulence, $t>2\tcross$ \citep[e.g.][]{FederrathKlessenSchmidt2009}, when the RMS Mach number has reached its statistical steady state of $M_\mathrm{rms}\!\sim\!3.9$.}
\label{fig:mach_evol_driving}
\end{figure*}

We simulate driven turbulence using \texttt{GADGET} following the prescription in \citet{JappsenEtAl2005}. The driving phase starts with an initially homogeneous density distribution with $\rho_0=3.3\times10^{-19}\,\mathrm{g\,cm^{-3}}$ at rest. As turbulence gets driven, the RMS Mach number increases with time. It levels off at a value of $M_\mathrm{rms}\!\sim\!3.9$ after about $4\,t_{\rm ff}$ (see Fig.~\ref{fig:mach_evol_driving}). This time corresponds to roughly $2.5$ crossing times, $\tcross$. We start the decay experiments at this time, i.e.~after turbulence has been driven for $2.5\,\tcross$. The particle distribution obtained at this time is interpolated onto a grid. This grid is used as the initial condition for the decay simulations using the grid codes. For the decay simulations using the SPH codes, we directly use the particle distribution obtained with \texttt{GADGET} at the same time ($t=2.5\,\tcross$). For the following calculations the time was reset to zero, and the turbulence decay was followed over roughly six crossing times with each of the grid and SPH codes.

\begin{figure*}
\begin{center}
\begin{tabular}{cc}
\includegraphics[width=0.36\linewidth]{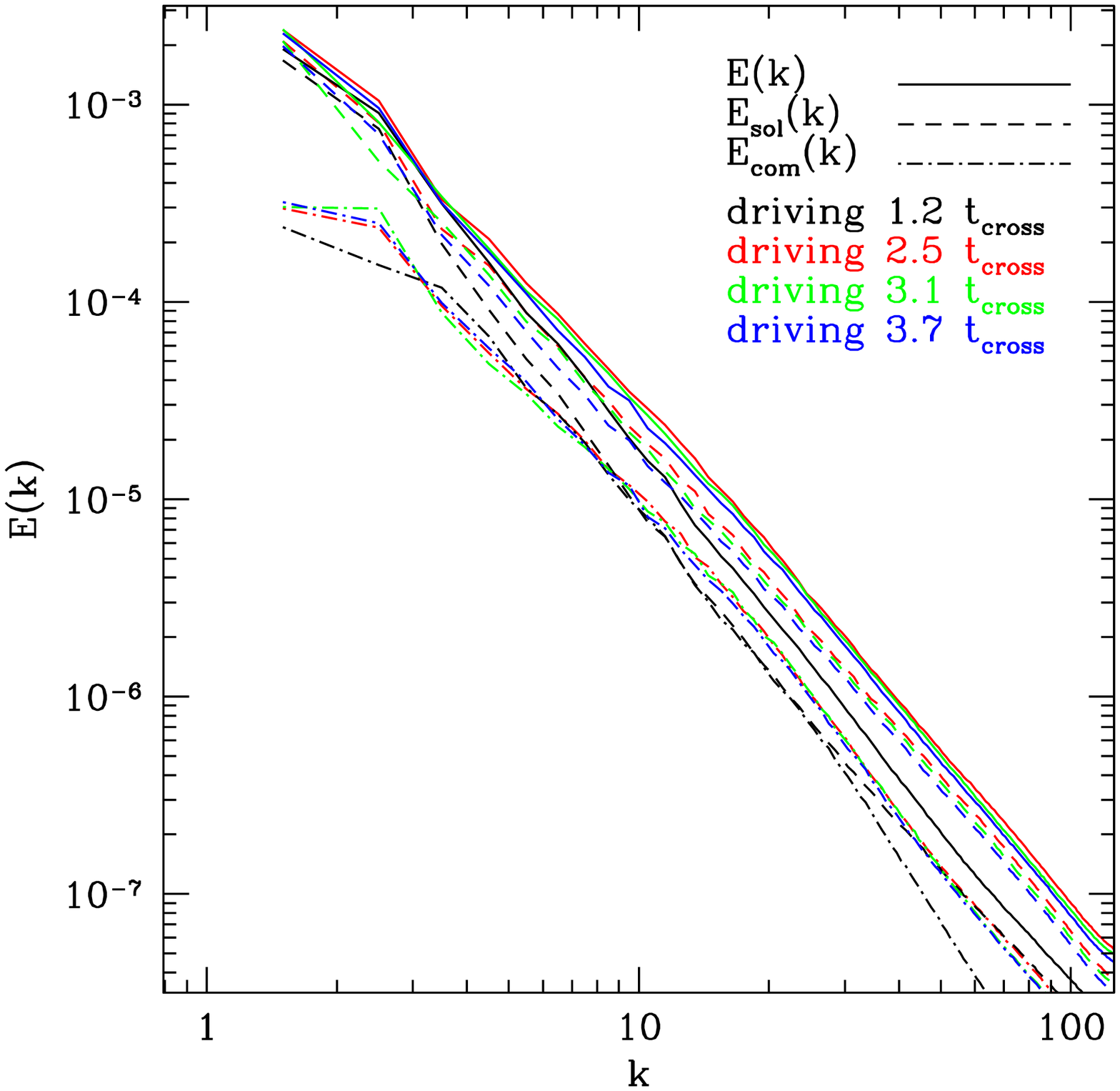} &
\includegraphics[width=0.36\linewidth]{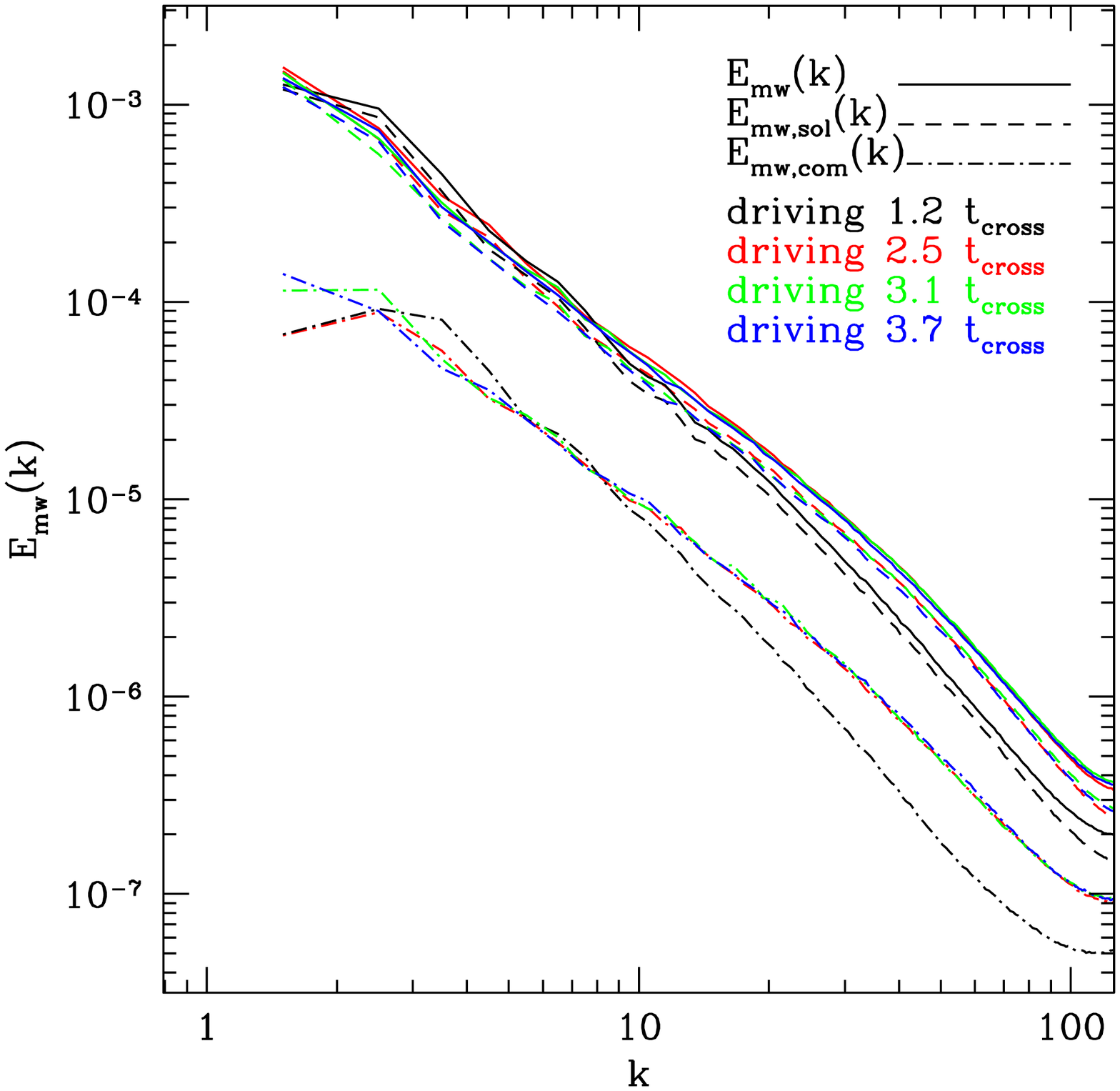} \\
\includegraphics[width=0.36\linewidth]{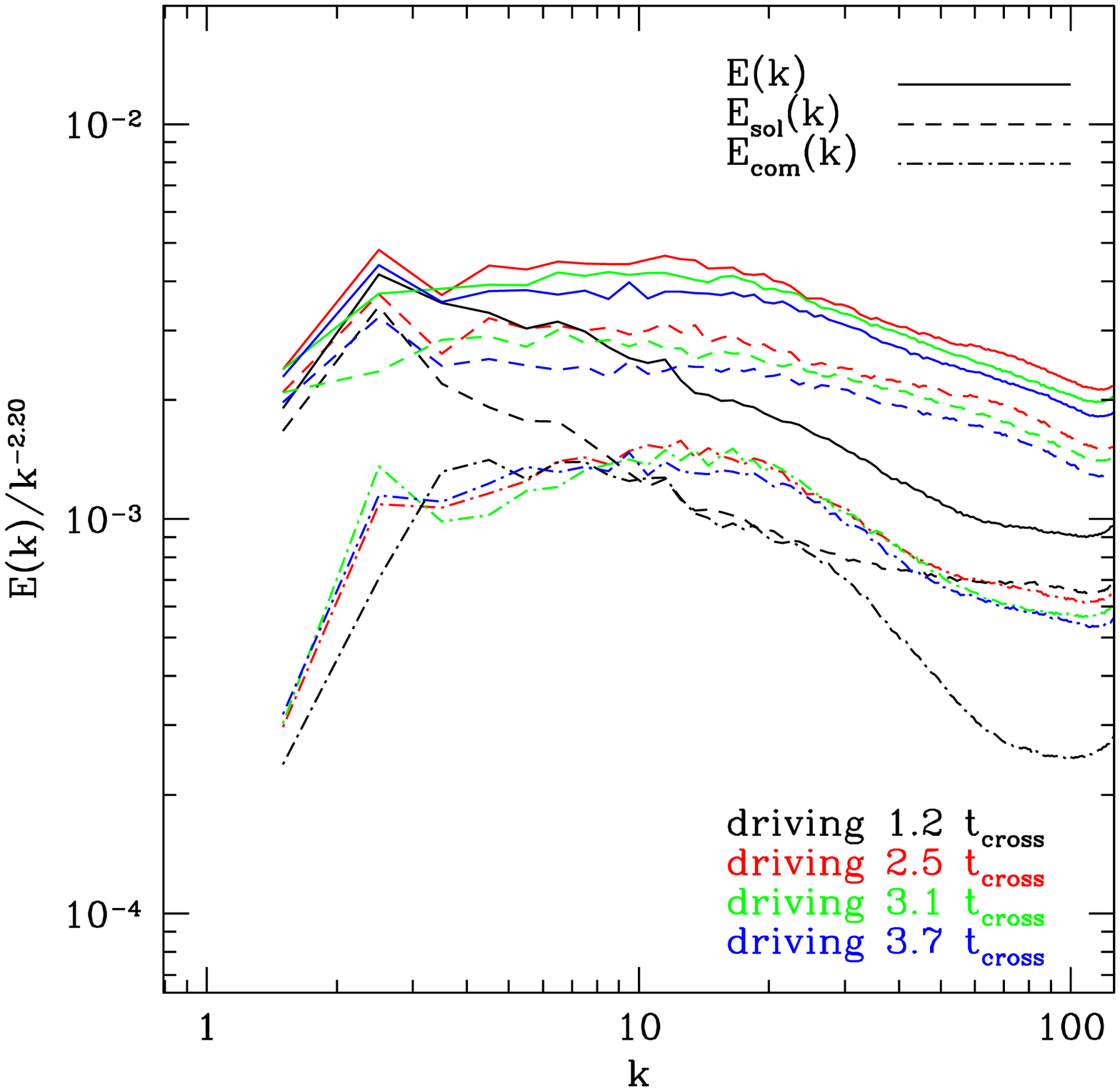} &
\includegraphics[width=0.36\linewidth]{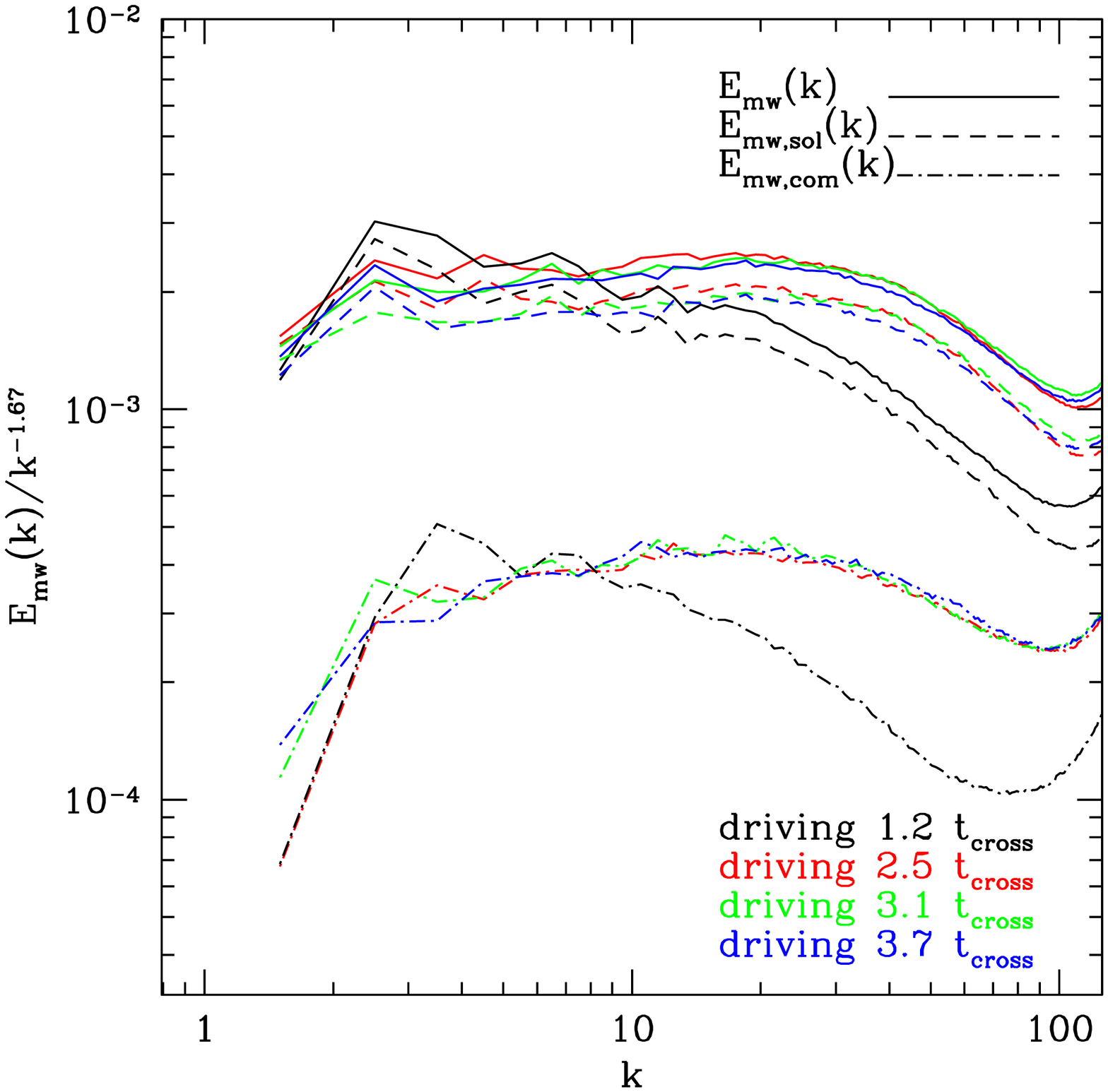}
\end{tabular}
\end{center}
\caption{\emph{Top:} Velocity power spectra obtained at four different snapshots during the driving phase (volume-weighted spectra on the left, and density-weighted spectra on the right panels). The spectrum of each snapshot is plotted with a different colour ($1.2\,\tcross$: black lines; $2.5\,\tcross$: red lines; $3.1\,\tcross$: green lines; $3.7\,\tcross$: blue lines). The decay experiments using the SPH and grid codes starts with the snapshot at $2.5\,\tcross$ when turbulence is fully established. The solid lines correspond to $E(k)=E_\mathrm{sol}+E_\mathrm{com}$ (volume-weighted, left), and $E_\mathrm{mw}(k)=E_\mathrm{mw,sol}+E_\mathrm{mw,com}$ (density-weighted, right), the dashed lines to the solenoidal (transverse) part, and the dash-dotted lines to the compressible (longitudinal) part of the spectra. \emph{Bottom:} The same spectra as on the top, but compensated with power-law slopes of 2.20 (left panel -- volume-weighted case) and 1.67 (right panel -- density-weighted case).}
\label{fig:spect_driving_phase}
\end{figure*}

Turbulence has been established when the decay experiments start. This is shown in Figure~\ref{fig:spect_driving_phase}, where we plot the velocity power spectra, calculated as explained in \S~\ref{sec:method_spectra_vw} and \S~\ref{sec:method_spectra_mw} for the volume- and density-weighted spectra, respectively, of four different snapshots taken along the driving phase. On the left panels, volume-weighted spectra are plotted whereas density-weighted spectra are shown on the right panels. Driving for $1.2\,\tcross$ (black lines) was not sufficient to produce a statistically fully established turbulent flow, as there was not enough time for turbulence (driven on large scales) to cascade down to the smallest spatial scales. However, the turbulence is fully established after about $2.5\,\tcross$: there is no significant variation of the velocity power spectra when we attempted to drive turbulence for longer times (cf.~the red, green and blue lines corresponding to driving times of 2.5, 3.1, and $3.7\,\tcross$, respectively). We conclude that starting the decay with the gas distribution obtained after $2.5\,\tcross$ of driving is a reasonable choice of initial conditions. \citet{SchmidtEtAl2009}, \citet{FederrathKlessenSchmidt2009}, and \citet{GloverFederrathMacLowKlessen2009} also conclude that after about $2\,\tcross$, supersonic turbulence has established a statistical invariant state. However, significant statistical fluctuations from snapshot to snapshot remain \citep{FederrathDuvalKlessenSchmidtMacLow2009}, which explains the slight changes visible in the velocity power spectra at $t\,=\,2.5$, 3.1, and $3.7\,\tcross$ in Figure~\ref{fig:spect_driving_phase}. The variations seen in Figure~\ref{fig:spect_driving_phase} for $t\gtrsim2.5\,\tcross$ are at most in the order of the typical snapshot-to-snapshot variations introduced by intermittent fluctuations, i.e., less than 10\% \citep[see also][Fig.~1]{KritsukEtAl2007}. The initial conditions used for our code comparison therefore constitute a statistically fully established supersonic turbulent density and velocity distribution.

\subsection{The result of driving: initial conditions for the decay experiments}
\begin{figure*}
\begin{center}
\begin{tabular}{cc}
\includegraphics[width=0.45\linewidth]{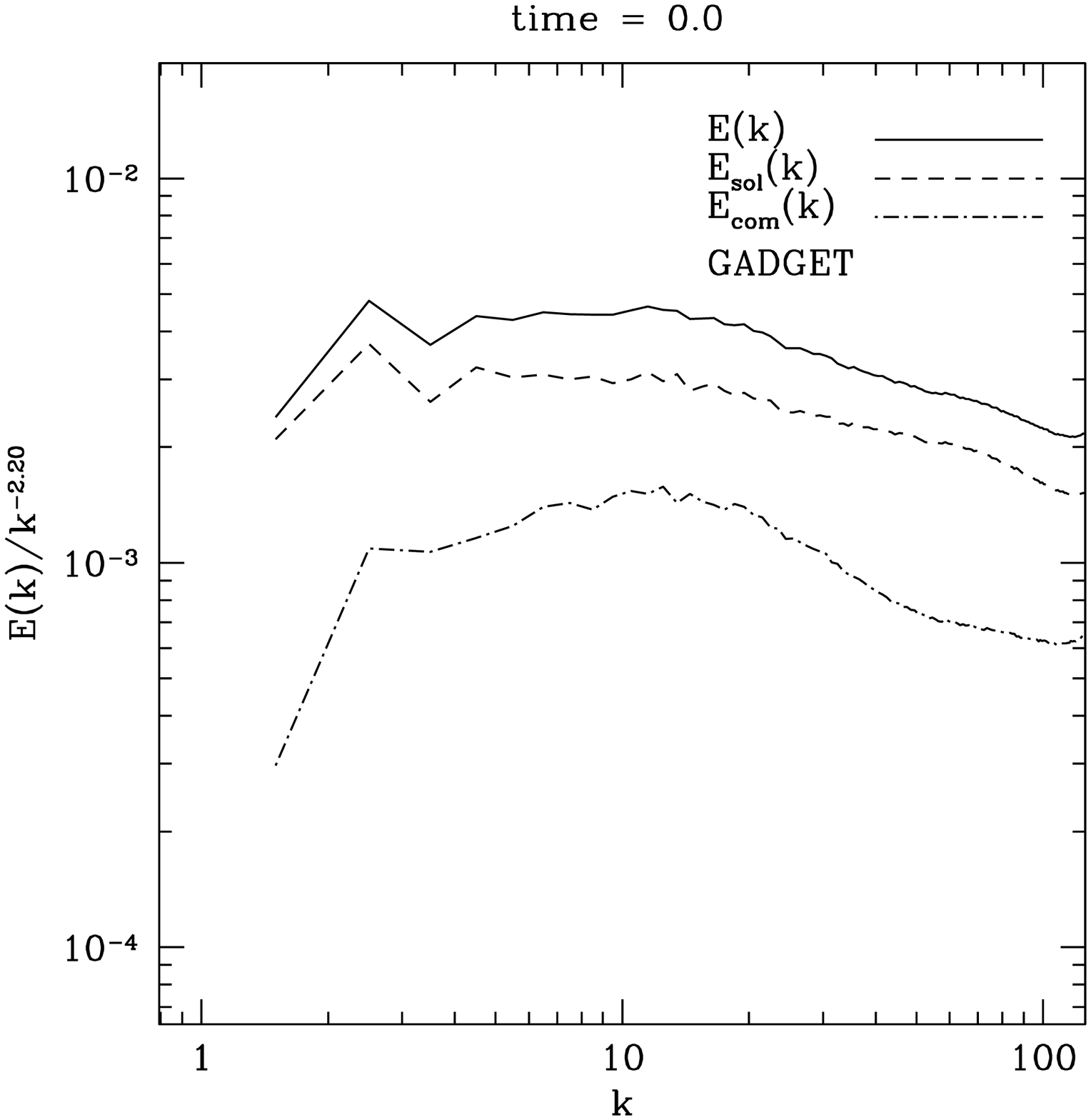} &
\includegraphics[width=0.45\linewidth]{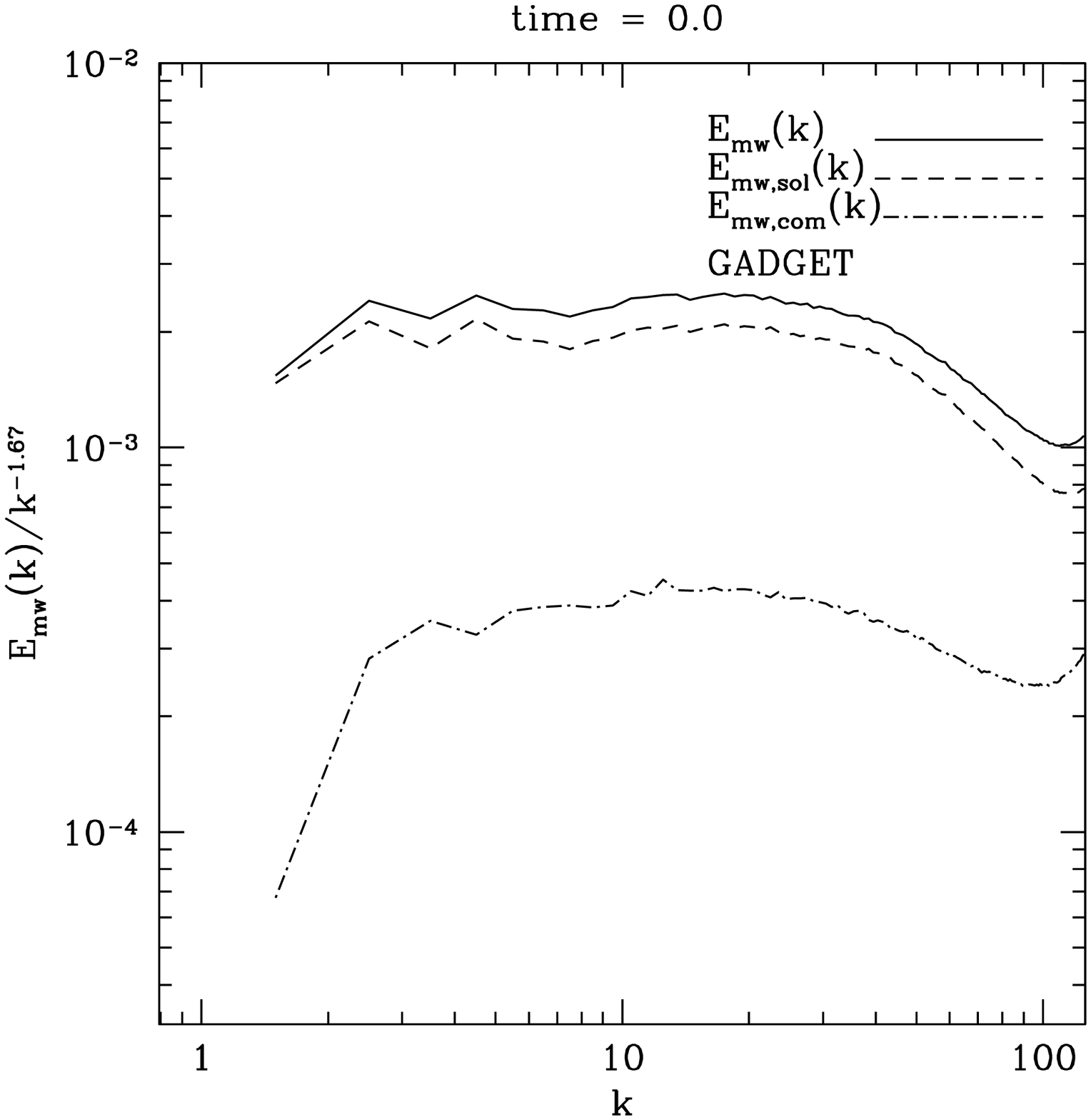}
\end{tabular}
\end{center}
\caption{Velocity power spectra of the initial conditions used for the decay experiments (volume-weighted spectrum on the left, and density-weighted spectrum on the right panel). The spectra were compensated with power-law slopes of 2.20 (left panel -- volume-weighted case) and 1.67 (right panel -- density-weighted case).}
\label{fig:spect_init}
\end{figure*}

In this section, we present the velocity power spectra of the initial conditions used for the decay experiments. These initial conditions have been produced with \texttt{GADGET} using the turbulence driving routine developed by \citet{MacLow1999}, and employed in \citet{JappsenEtAl2005}. They present the state of the system after $2.5\,\tcross$ of driving (see \S~\ref{driving_time}). On the left panel of Fig.~\ref{fig:spect_init}, we plot the volume-weighted velocity spectrum, with the density-weighted velocity spectrum shown on the right panel. The spectra were compensated with power-law slopes of 2.20 (left panel -- volume-weighted case) and 1.67 (right panel -- density-weighted case).


\subsubsection{Volume-weighted velocity power spectrum of the initial conditions}
From the volume-weighted velocity power spectrum computed with \texttt{GADGET} we derive a slope of about 2.2, which is obtained in the wavenumber range $4\!\lesssim\!k\!\lesssim\!12$. If any inertial-range scaling could be inferred at all due to our limited numerical resolution, it may only exist in the close vicinity of $k\sim8$. 

In the presence of a bottleneck effect \citep[e.g.,][]{DoblerEtAl2003,HaugenBrandenburg2004,SchmidtHillebrandtNiemeyer2006}, \citet{PorterWoodward1994} argued that the bottleneck affects all scales up to $k\sim N/32=8$ for grids of size $N=256$. \citet{SchmidtHillebrandtNiemeyer2006} suggested that the bottleneck peaks at $k \sim N/10 \sim 26$. These authors also argued that, in codes showing no bottleneck, numerical dissipation will start acting at wavenumbers smaller than $k \sim N/10$. Since our initial conditions do not seem to exhibit a bottleneck, dissipation will certainly start at $k\lesssim26$. Since a power law is established for scales $4\!\lesssim\!k\!\lesssim\!12$, and since this power law breaks down at $k\gtrsim12$, we argue that dissipation did not play a significant role for wavenumbers $k\lesssim12$ in the spectra of the initial conditions.

\citet{PadoanEtAl2007} and \citet{KritsukEtAl2007} performed high-resolution hydrodynamic simulations of driven turbulence using \texttt{ENZO} at resolutions of $1024^3$ grid cells. They obtained a significantly shallower slope of 1.9 for the volume-weighted velocity spectrum. \citet{FederrathDuvalKlessenSchmidtMacLow2009} showed that driving turbulence with the \texttt{FLASH} code at resolutions of $512^3$ and $1024^3$ grid cells results in a slope consistent with those of \citet{PadoanEtAl2007} and \citet{KritsukEtAl2007}. In contrast, at the resolution of $256^3$ (as used here), a steeper slope of the order of 2.1 was obtained \citep{FederrathDuvalKlessenSchmidtMacLow2009}, which is consistent with our result for the volume-weighted spectra. \citet{FederrathDuvalKlessenSchmidtMacLow2009} directly demonstrated the steepening of the velocity spectrum with decreasing numerical resolution (cf.~their Fig.~C.1). However, they also find that the slope of the velocity spectrum is converged to within less than 3\% going from $512^3$ to $1024^3$ in their numerical experiments. Considering the low resolution in the present study, it is therefore not surprising that we find a slope of about 2.2 for the volume-weighted velocity power spectrum. This result however can also be taken as an indication that the initial conditions used for our comparison experiments did not strongly depend on the method by which these initial conditions have been produced (i.e.~\texttt{GADGET}).

\subsubsection{Density-weighted velocity power spectrum of the initial conditions} \label{sec:driving_spect_rho}
\begin{figure*}
\begin{center}
\begin{tabular}{cc}
\includegraphics[width=0.45\linewidth]{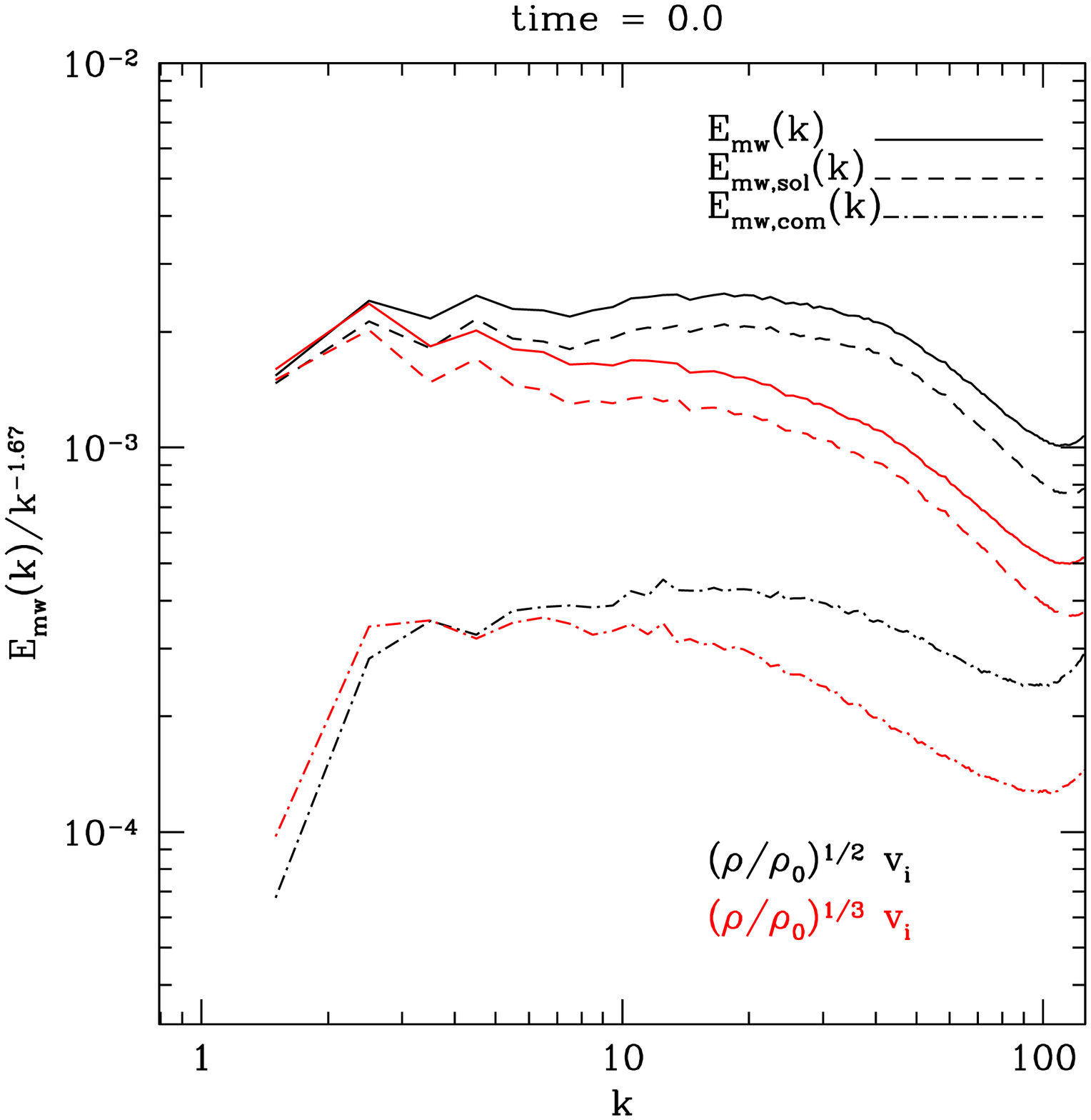} &
\includegraphics[width=0.45\linewidth]{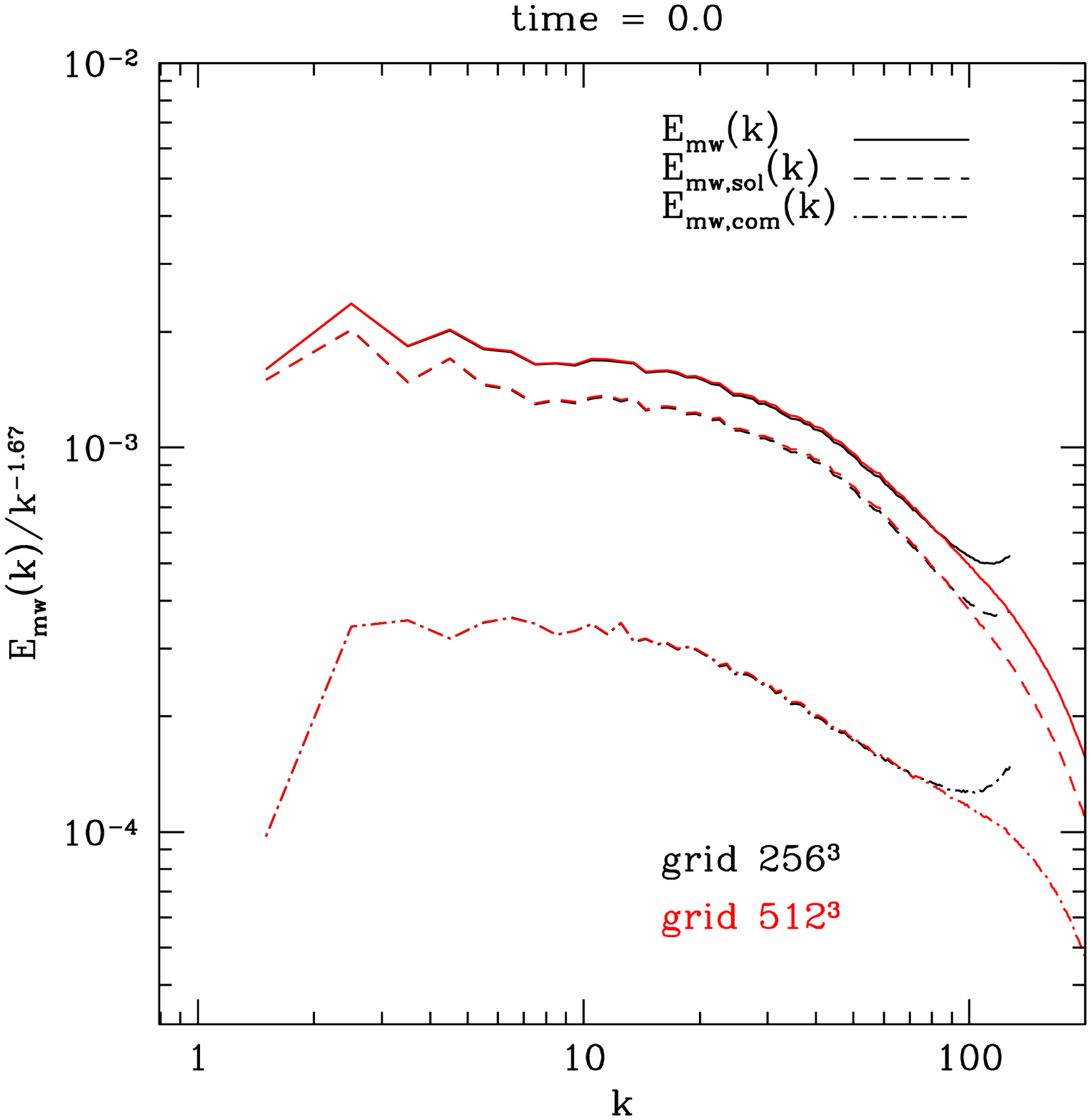}
\end{tabular}
\end{center}
\caption{\emph{Left panel:} Density-weighted velocity power spectra of the initial conditions using different velocity weights: ${\bf v}_\mathrm{mw}\,=\,(\rho/\rho_0)^{1/2}\,{\bf v}$ (black lines), and ${\bf v}_\mathrm{mw}\,=\,(\rho/\rho_0)^{1/3}\,{\bf v}$ (red lines). \emph{Right panel:} Density-weighted velocity power spectra [${\bf v}_\mathrm{mw}\,=\,(\rho/\rho_0)^{1/3}\,{\bf v}$] of the initial conditions interpolated to grids of $256^3$ cells (black lines) and $512^3$ cells (red lines).}
\label{fig:spect_rho3weights}
\end{figure*}

From the density-weighted velocity power spectrum we obtain a scaling close to the \citet{Kolmogorov1941c} scaling with a slope of about $5/3$. This slope is found within $5 \leq k \leq 20$. We have used ${\bf v}_\mathrm{mw} = (\rho/\rho_0)^{1/2}\,{\bf v}$ (see \S~\ref{sec:method_spectra_mw}) instead of ${\bf v}_\mathrm{mw}=(\rho/\rho_0)^{1/3}\,{\bf v}$, which was used by \citet{KritsukEtAl2007}, \citet{KowalLazarian2007}, \citet{SchmidtFederrathKlessen2008}, and \citet{FederrathDuvalKlessenSchmidtMacLow2009}. The $(\rho/\rho_0)^{1/2}$ weights correspond to a quantity that has physical reference to kinetic energy $(1/2)\,\rho\,|{\bf v}|^2$, while the $(\rho/\rho_0)^{1/3}$ weights correspond to a constant kinetic energy dissipation rate within the inertial range \citep{KritsukEtAl2007}. For \texttt{GADGET} only, we additionally compute energy spectra with the $(\rho/\rho_0)^{1/3}$ weights. The comparison of the $(\rho/\rho_0)^{1/2}$ to the $(\rho/\rho_0)^{1/3}$ weights is shown in Figure~\ref{fig:spect_rho3weights} (left panel). We find a steeper slope of about $1.8$ for the $(\rho/\rho_0)^{1/3}$ weights. The fact that \citet{KritsukEtAl2007} obtain Kolmogorov-type scaling using the $(\rho/\rho_0)^{1/3}$ weights is a consequence of their volume-weighted velocity power spectrum being shallower than ours with slopes of 1.9 and 2.2, respectively. We find this steeper slope of about 2.2 due to our limited numerical resolution as discussed in the previous section. Therefore, the fact that our density-weighted spectra show scaling close to Kolmogorov scaling is a result of the rather small numerical resolution adopted in this comparison. Clearly, the slopes of the density-weighted spectra depend not only on the velocity statistics, but also on the convolution of density and velocity statistics.

However, we argue that the density information should be taken into consideration in the statistical analysis of compressible turbulence, as most of the mass ends up in small volumes through shocks. This fact is neglected by statistical measures that take into account volume-weighted velocities only \citep[for instance, some models of the mass distribution of molecular cloud cores and stars are based on the volume-weighted velocity power spectrum, e.g.,][]{PadoanNordlund2002}.

\subsubsection{The effective SPH resolution} \label{lagrangian}
We would like to comment here on the Lagrangian nature of the SPH method. In Figure~\ref{fig:spect_init}, there is a rise in power on scales $k \gtrsim 100$, particularly prominent in the density-weighted case (right panel). This is a consequence of the adaptivity in resolution that is intrinsic to the SPH method: as the SPH particles move with the flow, the build-up of high densities is accompanied by an increase in the number of particles (sampling-points) for a fixed volume element. Therefore, as high densities build up on small scales due to the shocks developed in supersonic turbulence, the SPH particle concentration increases on these scales. In particular, in driven turbulence the effective SPH resolution in high-density regions eventually becomes superior to the resolution initially employed. Hence, the extra power developed on scales $k \gtrsim 100$ is a result of the interpolation of the SPH particle distribution onto a grid of resolution lower than the effective SPH resolution\footnote{Note that the total number of sampling points for the SPH runs ($215^3$ particles) was smaller than the number of sampling points employed for our grids ($256^3$ grid cells).}. When the SPH particle distribution is interpolated onto a larger grid ($N=512^3$) this rise in power is no longer observed (right panel of Fig.~\ref{fig:spect_rho3weights}). In other words, because of the finite extent of the grid we used for our comparison experiments ($N=256^3$), all SPH information on scales smaller than $k_{\rm max}$ gets interpolated into the $k_{\rm max}$ bin. This appears as a rise in the power on the smallest scales of our grid.

\section{Results of the turbulence decay code comparison} \label{results}

\subsection{Volume-weighted and density-weighted velocity power spectra} \label{spectra}

\begin{figure*}
\begin{center}
\begin{tabular}{cc}
\includegraphics[width=0.31\linewidth]{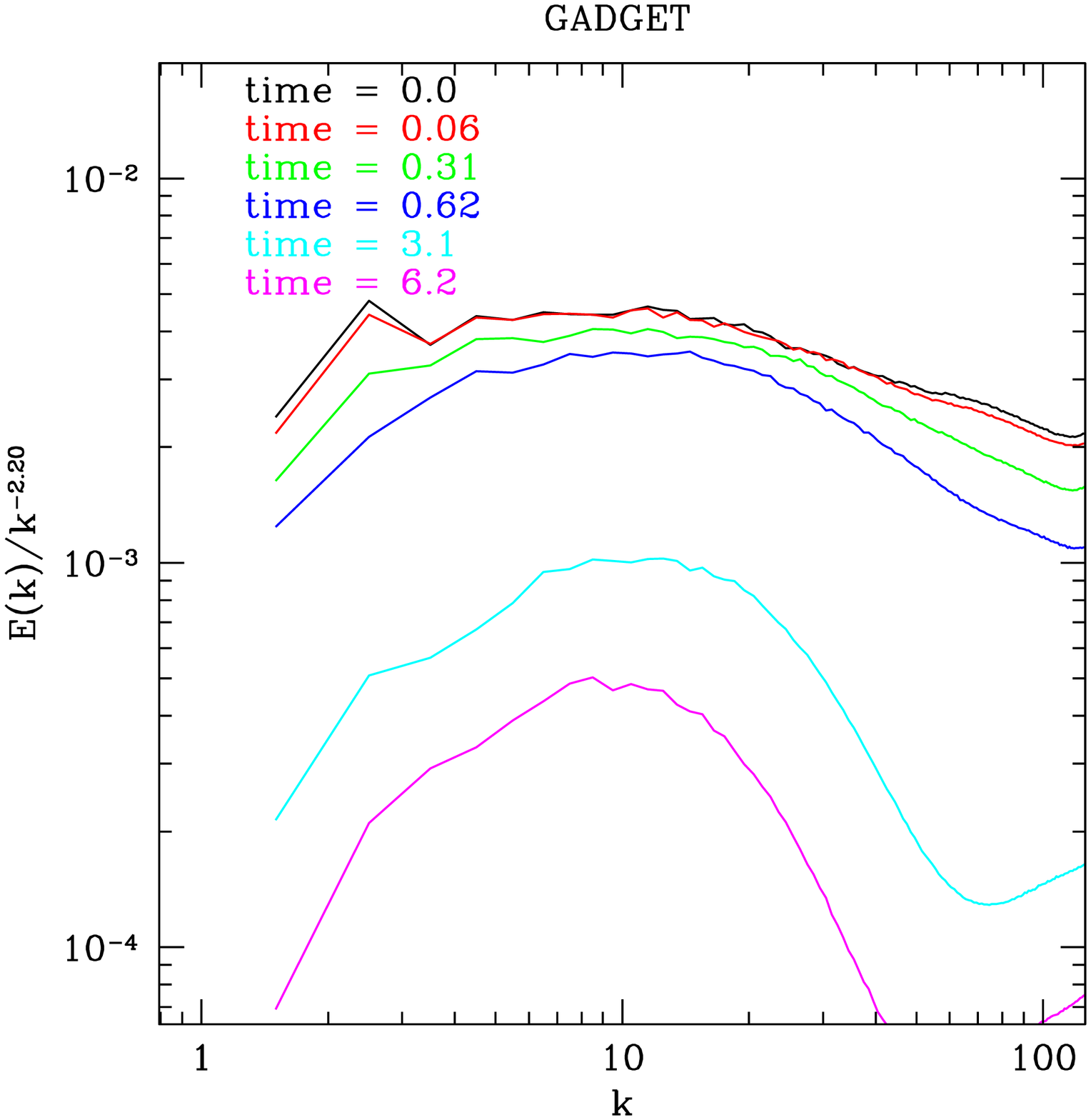} &
\includegraphics[width=0.31\linewidth]{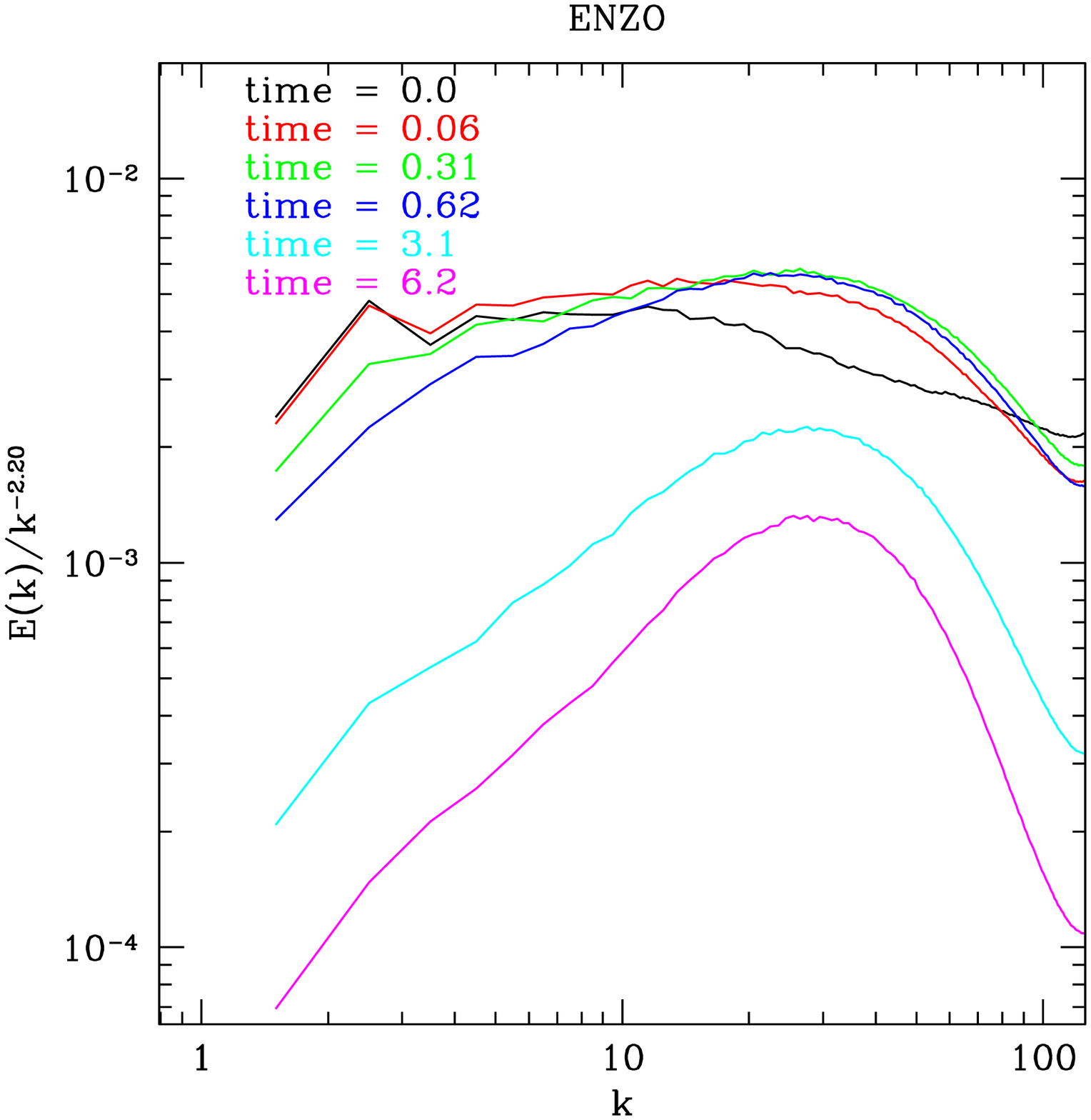} \\
\includegraphics[width=0.31\linewidth]{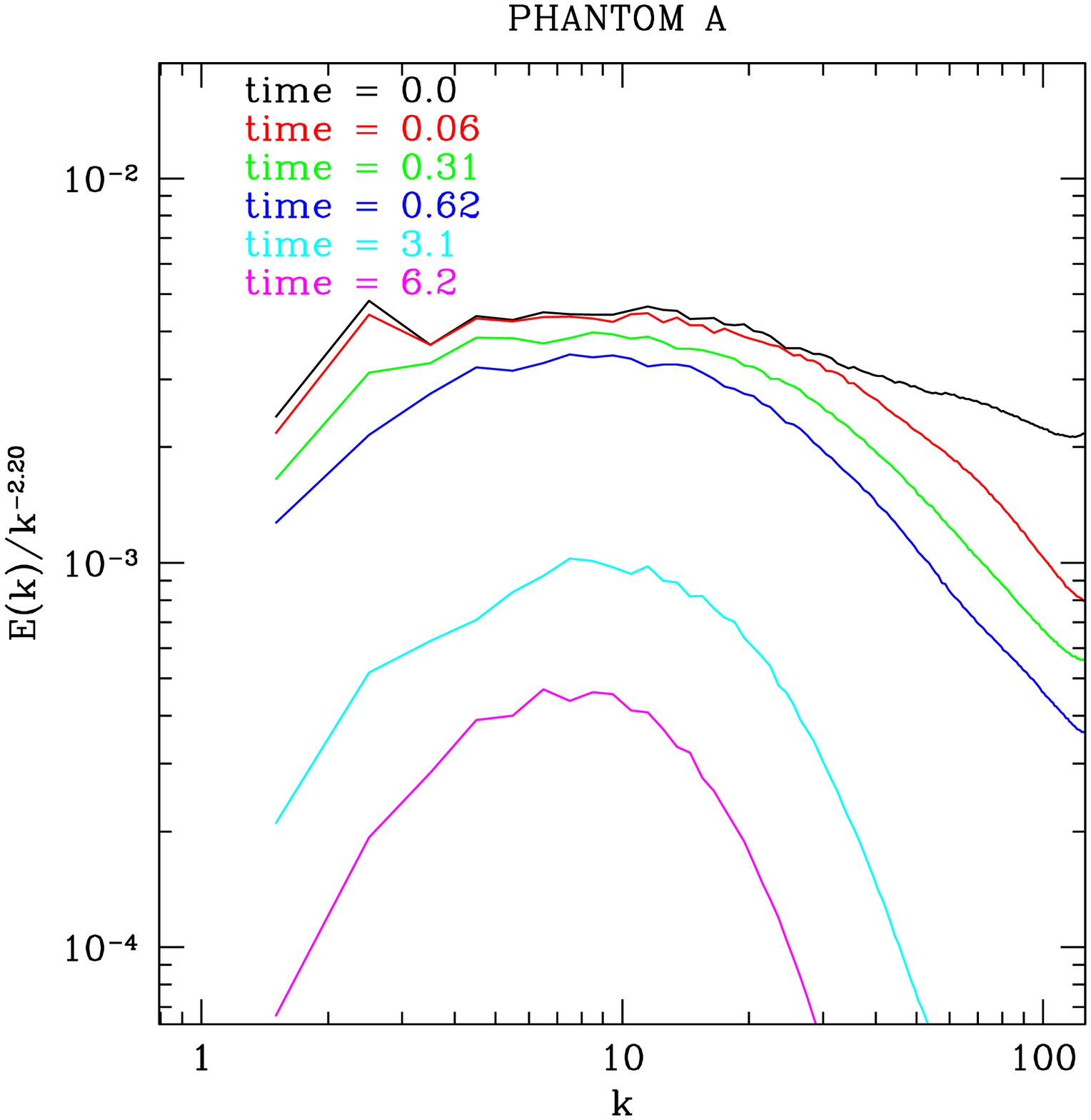} &
\includegraphics[width=0.31\linewidth]{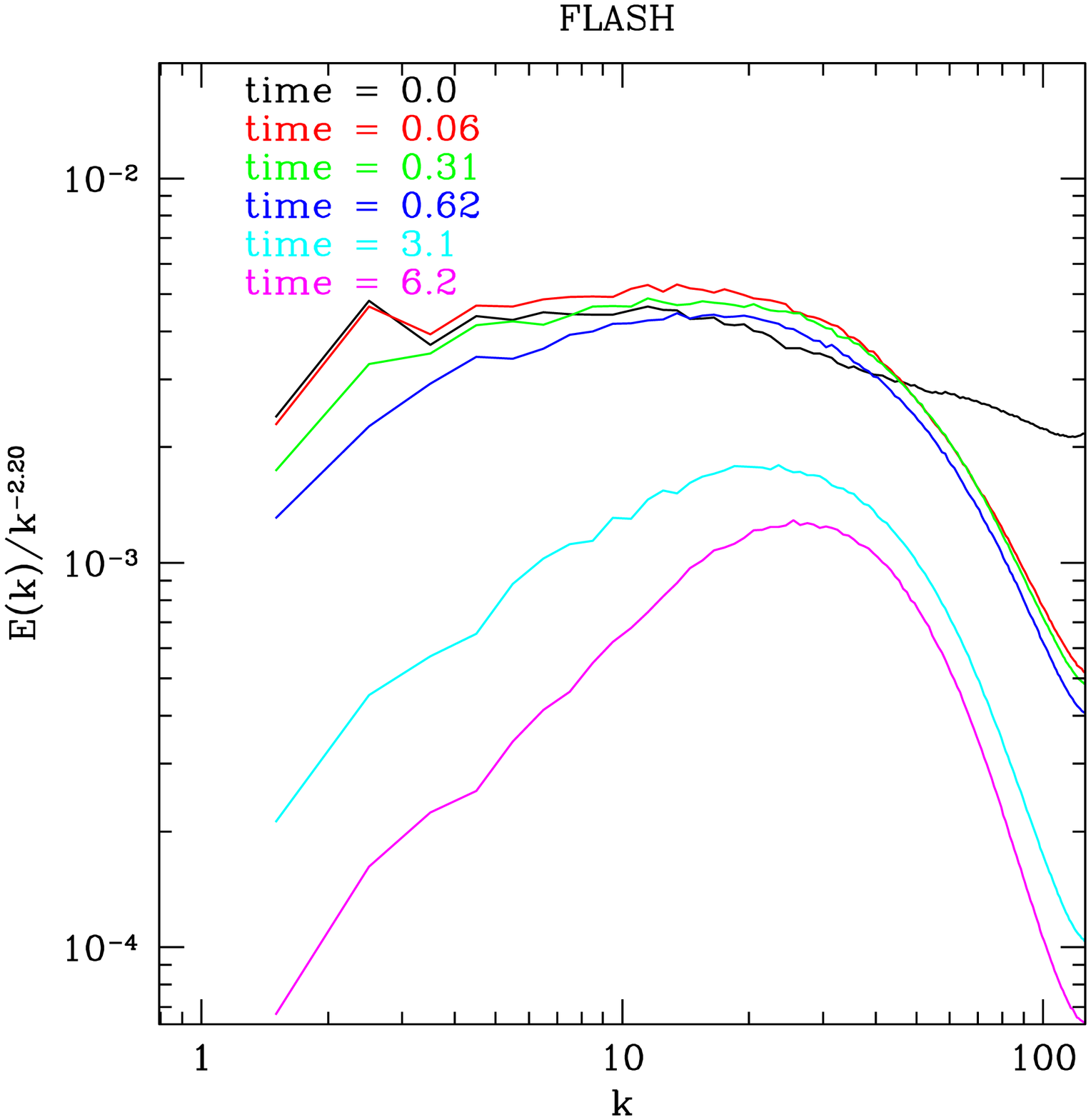} \\
\includegraphics[width=0.31\linewidth]{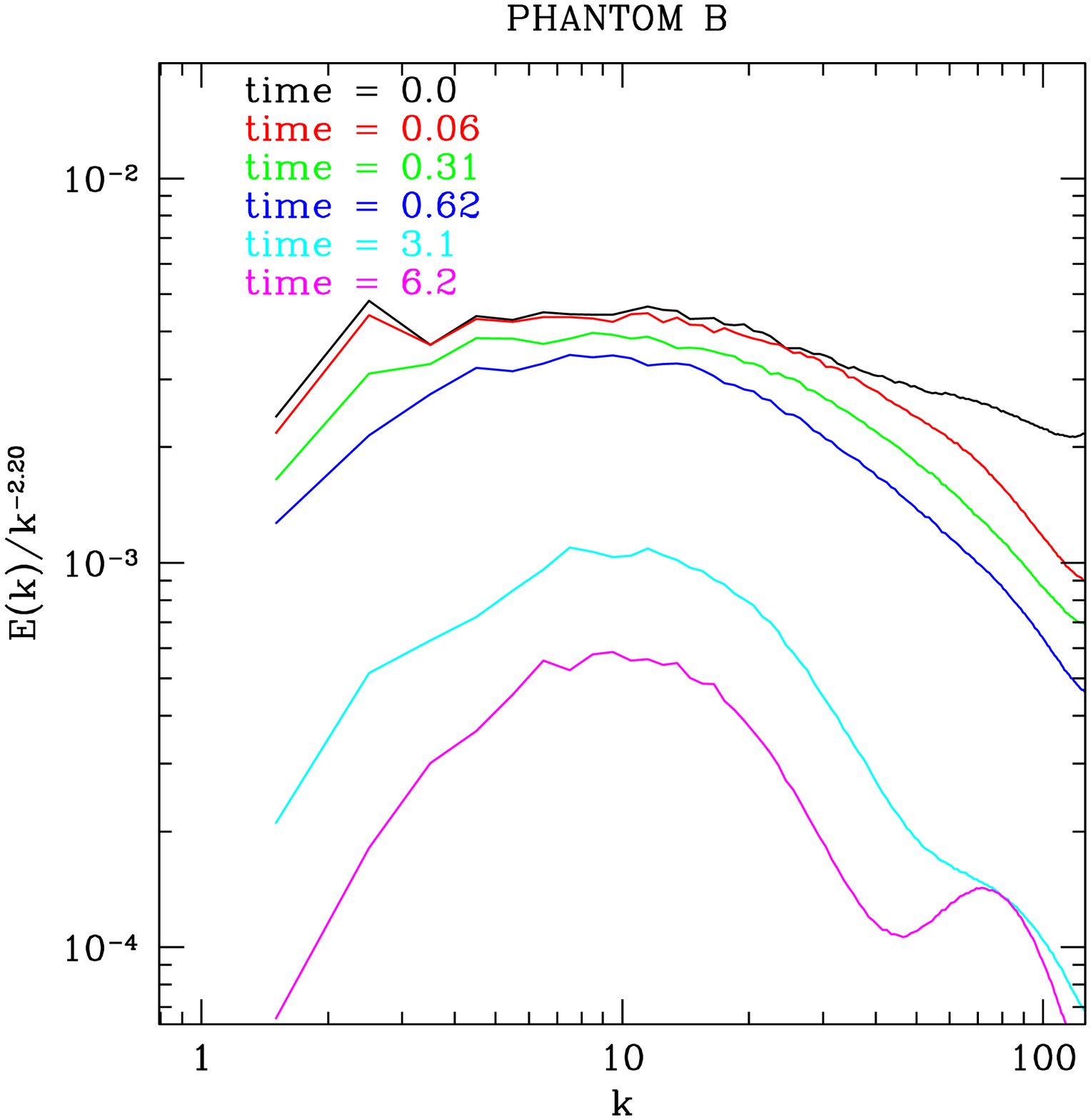} &
\includegraphics[width=0.31\linewidth]{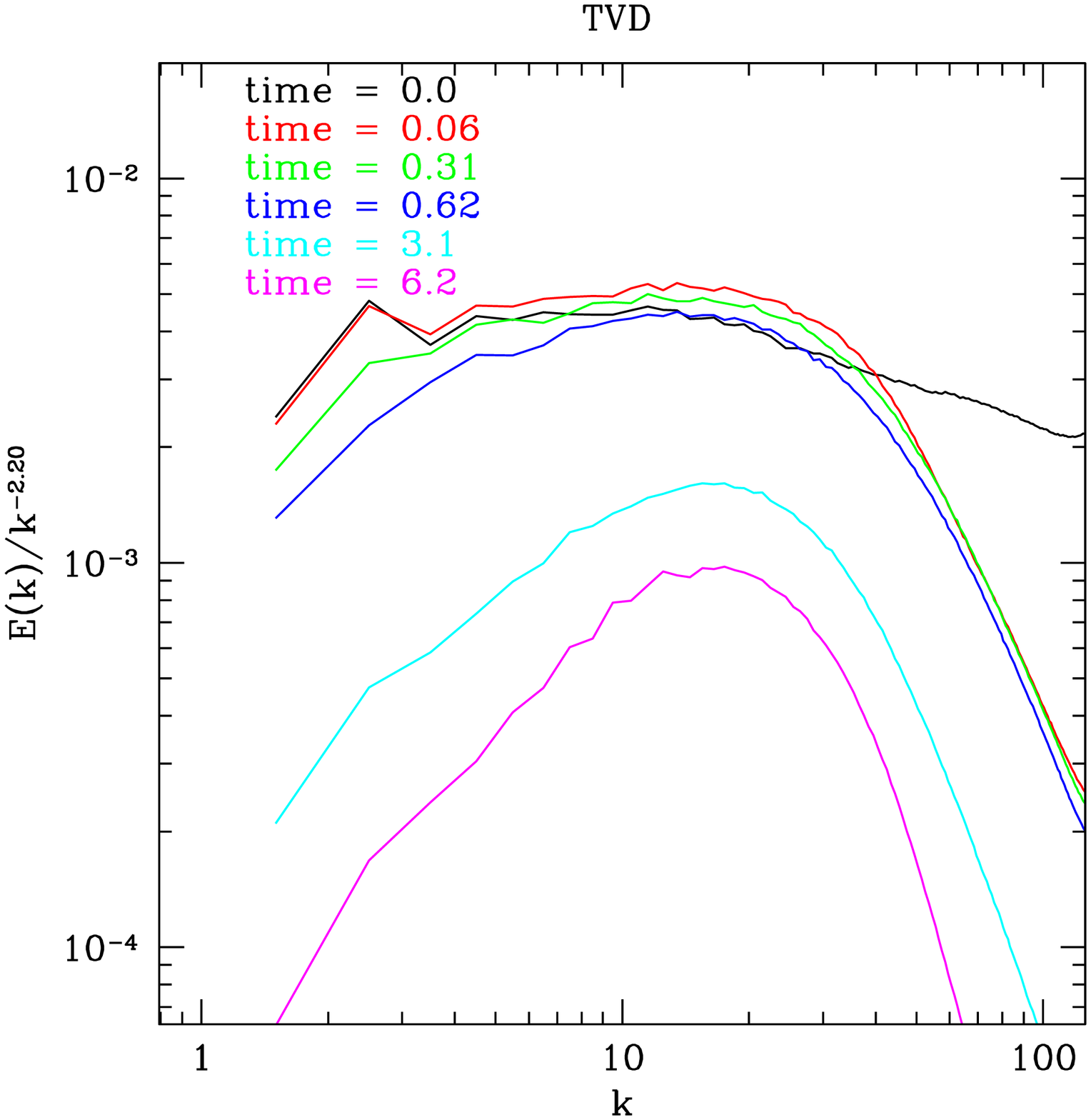} \\
\includegraphics[width=0.31\linewidth]{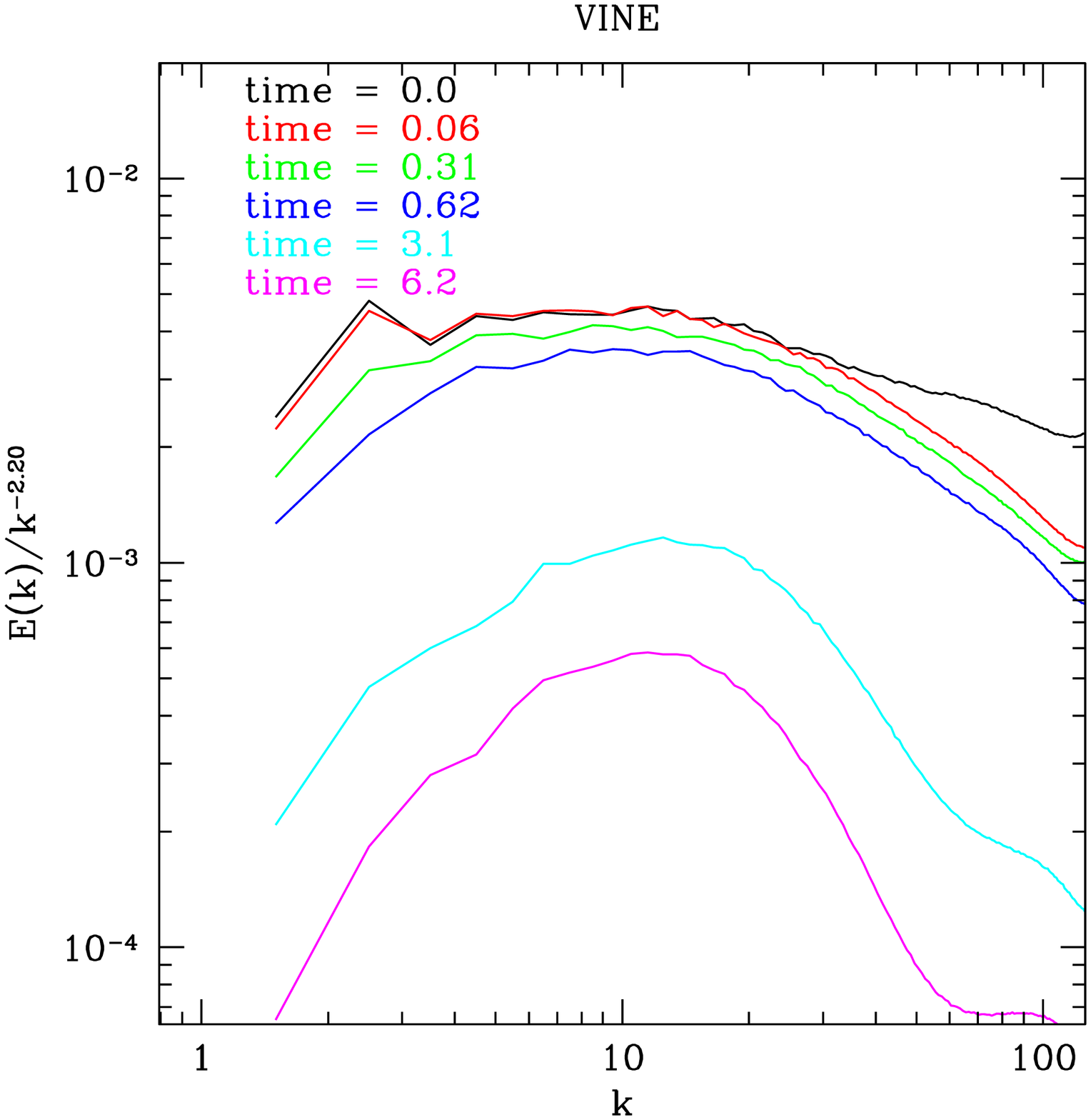} &
\includegraphics[width=0.31\linewidth]{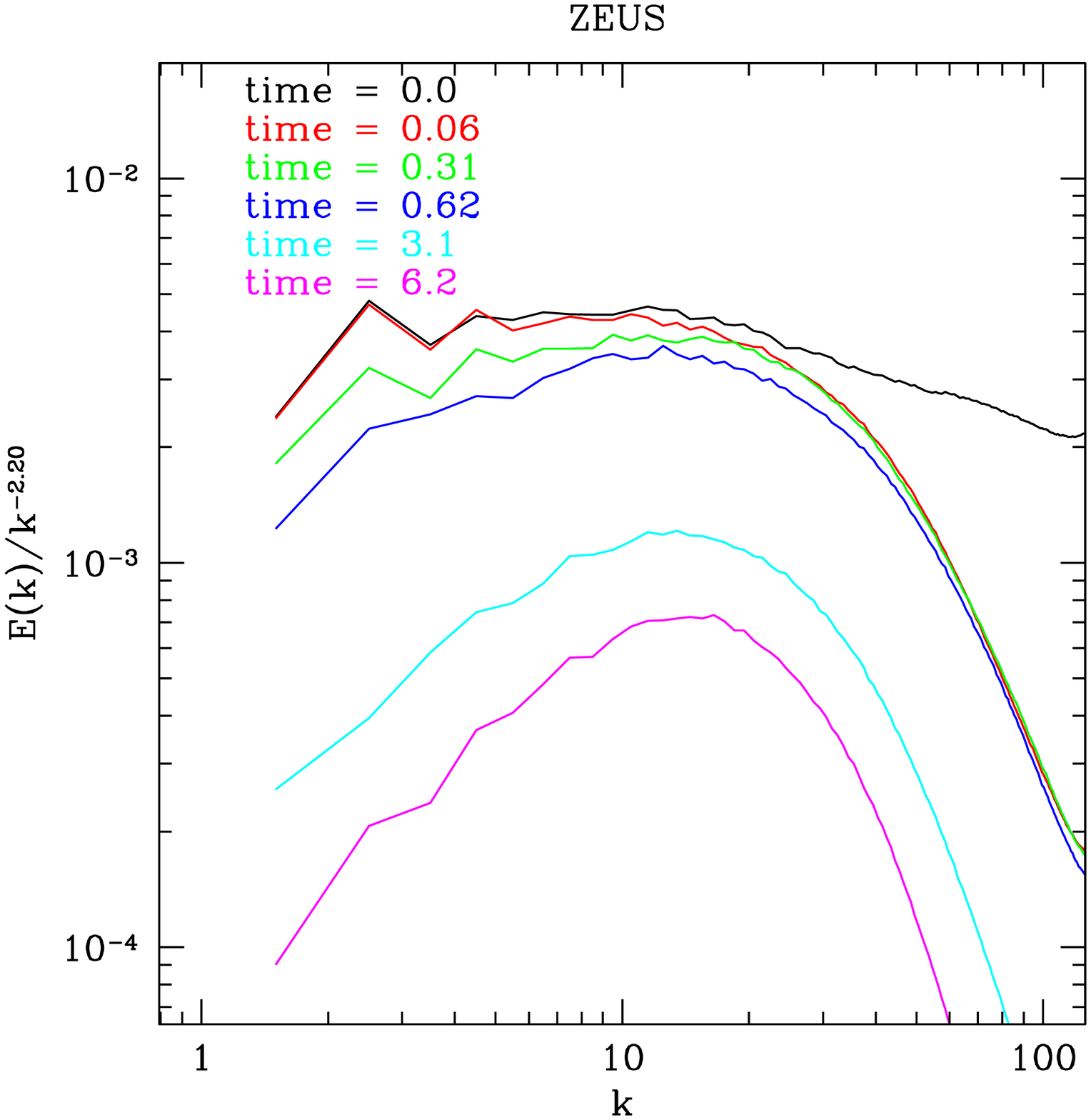}
\end{tabular}
\end{center}
\caption{Time evolution of the volume-weighted velocity power spectra (compensated with power-law slopes of 2.20) of the SPH codes \texttt{GADGET}, \texttt{PHANTOM} (runs: \texttt{A}, \texttt{B}), and \texttt{VINE} (left column) and of the grid codes \texttt{ENZO}, \texttt{FLASH}, \texttt{TVD}, and \texttt{ZEUS} (right column) for the following snapshots: $t\,=\,0.0$, $0.06$, $0.31$, $0.62$, $3.1$, and $6.2\;\tcross$.}
\label{fig:spect_evol_vol}
\end{figure*}

\begin{figure*}
\begin{center}
\begin{tabular}{cc}
\includegraphics[width=0.31\linewidth]{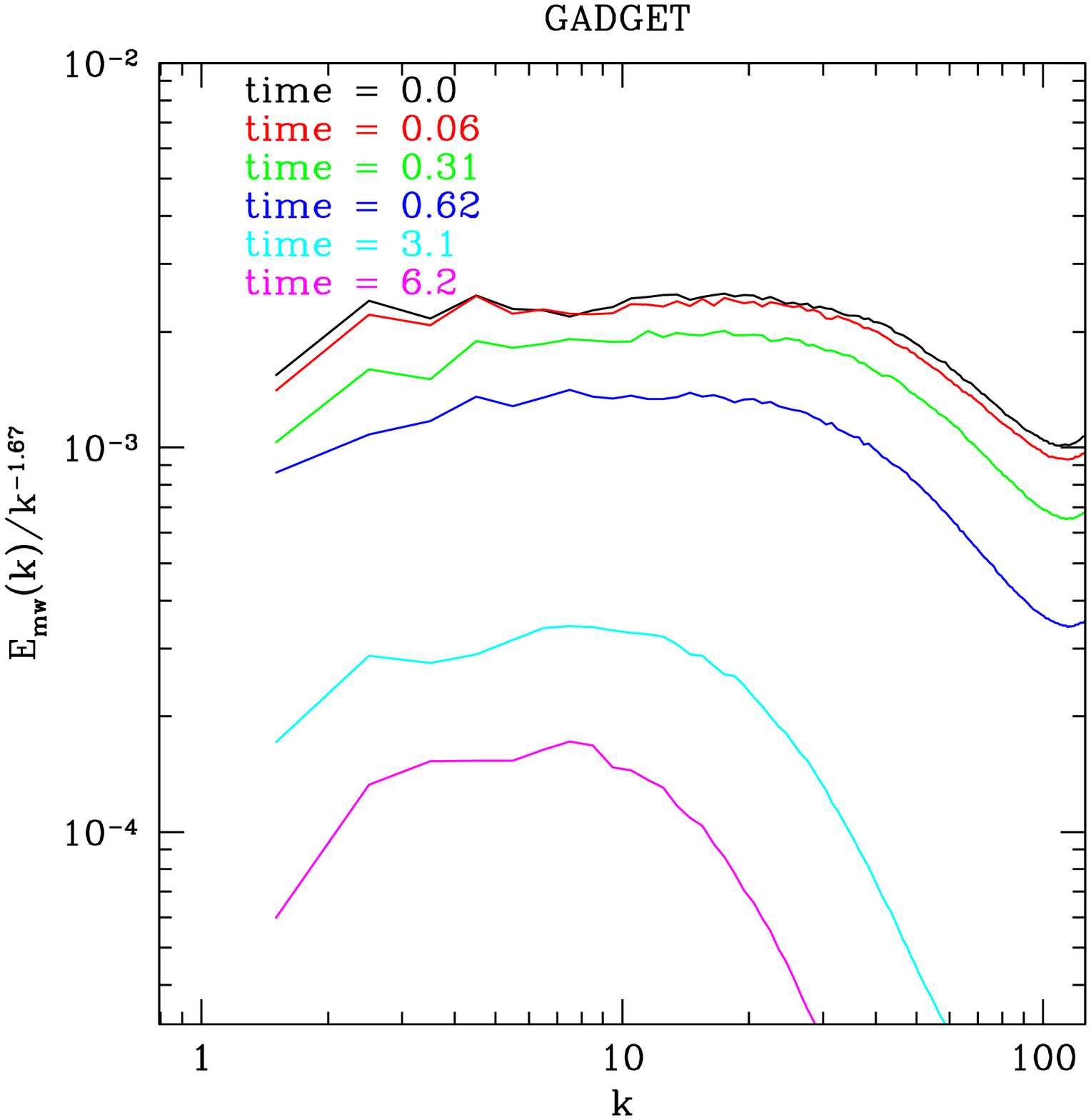} &
\includegraphics[width=0.31\linewidth]{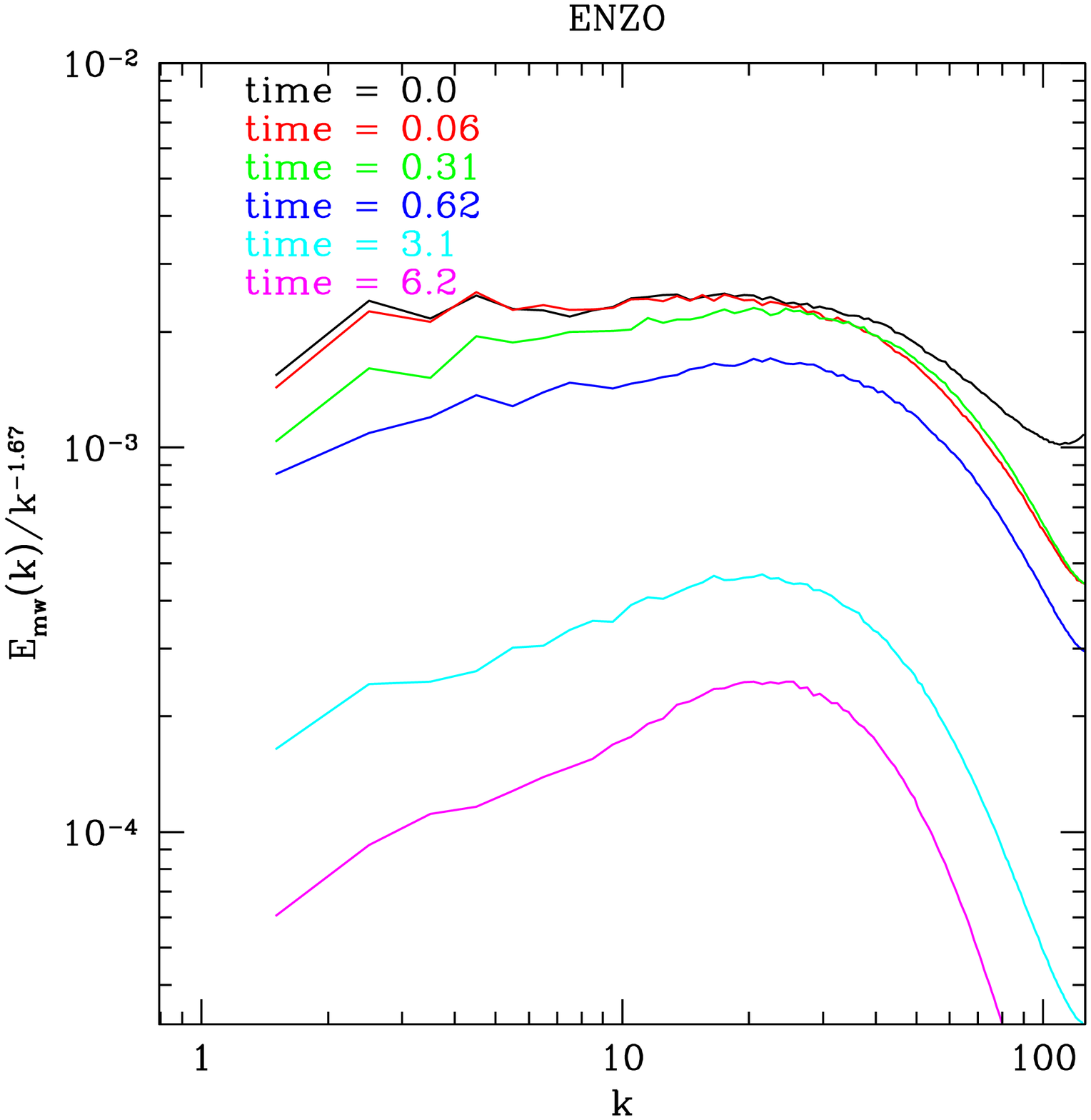} \\
\includegraphics[width=0.31\linewidth]{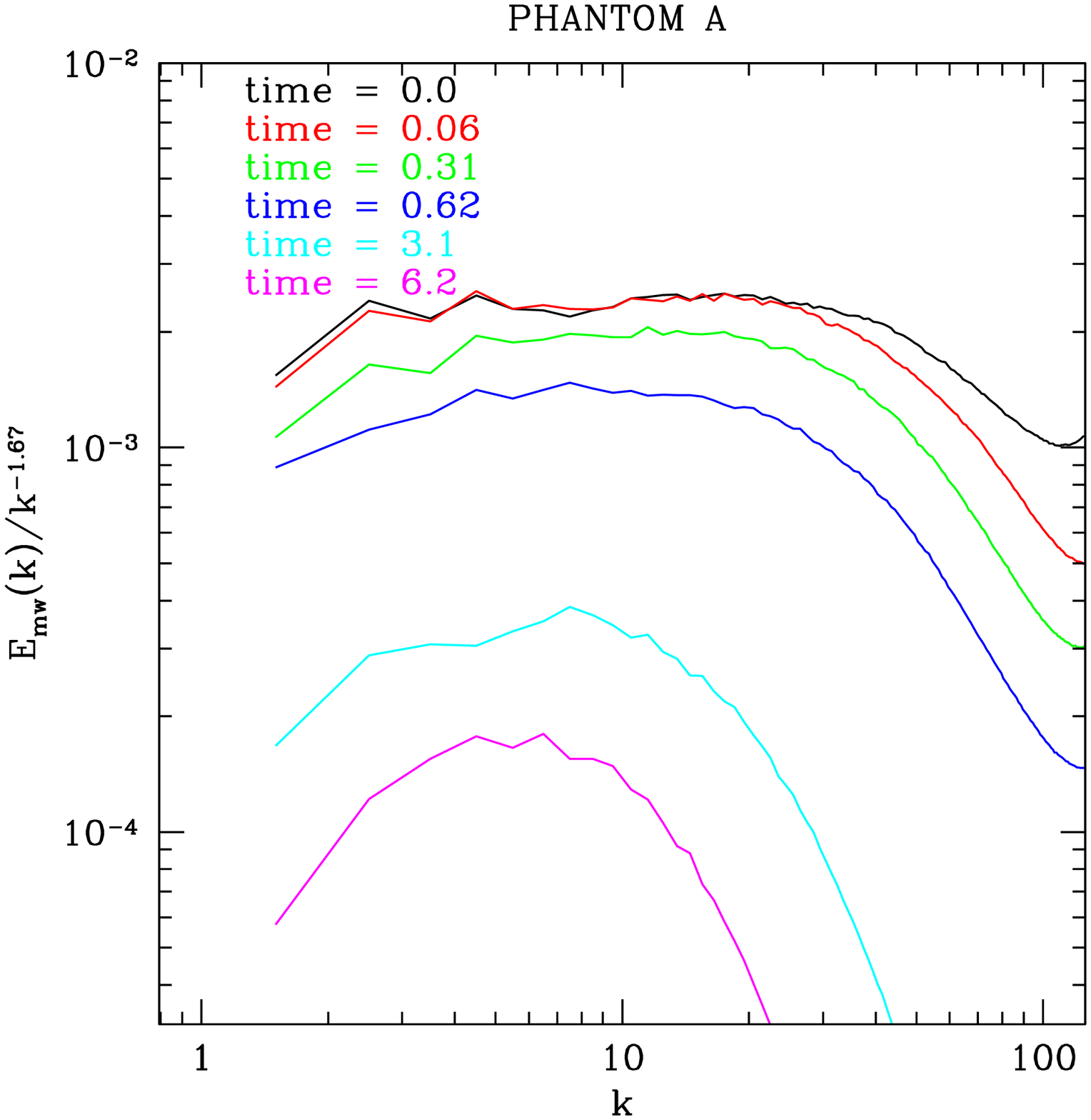} &
\includegraphics[width=0.31\linewidth]{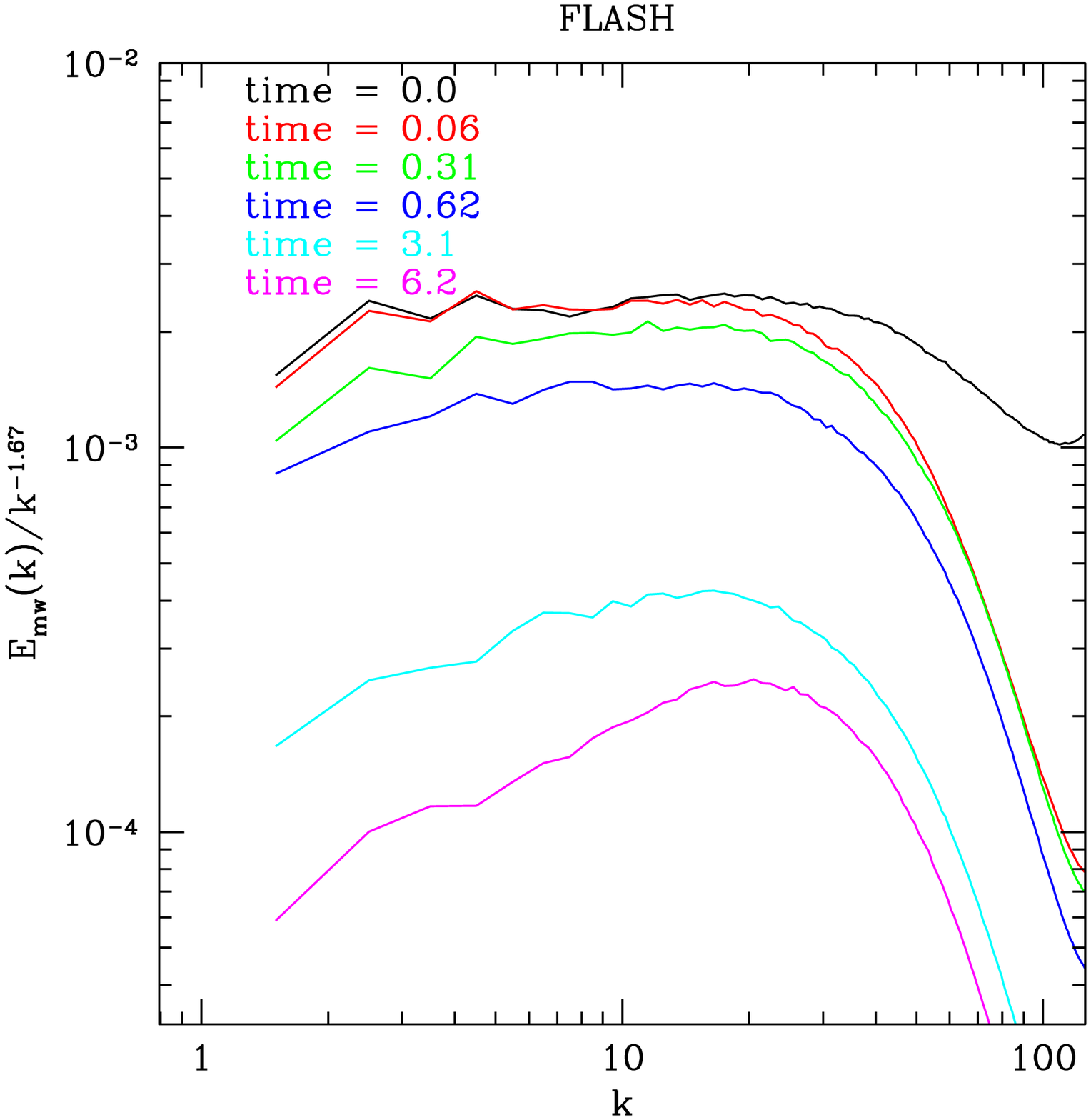} \\
\includegraphics[width=0.31\linewidth]{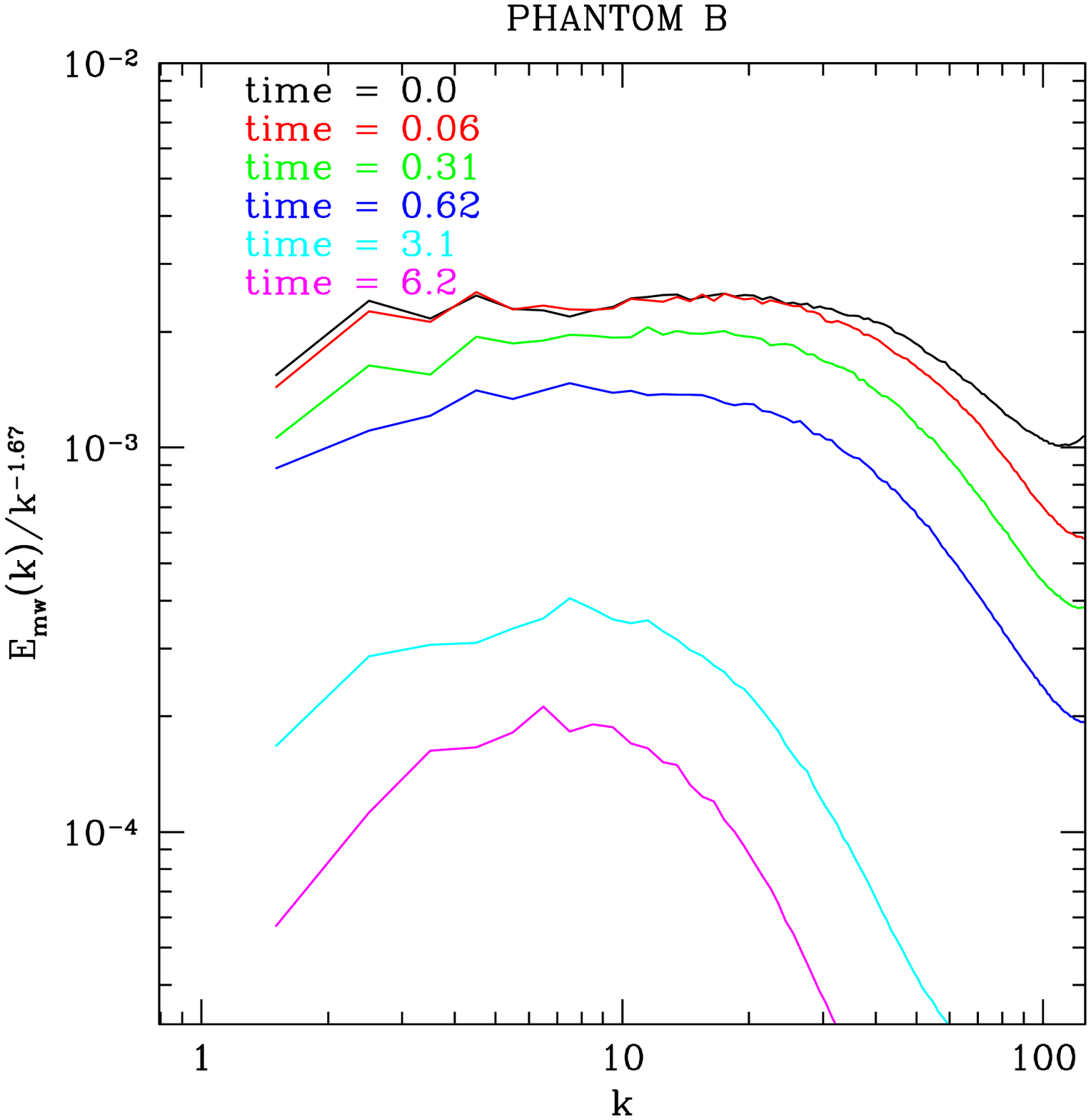} &
\includegraphics[width=0.31\linewidth]{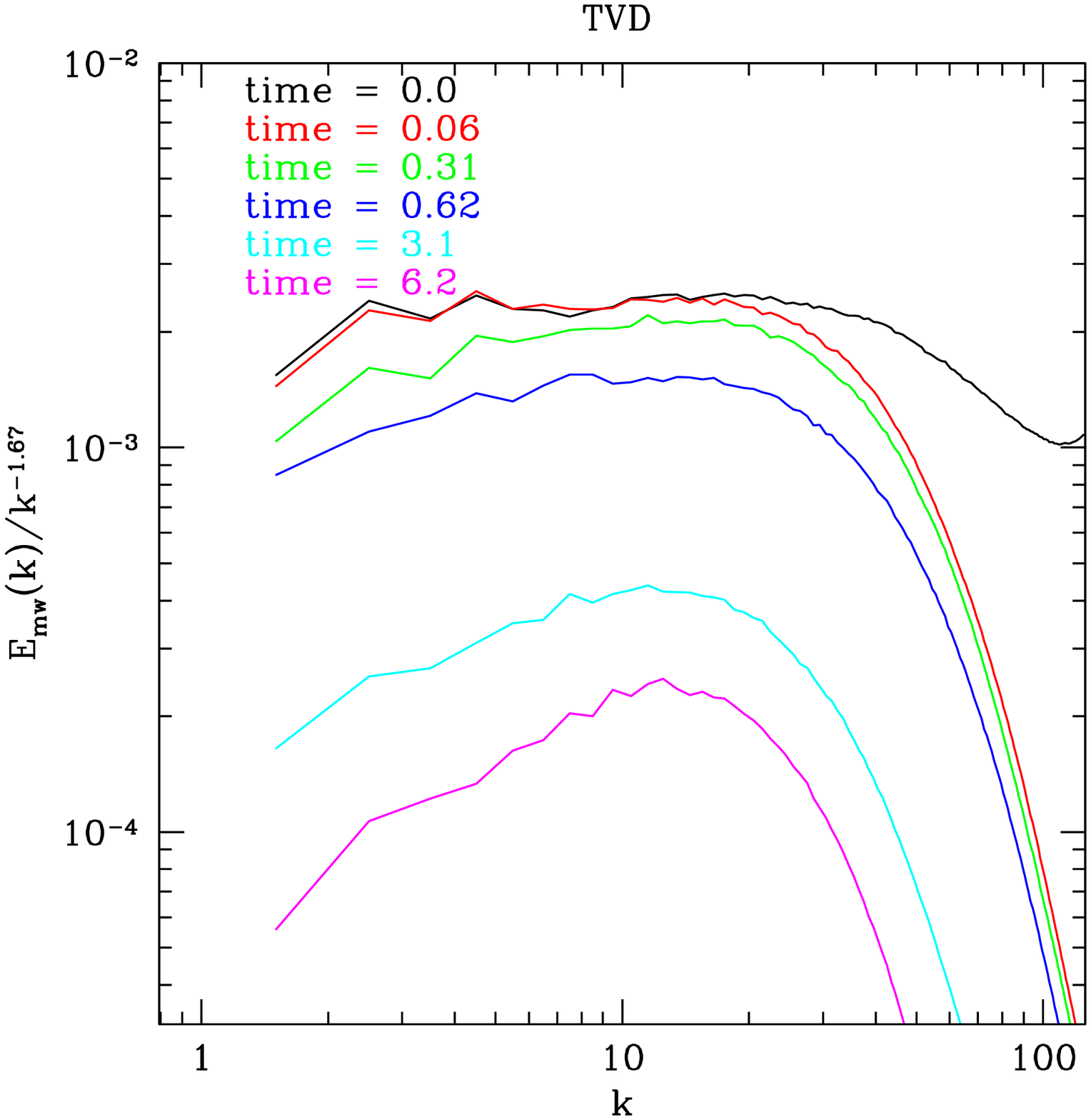} \\
\includegraphics[width=0.31\linewidth]{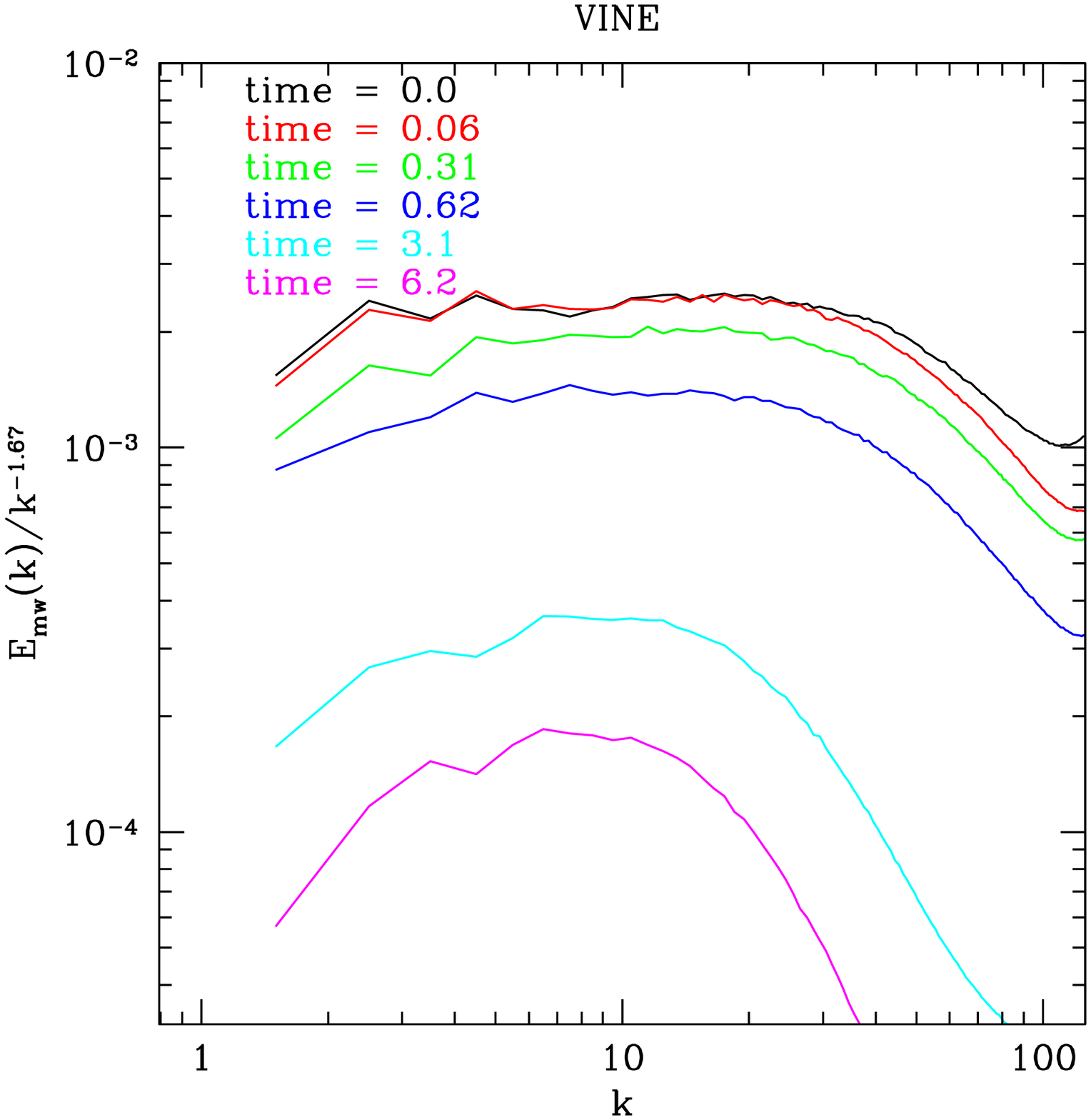} &
\includegraphics[width=0.31\linewidth]{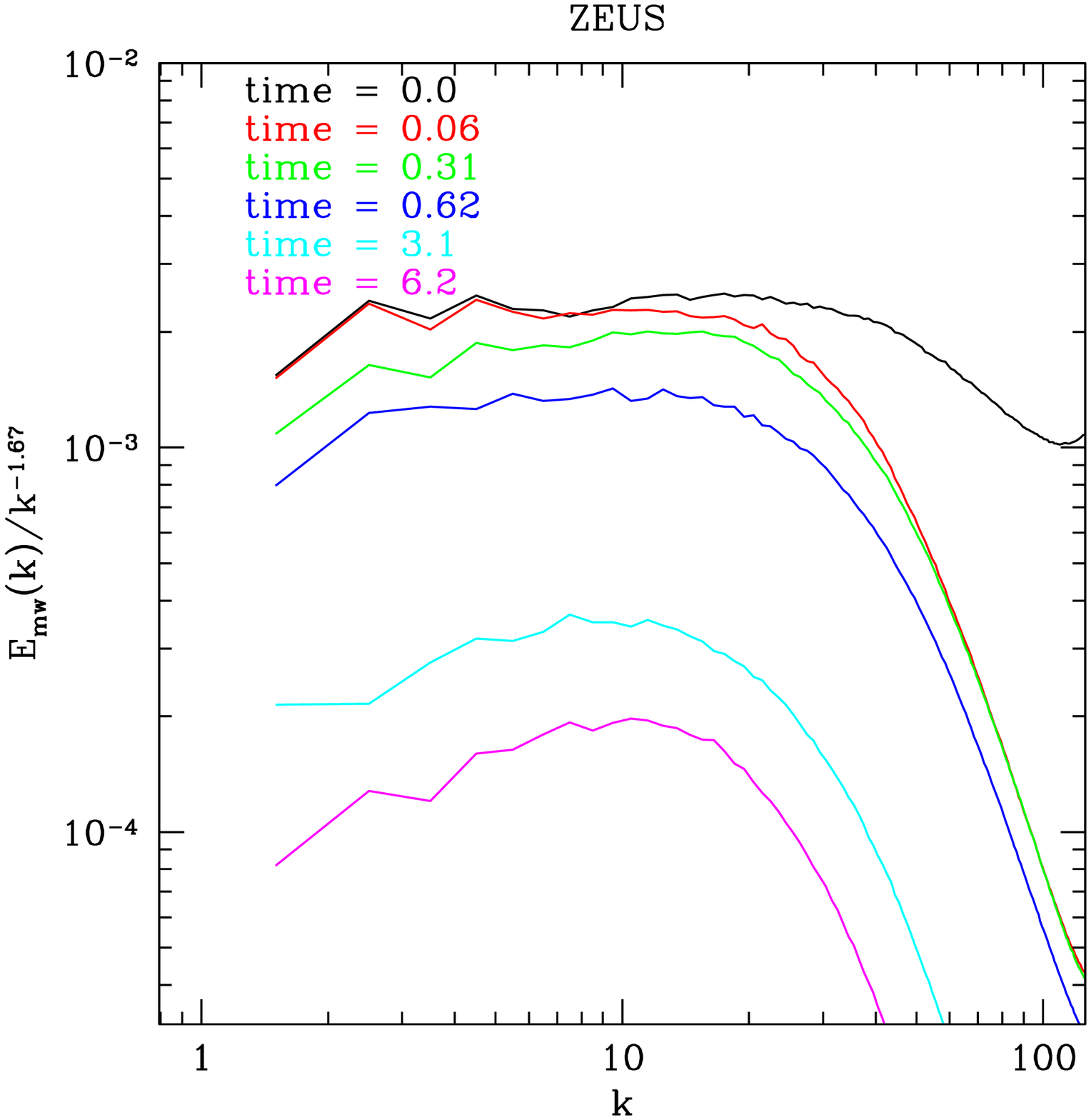}
\end{tabular}
\end{center}
\caption{Same as Fig.~\ref{fig:spect_evol_vol}, but the density-weighted velocity power spectra (compensated with power-law slopes of 1.67) are shown.}
\label{fig:spect_evol_rho}
\end{figure*}

Figures~\ref{fig:spect_evol_vol} and~\ref{fig:spect_evol_rho} present the time evolution of the volume- and density-weighted velocity power spectra obtained with the various codes employed in this work. Data from the following snapshots are shown: initial conditions at $t\,=\,0.0\,\tcross$ (black lines), $t\,=\,0.06\,\tcross$ (red lines), $t\,=\,0.31\,\tcross$ (green lines), $t\,=\,0.62\,\tcross$ (blue lines), $t\,=\,3.1\,\tcross$ (cyan lines), and $t\,=\,6.2\,\tcross$ (magenta lines). Figures~\ref{fig:spect_codes_vol} and~\ref{fig:spect_codes_rho} show the volume- and density-weighted velocity power spectra of each code plotted on top of each other in a single plot as a function of time, so that the spectra obtained for each snapshot can be directly compared across all codes for each time.

The grid codes dissipate the power produced on small spatial scales ($k \gtrsim 100$) of the initial conditions faster than the SPH codes. This is a result of the SPH interpolation onto the grid (see \S~\ref{lagrangian}). Due to their Lagrangian nature the SPH runs (\texttt{GADGET}, \texttt{PHANTOM~A}, \texttt{PHANTOM~B}, and \texttt{VINE}), maintained power at $k\gtrsim20$ for a longer time. All grid codes have lost slightly more power for $k\gtrsim20$ than the SPH codes, immediately after the decay simulations start (i.e.~from the first snapshot at $0.06\,\tcross$). The differences seen at early times on the small scales of the SPH power spectra is a result of the different methods that each of the SPH codes adopt for calculating particle smoothing lengths and/or the use of different smoothing kernels, and different implementations of artificial viscosity.

The volume-weighted velocity power spectra obtained with the SPH codes and with the grid code \texttt{ZEUS} are quite similar for $k\lesssim20$ at $t=0.06\,\tcross$, with power law slopes of about 2.2. \texttt{ENZO}, \texttt{FLASH}, and \texttt{TVD} have a slightly shallower slope of about 2.1. This slope agrees with the low-resolution models in \citet{FederrathDuvalKlessenSchmidtMacLow2009} using $256^3$ grid cells. At $t=0.31$, and $0.62\,\tcross$, also \texttt{ZEUS} develops a slightly shallower slope that roughly agrees with the slopes obtained using the other grid codes. The slopes have droped to about 1.95 and 1.9 at $t=0.31$, and $0.62\,\tcross$, respectively. The wavenumber ranges over which these slopes are maintained are slightly smaller for \texttt{ZEUS} than for \texttt{TVD} and \texttt{FLASH} (up to $k\sim12$), while \texttt{ENZO} maintains the slopes up to $k\sim18$. The SPH codes again have a slightly steeper slope by about 0.1 than the grid codes, and the wavenumber range over which this slope is maintained is comparable with the range obtained using the \texttt{ZEUS} code.

The density-weighted velocity power spectra are shallower than the volume-weighted spectra for all codes with slopes of about 1.6 at $t=0.06\,\tcross$. This much shallower slope is a result of the low resolution of our numerical experiments as discussed in \S~\ref{sec:driving_spect_rho}. Similar to the results obtained from the volume-weighted spectra, all grid codes dissipate the initial power on scales $k\gtrsim20$ faster than the SPH codes with \texttt{ZEUS} having dissipated most. However, there is an important exception to this result concerning the grid codes: the density-weighted velocity power spectrum produced by the grid code \texttt{ENZO} is almost identical to the power spectra produced with the SPH codes at $t\,=\,0.06\,\tcross$, while \texttt{FLASH}, \texttt{TVD}, and \texttt{ZEUS} have lost a considerable amount of their power at $k\gtrsim20$. The power spectrum obtained with the \texttt{ZEUS} code shows the break into the dissipation range already at $k\!\sim\!10$ and produces a slightly steeper slope of about 1.65 than all other codes. At later times ($t\,=\,0.31\,\tcross$, and $t\,=\,0.62\,\tcross$) all codes produced similar density-weighted power spectra for $k\lesssim20$, while the \texttt{ENZO} code develops a clear bottleneck \citep[see e.g.][]{DoblerEtAl2003,HaugenBrandenburg2004,SchmidtHillebrandtNiemeyer2006}, which manifests itself in the excess power seen at $k\gtrsim10$. Since \texttt{ENZO} was run here with a PPM diffusion parameter set to $K=0.0$ \citep{ColellaWoodward1984}, the bottleneck effect is quite strong \citep[see also][]{KritsukEtAl2007,FederrathDuvalKlessenSchmidtMacLow2009}. Although the \texttt{FLASH} code uses the same numerical technique as \texttt{ENZO} it does not show such a pronounced bottleneck effect, because \texttt{FLASH} was used with the recommended PPM diffusion parameter of $K=0.1$ \citep{ColellaWoodward1984}.

As the decay progresses ($t=3.1$, and $6.2\,\tcross$), power gets dissipated differently by the various codes at $k\gtrsim8$. However, on large scales ($k\lesssim8$), all codes used in the present study show very similar volume- and density-weighted velocity power spectra with slight variations that can be attributed to statistical fluctuations \citep{KritsukEtAl2007,FederrathDuvalKlessenSchmidtMacLow2009}. This is an important result, as it shows that all codes, despite having different dissipation mechanisms acting on small scales at $k\gtrsim8$, on large spatial scales the results of the decaying turbulence experiments presented here are quite robust. They do not show considerable systematic differences for the codes employed here at the resolutions studied. Moreover, it is important to note that the dissipation ranges at $k\gtrsim8$ are not just different when we compare grid codes with SPH codes, but they are also different among the SPH codes and among the grid codes. Thus, we conclude that the dissipation range is strongly dependent on the code being used, while scales with $k\lesssim8$ are similarly well reproduced by all the hydrodynamic codes employed here.

The overall performance of the codes, as seen through the analysis of their velocity power spectra at times $t\,=\,3.1\,\tcross$, and $t\,=\,6.2\,\tcross$ shows that the numerical viscosity of grid codes is generally lower than that of SPH codes, and details of the method used will determine the detailed ranking. For example, using $K=0$ in PPM, as done for the \texttt{ENZO} run, yields a lower dissipation value, although with a stronger bottleneck effect than the $K=1$ value used for the \texttt{FLASH} run.  Similarly, we find that the viscosity implementation by \citet{MorrisMonaghan1997} used in the \texttt{PHANTOM~B} and \texttt{VINE} runs is superior to that of \texttt{GADGET} and the \texttt{PHANTOM~A} run.


Considering the resolution of $N^3=256^3$ cells used in the present study ($215^3$ SPH particles interpolated to a $256^3$ grid), the fact that the velocity power spectra are different at $k\gtrsim8$ implies that one should be cautious with the interpretation of results obtained with power spectra at wavenumbers $k\gtrsim N/32$. This means that length scales smaller than about 32 grid cells for grid codes (and SPH codes using a similar number of resolution elements interpolated to a grid of equivalent size), are affected by the individual dissipation mechanisms acting in hydrodynamical codes. In contrast, the results of the various codes are robust for $k\!\lesssim\!N/32$. This is encouraging, because the results of all the hydrodynamical codes used here agree well in this regime, and one is free to choose a code for modelling supersonic turbulence as long as only results for scales $k\!\lesssim\!N/32$ are considered. However, this also means that one needs resolutions of at least $1024^3$ grid cells to obtain roughly one full decade in length scale over which a power law could be fitted to turbulent velocity spectra. In practice, this range turns out to be even smaller than one decade in length scales at a resolution of $1024^3$ grid cells \citep{KritsukEtAl2007,KleinEtAl2007,FederrathDuvalKlessenSchmidtMacLow2009}.

\subsection{Kinetic energy dissipation rates}
Tables~\ref{tab:e_vw} and~\ref{tab:e_mw} list the integrals over the volume- and density-weighted velocity spectra, respectively for each code and run at each of the snapshots presented here. Up to $t=0.31\,\tcross$, all SPH codes dissipate volume-weighted power slightly faster than the grid codes, while for the density-weighted power, both SPH and grid codes dissipate kinetic energy at roughly the same rate. At $t=0.62\,\tcross$, however, \texttt{ZEUS} has dissipated about 15\% more power in velocity fluctuations than the other codes, while all other grid codes have still dissipated less than the SPH codes. The density-weighted integral gives more similar results for all codes at all times analysed. At $t=3.1$, and $t=6.2\,\tcross$, all codes have roughly dissipated the same amount of volume- and density-weighted power, except for the \texttt{PHANTOM~B} run having kinetic energies about 16\% larger than all other codes.

\subsection{Time evolution of the RMS Mach number}
The time evolution of the RMS Mach number is shown in Figure~\ref{fig:mach_evol}. The dotted line shows the expected power-law decay rate $M_\mathrm{rms}\propto t^{-1/2}$ for supersonic turbulence \citep{MacLowEtAl1998,StoneOstrikerGammie1998,MacLow1999}, starting at an RMS Mach number of $M_\mathrm{rms}\sim3.9$:
\begin{equation}
M_\mathrm{rms}(t) = 3.9\,\left(\frac{t}{\tcross}+1\right)^{-1/2}\,.
\end{equation}
Clearly, the RMS turbulent flow remains supersonic (i.e. $M_\mathrm{rms}>1$) for all times analysed in the present study. However, turbulent velocity fluctuations become smaller on smaller scales. The transition scale separating supersonic motions on large scales and subsonic motions on small scales is called the sonic scale $k_s$ \citep[e.g.,][]{VazquezBallesterosKlessen2003,FederrathDuvalKlessenSchmidtMacLow2009}. We computed an estimate of the sonic scale using the definition of equation~(\ref{eq:sonic_scale}).

\subsection{Time evolution of the sonic scale}
Table~\ref{tab:sonic_scale} lists the evolution of the sonic scale for all codes/runs and all snapshots. The sonic scale decreases fastest for \texttt{PHANTOM~A}, \texttt{PHANTOM~B} and \texttt{ZEUS}, and slowest for \texttt{ENZO}. \texttt{GADGET}, \texttt{VINE}, \texttt{FLASH} and \texttt{TVD} show quite similar results for the evolution of the sonic scale. The sonic scale does not differ considerably among the various codes at later times ($t=3.1\,\tcross$ and $t=6.2\,\tcross$). However, during the initial stages of the decay, there are differences up to 30\%. This is partly a result of our computation of the sonic scale (cf.~\S~\ref{sec:method_spectra_vw}). Since the different codes have quite different dissipation properties, the sonic scale is affected accordingly (cf.~\S~\ref{spectra}). Also note that the sonic scale is given as an integer. Thus, the fact that, for instance, \texttt{ENZO} maintains $k_\mathrm{s}=10$ until $t=0.62\,\tcross$ does not imply that its sonic scale stays exactly the same for all times $t<0.62\,\tcross$, but will have also decreased slightly. However, our grid of wavenumbers is binned such that only integer values of $k$ are permitted, and thus, rounding errors introduce uncertainties of about 10\% in the $k_\mathrm{s}$-values at early times of the decay ($t\lesssim1\tcross$).

\subsection{Probability distribution functions}

\subsubsection{Probability distribution functions of the gas density}
Figure~\ref{fig:densityPDFs} shows volume-weighted probability distribution functions (PDFs) of the gas density. Each panel shows the comparison of the density PDFs for all codes at $t\,=\,0.0$, $0.06$, $0.31$, $0.62$, $3.1$, and $6.2\;\tcross$ after they have been interpolated to grids of $256^3$ cells. The density PDFs were computed from the logarithm of the density $s=\ln{(\rho/\rho_0)}$. The PDF $p(s)$ is expected to follow roughly a Gaussian distribution
\begin{equation}
p(s)\,\mathrm{d}s=\frac{1}{\sqrt{2\pi\sigma_s^2}}\,\exp\left[-\frac{(s-s_0)^2}{2\sigma_s^2}\right]\,\mathrm{d}s \label{eq:log-normal}
\end{equation}
where $\sigma_s$ is the logarithmic density dispersion and $s_0$ is the mean value of $s$ \citep[e.g.,][]{Vazquez1994,PadoanNordlundJones1997,StoneOstrikerGammie1998,MacLow1999,NordlundPadoan1999,OstrikerGammieStone1999,Klessen2000,OstrikerStoneGammie2001,BoldyrevNordlundPadoan2002,LiKlessenMacLow2003,PadoanJimenezNordlundBoldyrev2004,GloverMacLow2007b,KritsukEtAl2007,BeetzEtAl2008,FederrathKlessenSchmidt2008,LemasterStone2008,SchmidtEtAl2009,FederrathDuvalKlessenSchmidtMacLow2009,GloverFederrathMacLowKlessen2009}.

The density PDFs of all codes show little variation around the peak of the distribution and at the high-density tail, and they are all roughly consistent with log-normal distributions, equation~(\ref{eq:log-normal}). The low density tails show stronger variations. This is because the low density tail is subject to stronger temporal variations caused by intermittent fluctuations \citep{KritsukEtAl2007,FederrathDuvalKlessenSchmidtMacLow2009}. In Table~\ref{tab:moments} we list the values of $s_0$ and $\sigma_s$ for each code. Note that for a log-normal volume-weighted density distribution, $s_0=-\sigma_s^2/2$. The means and standard deviations of the PDFs are similar for all codes and vary only by about 10\%, except for our \texttt{ZEUS} run at $t=0.62\,\tcross$, which has a mean value $|s_0|$ about 28\% larger than the average over all runs at that time. This appears slightly too high a variation to be attributed to temporal fluctuations. The difference of the density PDF around its peak obtained with the \texttt{ZEUS} run at $t=0.62\,\tcross$ is also visible in Figure~\ref{fig:densityPDFs} (bottom left panel). However, at later times, the \texttt{ZEUS} density PDFs are almost identical to the ones obtained with the other codes.

One would expect that SPH codes can resolve high-density regions better than grid codes. This is because the effective SPH resolution increases with increasing density (cf.~\S~\ref{lagrangian}). On the other hand, low-density gas will become less resolved as particles move towards higher density regions and away from low-density regions. However, the density PDFs of grid codes and SPH codes agree very well after interpolation of the particle data to grids. This may be caused by our interpolation procedure. We therefore would like to test the density PDF obtained from the SPH particle density directly without interpolation to a grid. The density PDF of SPH particles is naturally mass-weighted. Therefore, in order to obtain the volume-weighted density PDF directly from the particles we must weight the contribution of each particle into a density bin by the inverse of its density (for equal mass particles). Note that in general, volume-weighted PDFs, $p_v$, are related to mass-weighted PDFs, $p_m$, by $p_v = p_m/\rho$ \citep[e.g.][]{OstrikerStoneGammie2001,LiKlessenMacLow2003}. This conversion is necessary as particles and grid cells do not sample the computational volume in the same manner, i.e.~the particles are concentrated in high-density regions, whereas the grid cells sample the volume homogeneously. The same applies for the distribution of Lagrangian tracer particles used in grid simulations \citep{FederrathGloverKlessenSchmidt2008,PriceFederrath2009}. Therefore, we must compensate the particle PDF for the fact that most particles are located in high-density regions.

Figure~\ref{fig:densityPDFsInterpolation} shows the density PDF of the \texttt{GADGET} run as obtained both after interpolation to a grid (black line), and directly obtained from the SPH particle density (red line). Additionally, we show the SPH density PDF obtained from the \texttt{VINE} run for comparison. The SPH density PDFs for \texttt{GADGET} and \texttt{VINE} were transformed into volume-weighted PDFs using $p_v = p_m/\rho$ to allow for a direct comparison of the grid and the SPH density PDFs. The grid-interpolated and the SPH density PDFs show no significant differences: the grid-interpolated PDF exhibits less scatter in the low-density regime than the SPH particle PDF, indicating that the grid-interpolated PDF samples the low-density regime slightly better than the particle-based PDF. On the other hand, the particle-based density PDF shows better sampling of the high-density tail and extends to slightly higher densities and lower probability densities. This is to be expected because the effective resolution is larger for the particle-based distribution at high densities, and SPH density peaks are smoothed to slightly smaller densities by the interpolation technique, equation~(\ref{eq:gridinterpolation}). This analysis offers an additional illustration of the fact that the adaptivity of the SPH method is suppressed when the SPH information is interpolated onto a grid of resolution smaller than the effective SPH resolution (cf.~\S~\ref{lagrangian}). 

It is an encouraging result that all codes are able to reproduce the general form of the density PDF quite well. This is important, because the turbulent density PDF is an essential ingredient to many models of star formation: to understand the mass distribution of cores in molecular clouds and stars \citep{PadoanNordlund2002,HennebelleChabrier2008,HennebelleChabrier2009}, the star formation rate \citep{KrumholzMcKee2005,PadoanNordlund2009}, the star formation efficiency \citep{Elmegreen2008}, and the Kennicutt-Schmidt relation on galactic scales \citep{Elmegreen2002,Kravtsov2003,Tassis2007}.

\subsubsection{Probability distribution functions of the rate of strain}
Figure~\ref{fig:strainPDFs} presents the trace free rate of strain PDFs computed as explained in \S~\ref{sec:method_PDFs}. Each panel shows the comparison of the PDFs of all codes at $t\,=\,0.0$, $0.06$, $0.31$, $0.62$, $3.1$, and $6.2\;\tcross$. Within the first $\tcross$ (top row and bottom-left panel of Fig.~\ref{fig:strainPDFs}), all SPH codes and \texttt{ZEUS} have PDFs that are slightly narrower than those of the remaining grid codes (top-right and bottom-left panels of Fig.~\ref{fig:strainPDFs}). At later times (bottom-middle and bottom-right panels of Fig.~\ref{fig:strainPDFs}), all codes appear to have distributions with similar widths, but with the PDFs of the SPH codes and the \texttt{ZEUS} code peaking at almost half the $|S^*|$ value of the other grid codes. In particular, \texttt{GADGET}, \texttt{PHANTOM~A}, \texttt{PHANTOM~B}, \texttt{VINE} and \texttt{ZEUS} have their peaks at $\log{|S^*|} \lesssim 0$, while \texttt{ENZO} and \texttt{FLASH} have their peaks at $\log{|S^*|} \gtrsim 0$. \texttt{TVD} appears to peak in between. At $t\,=\,6.2\,\tcross$ the velocity power spectra of the codes are maintained up to $k$-values that can be ordered as follows (from higher to lower values): \texttt{ENZO}, \texttt{FLASH}, \texttt{TVD}, \texttt{ZEUS}, \texttt{VINE}, \texttt{PHANTOM~B}, \texttt{GADGET}, \texttt{PHANTOM~A} (cf.~the bottom-right panels of Fig.~\ref{fig:spect_codes_vol} and~\ref{fig:spect_codes_rho} in \S~\ref{spectra}). This is the order of the $|S^*|$-value of the peaks of the rate of strain PDFs at this time (bottom-right panel of Fig.~\ref{fig:strainPDFs}).

\subsubsection{Probability distribution functions of the vorticity}
In Figure~\ref{fig:vorticityPDFs}, we present the vorticity PDFs. The vorticity was computed as explained in \S~\ref{sec:method_PDFs}. Each panel shows the comparison of the vorticity PDFs of all codes at $t\,=\,0.0$, $0.06$, $0.31$, $0.62$, $3.1$, and $6.2\;\tcross$. The vorticity and the trace free rate of strain PDFs show a similar behaviour: within the first $\tcross$ (top row and bottom-left panel of Fig.~\ref{fig:vorticityPDFs}), the SPH codes and \texttt{ZEUS} show narrower distributions than the remaining grid codes. At later times (bottom-middle and bottom-right panels of Fig.~\ref{fig:vorticityPDFs}), the vorticity distributions of the SPH codes and \texttt{ZEUS} peak at almost half the vorticity value of the other grid codes. Again, the peaks of \texttt{PHANTOM~B}, \texttt{VINE}, \texttt{TVD}, and \texttt{ZEUS} are bracketed by the peaks of the remaining codes, with \texttt{GADGET} and \texttt{PHANTOM~A} peaking at the lowest values of the vorticity, and \texttt{ENZO} and \texttt{FLASH} peaking at the highest vorticity.

Since the vorticity is related to the ability of the codes to model turbulent eddies, these comparisons indicate that SPH codes exhaust their ability earlier than grid codes, most likely because of their excess viscosity acting on the smallest of these eddies and erasing them. As in the case of the rate of strain above, the order of the vorticity values of the peaks of the vorticity PDFs is the same as the order with which the codes maintain their velocity power spectra in the high-$k$ regime (cf.~the bottom right panel of Fig.~\ref{fig:spect_codes_vol} and~\ref{fig:spect_codes_rho}, and the bottom-right panel of Fig.~\ref{fig:strainPDFs}).

\subsubsection{Probability distribution functions of the velocity}
Figure~\ref{fig:velocityPDFs} shows the PDFs of the velocity component $v_{\rm z}$. Each panel shows the comparison of the PDFs of all codes at times $t\,=\,0.0$, $0.06$, $0.31$, $0.62$, $3.1$, $6.2\,\tcross$. For all these snapshots, the velocity PDFs are very similar for all codes. Apart from statistical fluctuations showing up in the wings of the distributions, the velocity PDFs are roughly Gaussian distributions. As the standard deviation of the density PDFs (cf.~Fig.~\ref{fig:densityPDFs}), the standard deviation of the velocity PDFs decreases with time, which simply reflects the turbulence decay (cf.~Fig.~\ref{fig:mach_evol}).

\section{Computational efficiency of the codes} \label{sec:efficiency}
Table~\ref{tab:code_efficiency} provides a summary of the computational efficiency of our codes/runs for the present setup. We remind the reader that in the current study all SPH codes used $215^3$ resolution elements, while the grid codes used $256^3$ resolution elements. We compensated for this discrepancy in the number of resolution elements, as well as compensated roughly for the different CPU clock rates in the last row of Table~\ref{tab:code_efficiency}. However, the various runs were performed on different parallel machines throughout the world, some with, others without optimisations. Furthermore, different supercomputers use different hardware solutions for the parallelisation. Thus, the given numbers should only be taken as a rough estimate that may be accurate to within factors of a few. However, we emphasise that we have performed both the \texttt{GADGET} and the \texttt{ENZO} run on exactly the same supercomputing platform with the same optimisations, such that we can compare the performance of these two runs directly.

The fastest of all runs was performed with the \texttt{TVD} grid code. It is roughly ten times faster than our \texttt{ENZO} run, about 15 times faster than \texttt{FLASH}, and about 27 times faster than the version of \texttt{ZEUS} used here. However, the official \texttt{ZEUS-MP} version is expected to be faster and to scale better. Due to its specific design and implementation, \texttt{PHANTOM} is the fastest of all the SPH codes employed, but still roughly 50\% slower than the slowest grid code (our version of \texttt{ZEUS}). \texttt{GADGET} and \texttt{VINE} are about 16 times, and about 30 times, respectively slower than \texttt{PHANTOM}. However, it must be remembered that the extra cost in SPH reflects the fact that resolution elements are placed to follow the mass, and thus preferentially to resolve high-density regions. Thus, additional information is calculated on small scales. This information however does not enter this comparison as the particles are interpolated onto a fixed grid. The extra cost for the SPH runs is similar to what can occur with grid-based Adaptive Mesh Refinement (AMR) \citep[e.g.][]{BergerColella1989}, because of the additional overhead to store and iterate over the AMR hierarchy \citep[see e.g.][]{SchmidtEtAl2009}. However, a quantitative analysis of the performance of AMR versus SPH is beyond the scope of this paper, and should be discussed elsewhere.

The fact that \texttt{GADGET} and \texttt{VINE} are more than one order of magnitude slower than any of the grid codes makes it computationally expensive to study supersonic turbulence at resolutions higher than about $256^3$ SPH particles. This may partly explain why no SPH calculations of supersonic turbulence have so far been using more than $512^3$ particles. The latter was only achieved with the \texttt{PHANTOM} code in \citet{PriceFederrath2009} and in the KITP code comparison project\footnote{\texttt{http://kitpstarformation07.wikispaces.com/ Star+Formation+Test+Problems}}.

\section{Conclusions} \label{sec:conclusions}
In this paper we report the comparison of the performance of four grid-based and three particle-based hydrodynamic codes on the modelling of supersonic turbulence decay. In particular, we have studied the decay of compressible, supersonic, isothermal turbulence in the absence of gravity within a periodic box using simulations with resolutions of $256^3$ grid cells for the grid codes, and $215^3$ particles for the SPH codes.

We have simulated driven turbulence with the SPH code \texttt{GADGET}. The SPH particle distribution at the end of the driving has been interpolated onto a grid that provides the initial conditions to the grid codes employed, namely \texttt{ENZO}, \texttt{FLASH}, \texttt{TVD} and \texttt{ZEUS}. We have also followed the decay of turbulence using several implementations of the SPH method, namely \texttt{GADGET}, \texttt{PHANTOM} (runs: \texttt{A}, \texttt{B}), and \texttt{VINE}, where \texttt{PHANTOM~B} used the \citet{MorrisMonaghan1997} viscosity switch, while \texttt{PHANTOM~A} was run without the switch.

The turbulence decay was followed for about six turbulent crossing times. During the whole decay phase considered here, the turbulent flow stays in the supersonic regime. The turbulent energy dissipation was measured for all codes. For times greater than about 3 crossing times (when any initial transient phase due to small variations in the initial conditions has disappeared, and all codes have developed their individual dissipation signatures) a comparison of volume- and density-weighted velocity spectra indicates that the numerical viscosity of grid codes is generally lower than that of SPH codes, with details of the method like the order of the code contributing secondarily.
We show that the differences between \texttt{ENZO} and \texttt{FLASH} are due to our choice of using different PPM diffusion parameters for the two codes, i.e., in this study \texttt{ENZO} was used with a PPM diffusion parameter of $K=0.0$, while \texttt{FLASH} was used with the PPM diffusion parameter set to $K=0.1$ as recommended by \citet{ColellaWoodward1984}. Switching-off PPM diffusion completely (as in our \texttt{ENZO} run) results in less dissipation, but produces a stronger bottleneck effect \citep[see also][]{KritsukEtAl2007,FederrathDuvalKlessenSchmidtMacLow2009}. Use of the \citet{MorrisMonaghan1997} viscosity implementation for SPH provides less dissipation as observed in our \texttt{PHANTOM~B} and \texttt{VINE} runs in comparison with the \texttt{GADGET} and \texttt{PHANTOM~A} runs. In general, the viscosity acts differently for the different grid- and SPH-codes at wavenumbers $k\gtrsim8$ at the resolutions studied here, shown by our analysis of velocity power spectra in \S~\ref{spectra}. However, all codes produced velocity spectra that are in good agreement for $k\lesssim8$.

Using Fourier spectra we also showed that the additional information that the SPH method can offer in high-density regions and/or on small scales will be suppressed if it is not interpolated onto a high enough resolution grid, as discussed in \S~\ref{lagrangian} and \S~\ref{spectra}.

The trace-free rate of strain and the vorticity PDFs confirm the ordering of the runs according to their dissipation given above. The density PDFs are very similar for all the runs performed in the present study. The means and standard deviations of the logarithmic density varied by less than 10\% for all codes at all times analysed, with one exception (the mean logarithmic density obtained in our \texttt{ZEUS} run at $t=0.62\,\tcross$ varied by about 30\%, which is also seen in its density PDF). For the SPH code \texttt{GADGET} we have shown that the density PDF obtained from the SPH particle distribution samples the high density tail slightly better, while our results indicate that using a grid, the low-density tail is slightly better sampled. However, the overall shape is very close to a log-normal distribution in density, and its mean and standard deviation are quite robust for all codes employed in the present study.

\begin{figure*}
\begin{center}
\begin{tabular}{ccc}
\includegraphics[width=0.29\linewidth]{figures/power_000_comp_vol.ps} &
\includegraphics[width=0.29\linewidth]{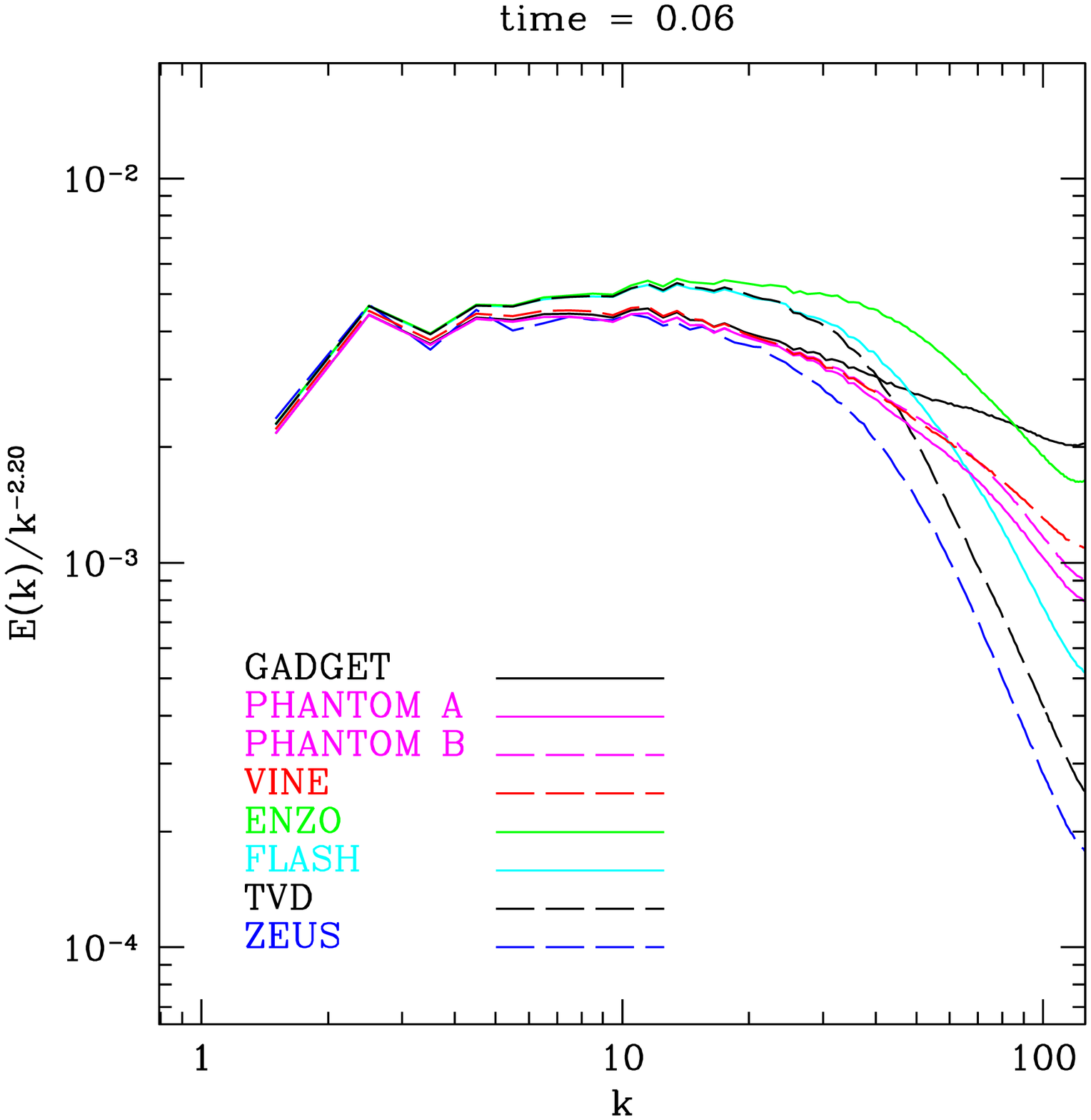} &
\includegraphics[width=0.29\linewidth]{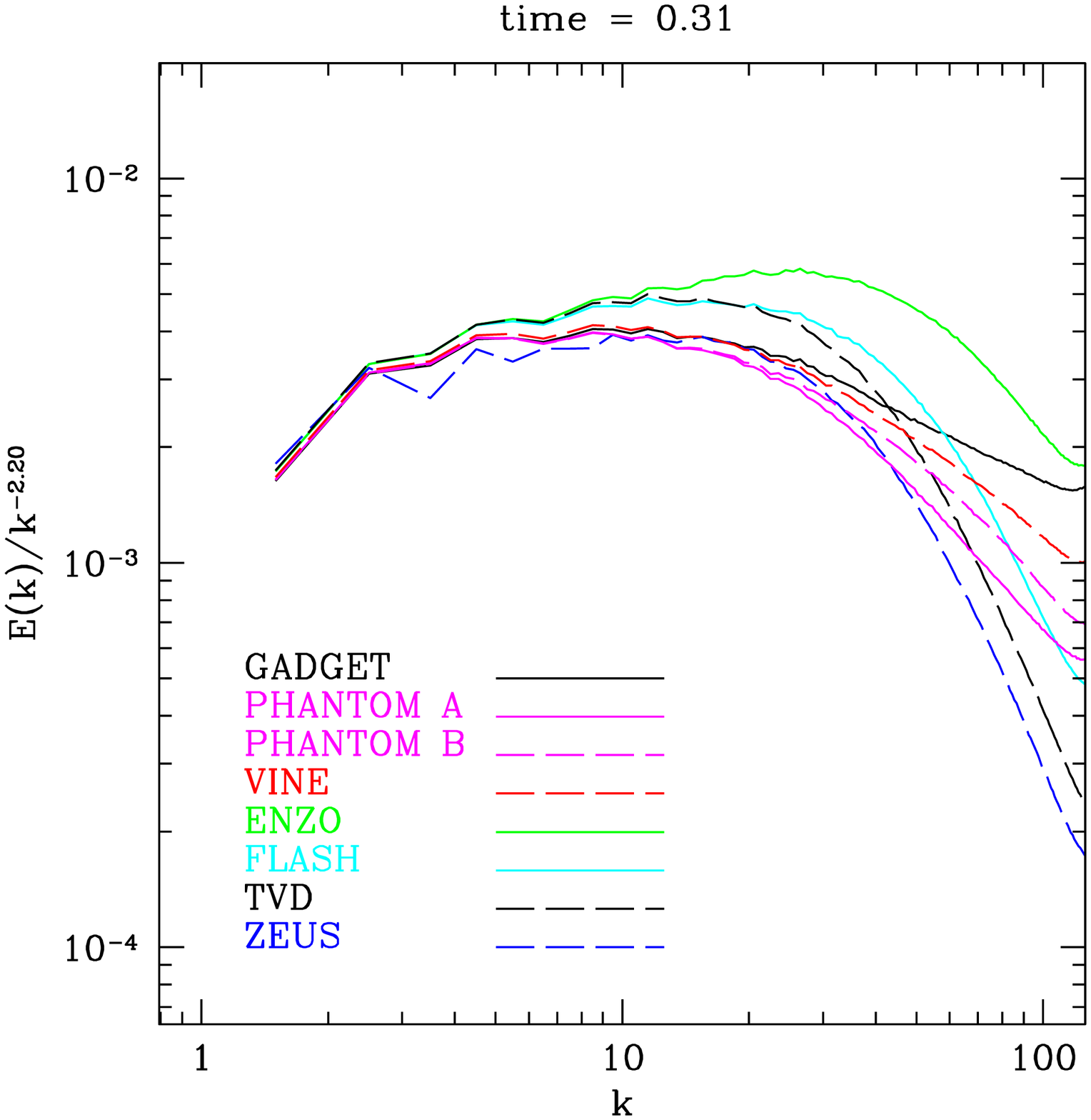} \\
\includegraphics[width=0.29\linewidth]{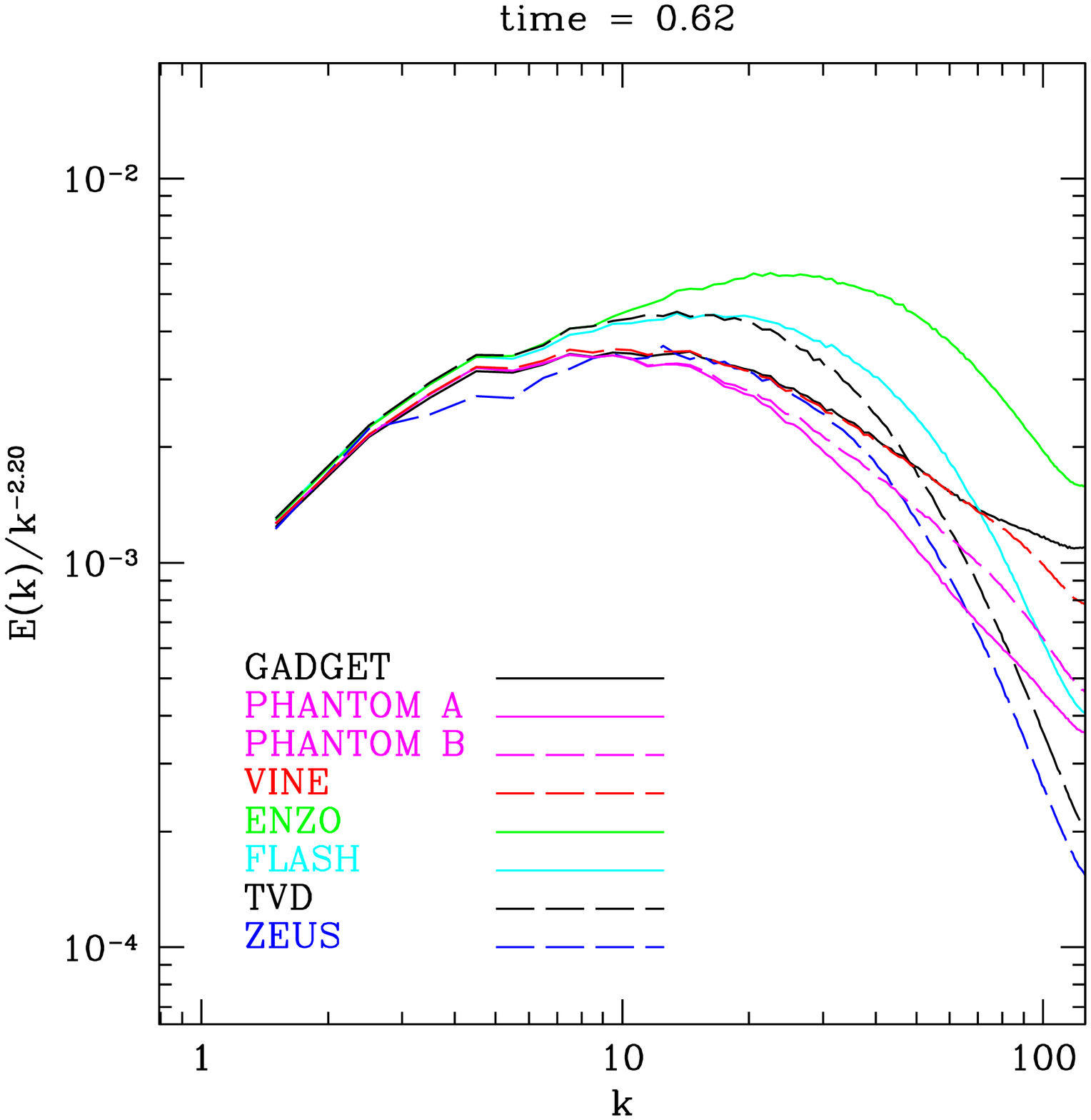} &
\includegraphics[width=0.29\linewidth]{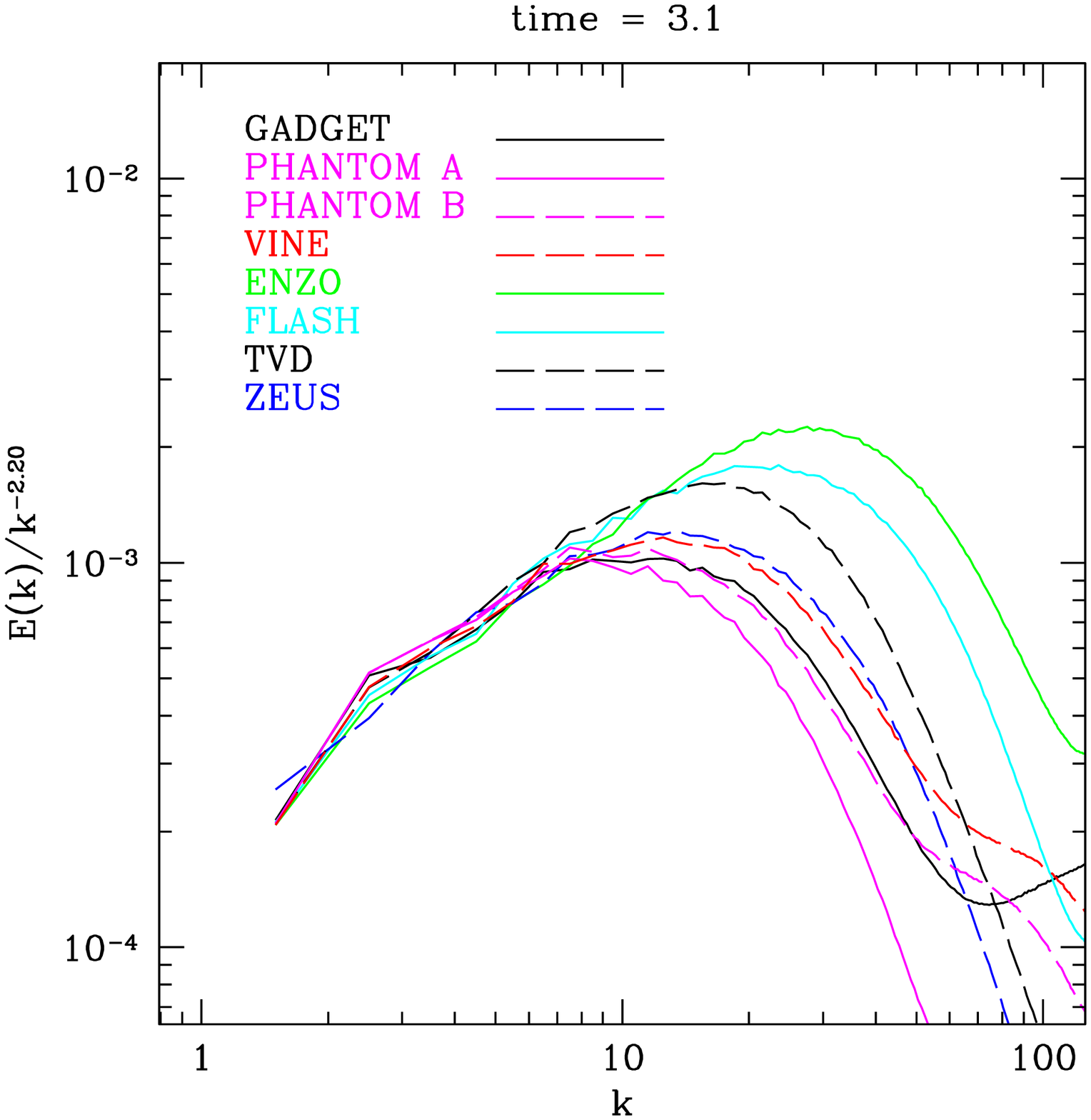} &
\includegraphics[width=0.29\linewidth]{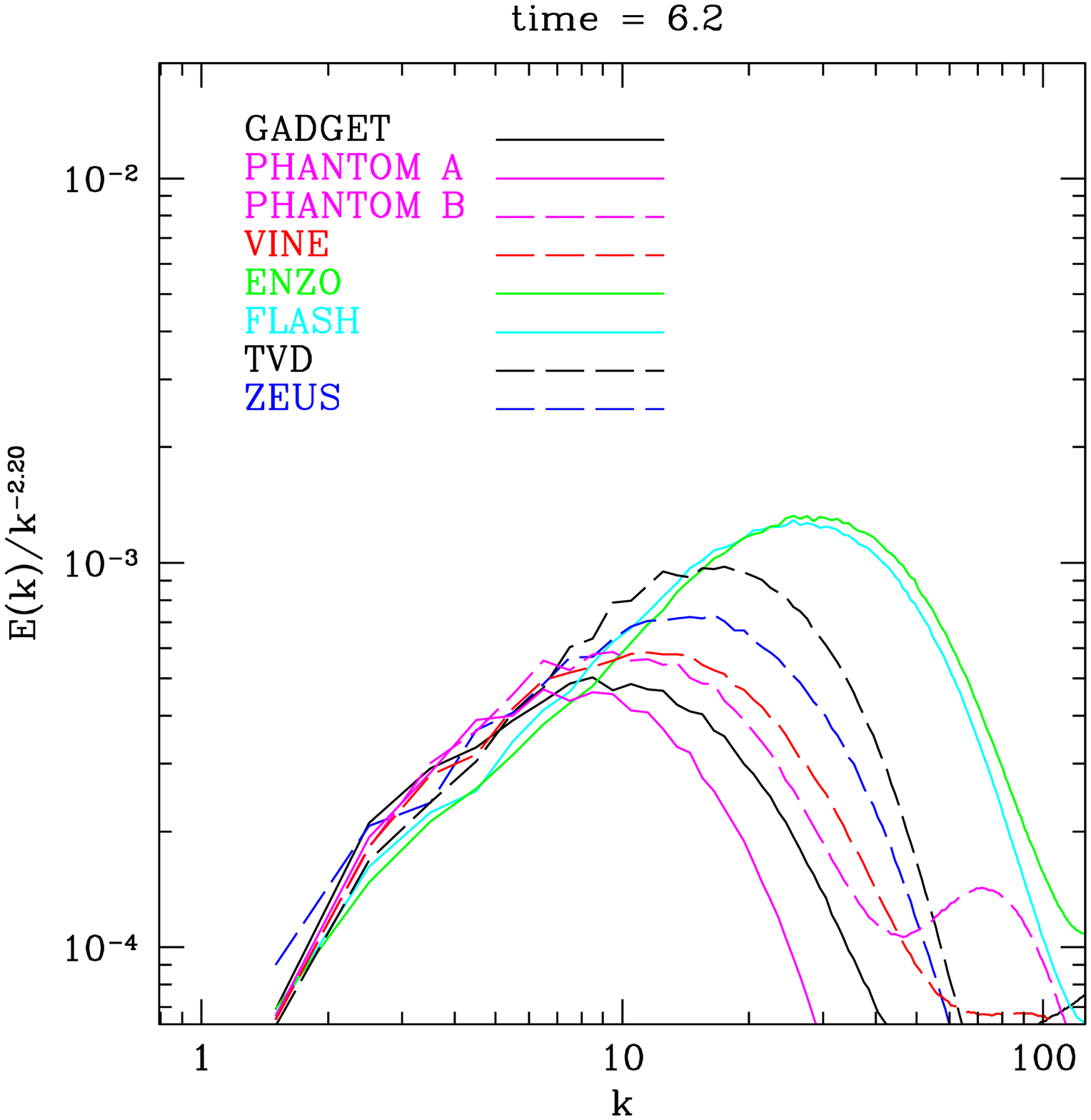}
\end{tabular}
\end{center}
\caption{Comparison of the volume-weighted velocity power spectra (compensated with power-law slopes of 2.20) of all codes at 
different times along the decay: $t\,=\,0.0$, $0.06$, $0.31$, $0.62$, $3.1$, and $6.2\;\tcross$.}
\label{fig:spect_codes_vol}
\end{figure*}

\begin{figure*}
\begin{center}
\begin{tabular}{ccc}
\includegraphics[width=0.29\linewidth]{figures/power_000_comp_rho.ps} &
\includegraphics[width=0.29\linewidth]{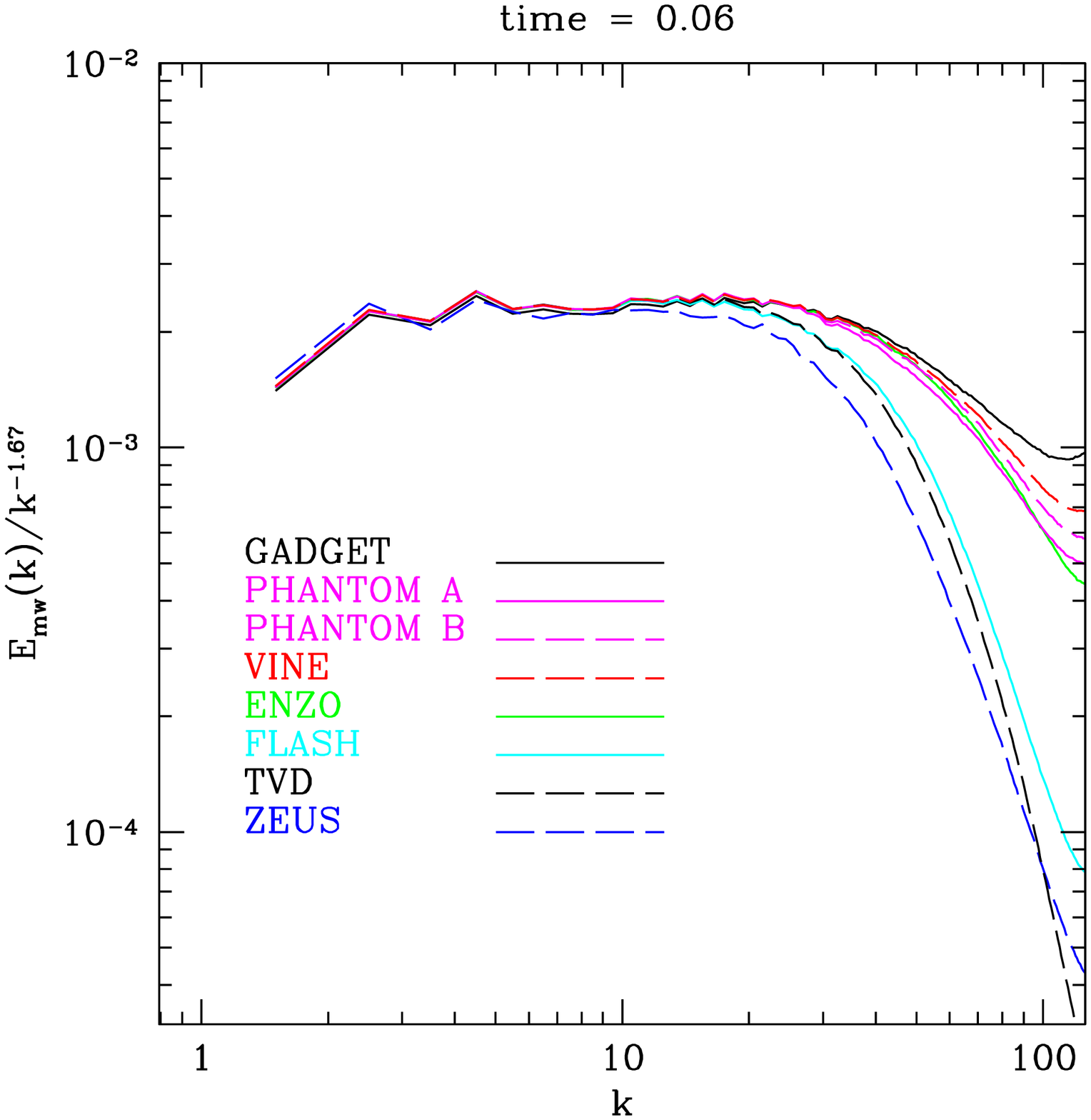} &
\includegraphics[width=0.29\linewidth]{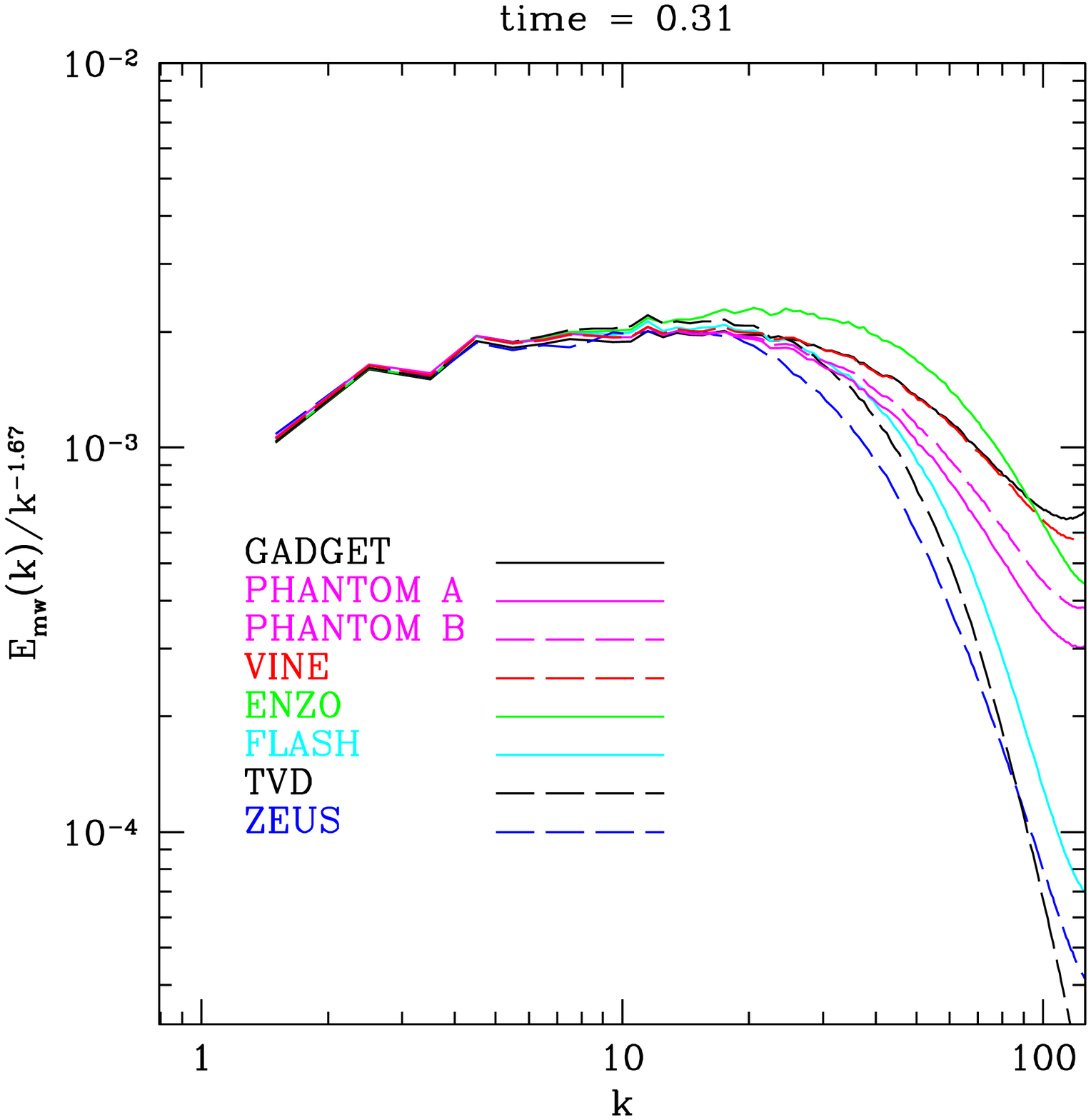} \\
\includegraphics[width=0.29\linewidth]{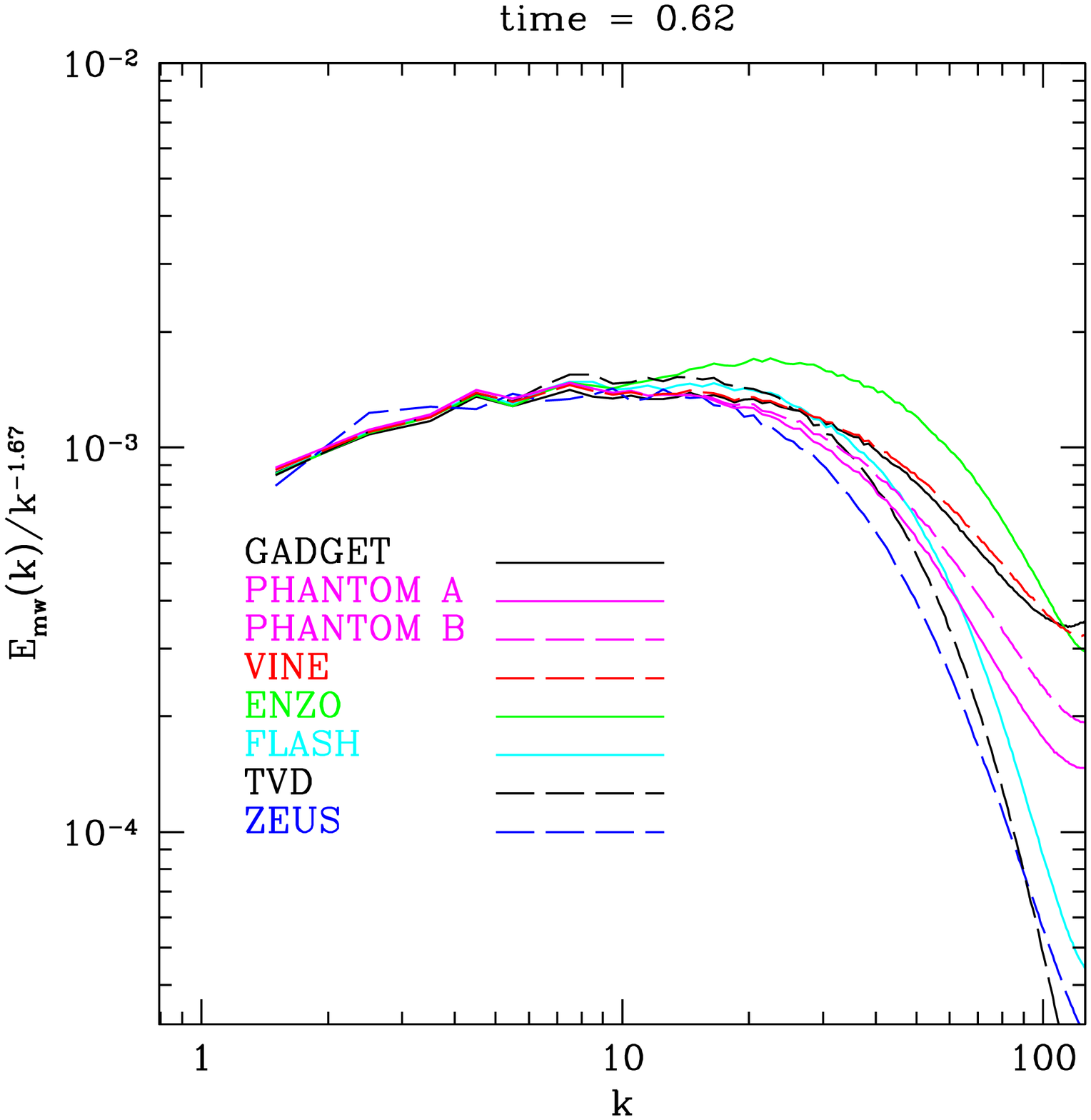} &
\includegraphics[width=0.29\linewidth]{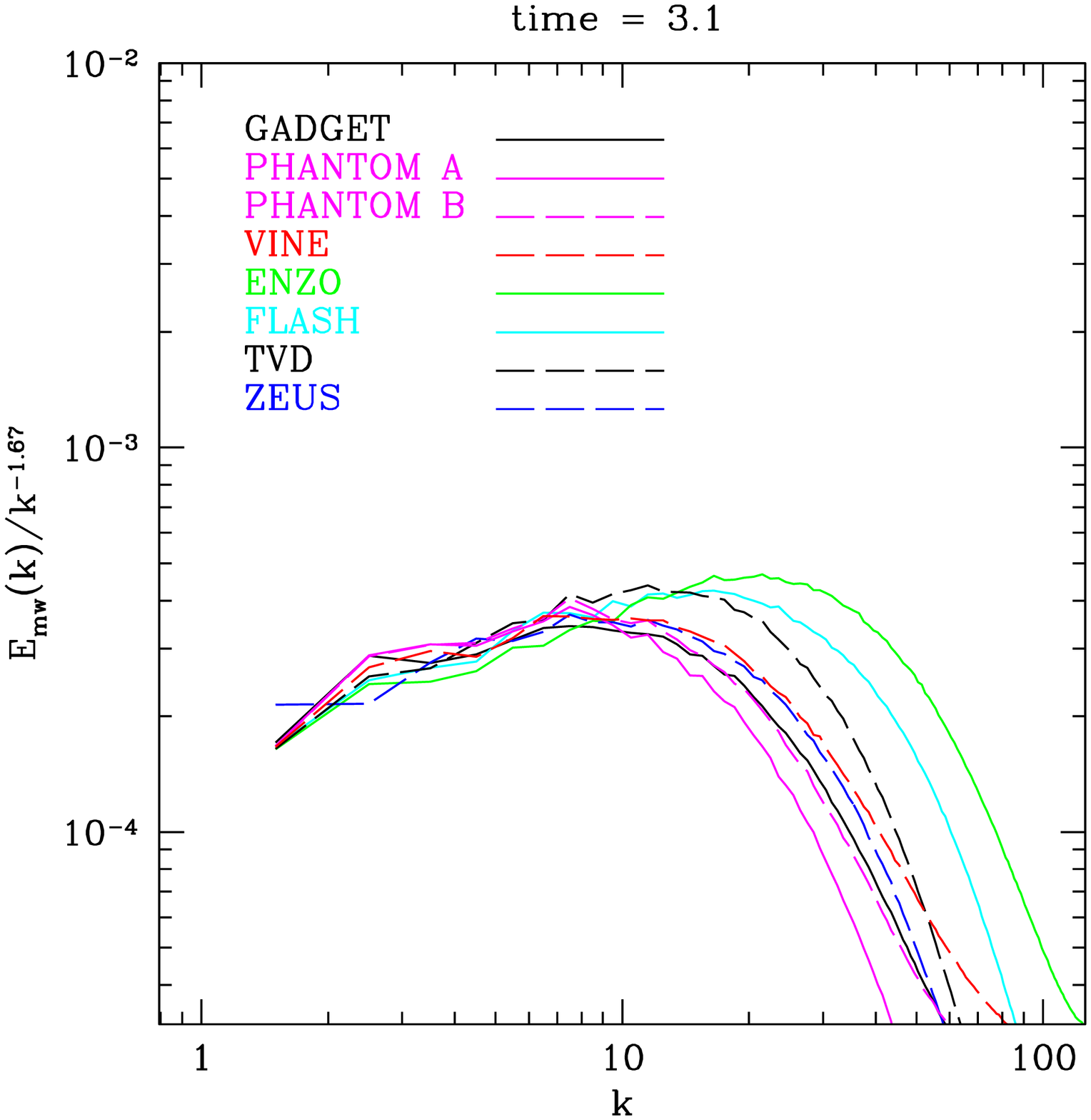} &
\includegraphics[width=0.29\linewidth]{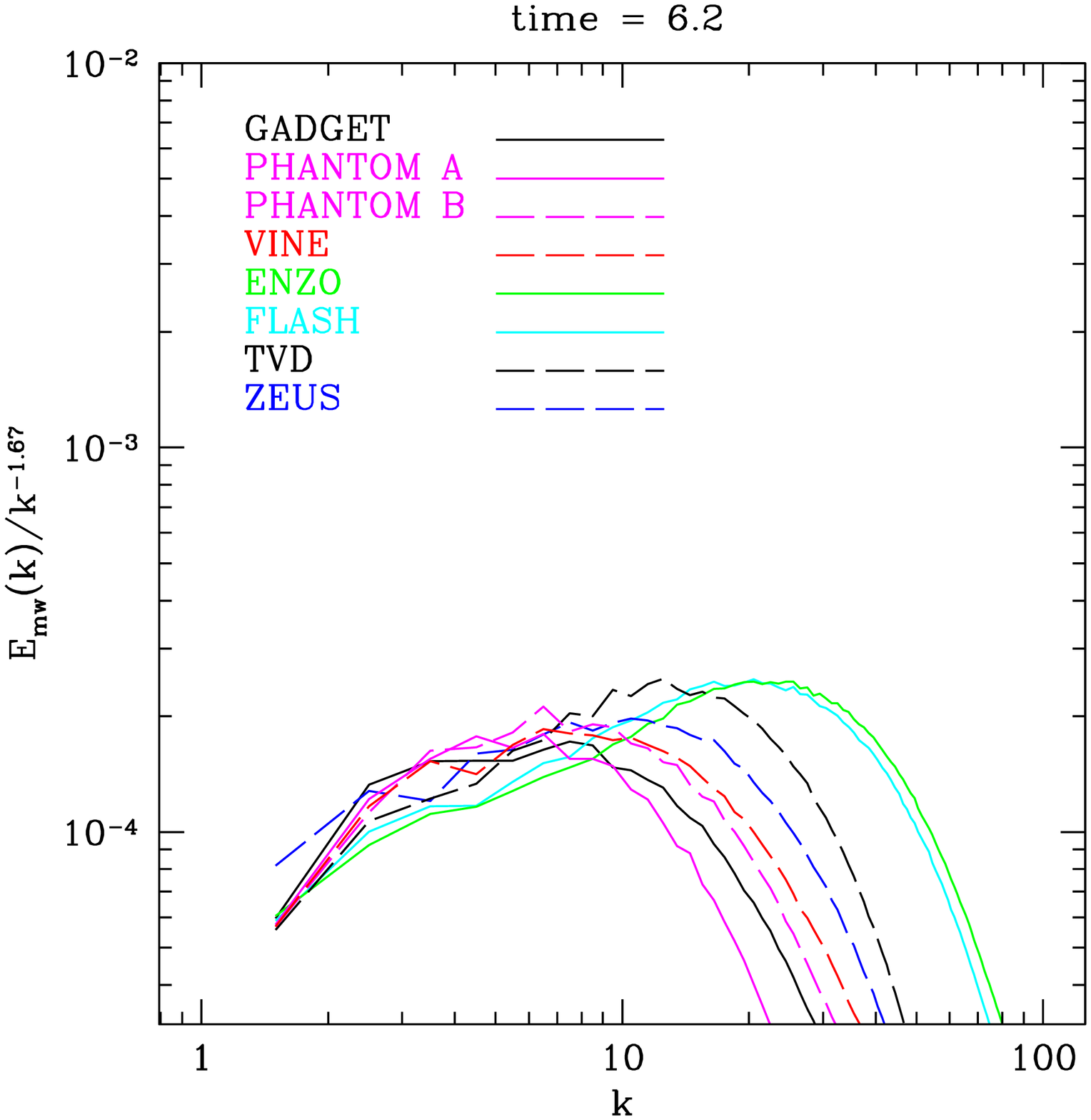}
\end{tabular}
\end{center}
\caption{Same as Fig.~\ref{fig:spect_codes_vol}, but the density-weighted velocity power spectra (compensated with power-law slopes of 1.67) are shown.}
\label{fig:spect_codes_rho}
\end{figure*}

\begin{figure*}
\centerline{\includegraphics[width=0.45\linewidth]{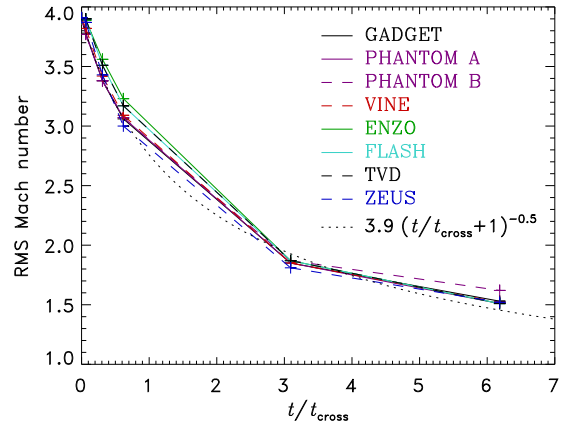}}
\caption{Evolution of the RMS Mach number as a function of time in units of the turbulent crossing time $\tcross$ for all codes/runs. The dotted line shows the expected power-law decay rate $M_\mathrm{rms}\propto t^{-1/2}$ for supersonic turbulence \citep{MacLowEtAl1998,StoneOstrikerGammie1998,MacLow1999}.}
\label{fig:mach_evol}
\end{figure*}

\begin{figure*}
\begin{center}
\begin{tabular}{ccc}
\includegraphics[width=0.29\linewidth]{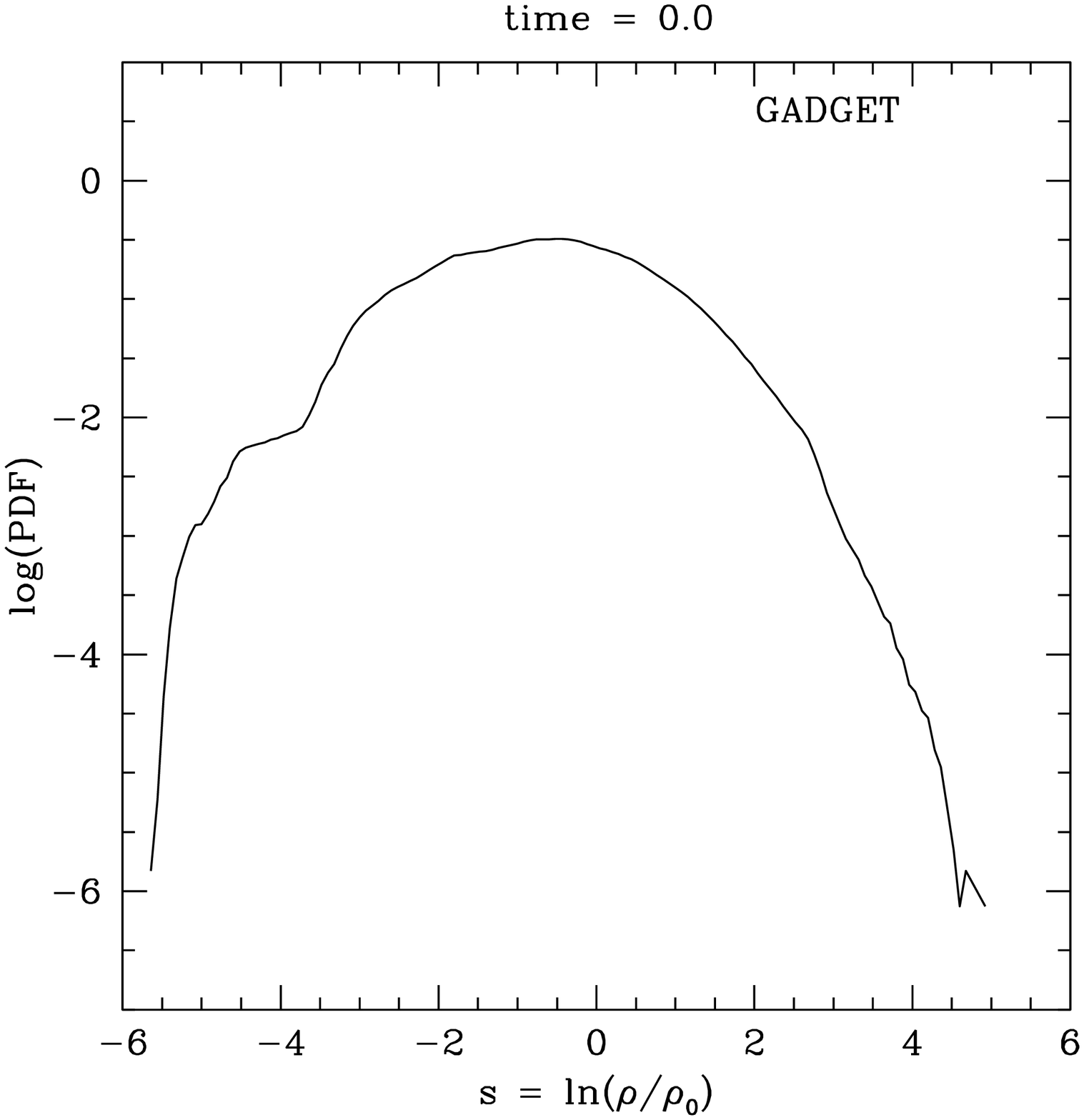} &
\includegraphics[width=0.29\linewidth]{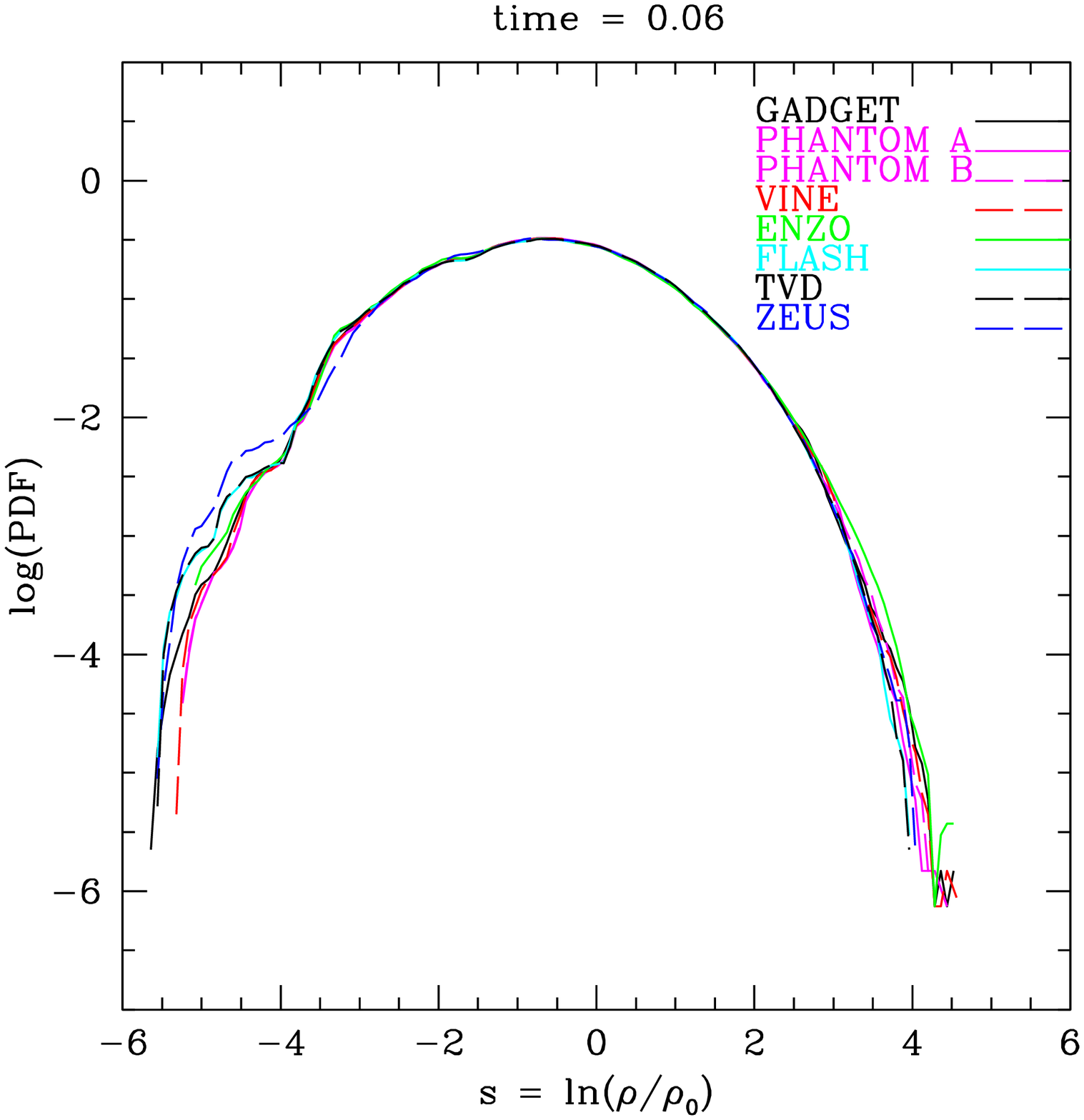} &
\includegraphics[width=0.29\linewidth]{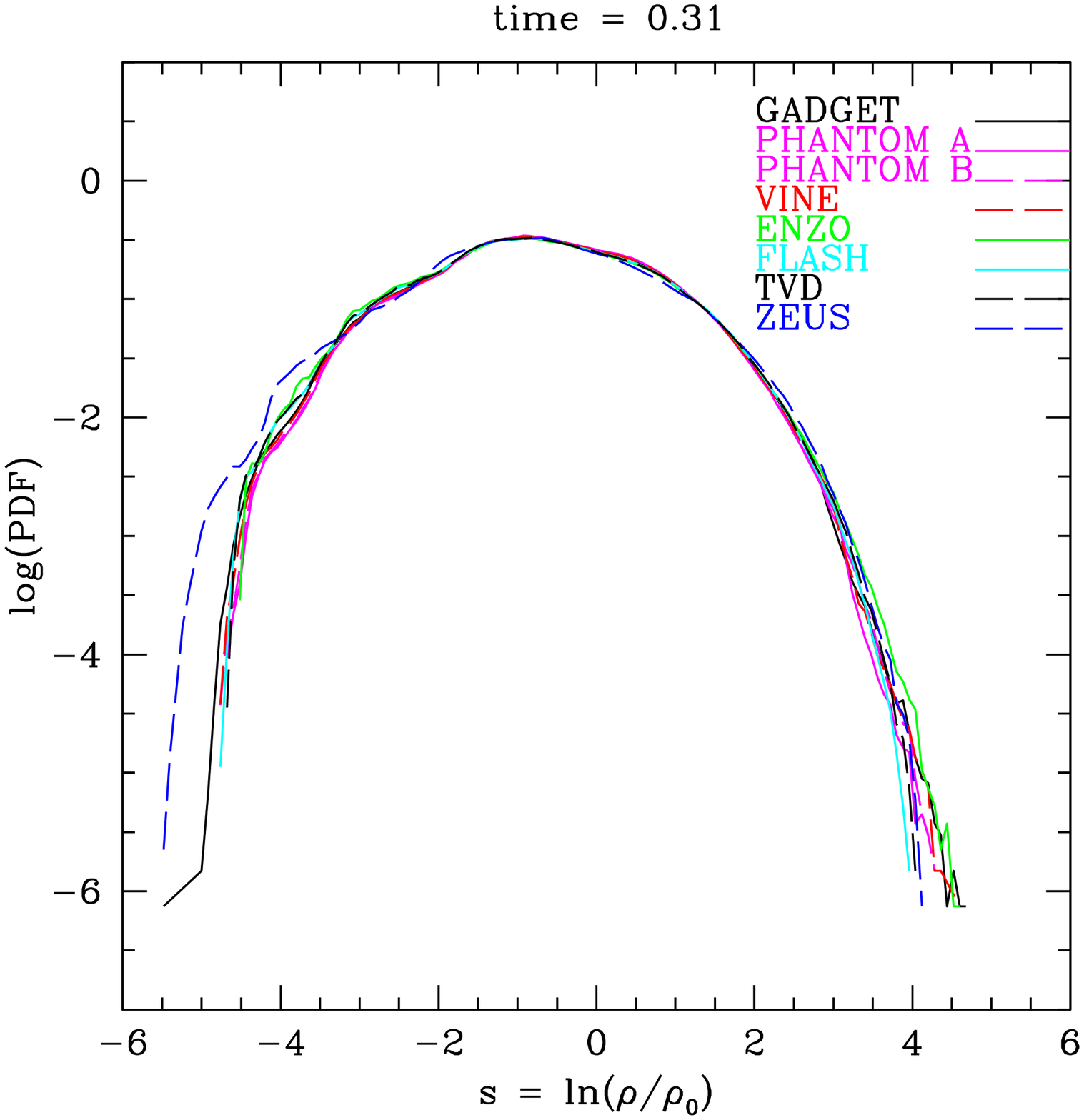} \\
\includegraphics[width=0.29\linewidth]{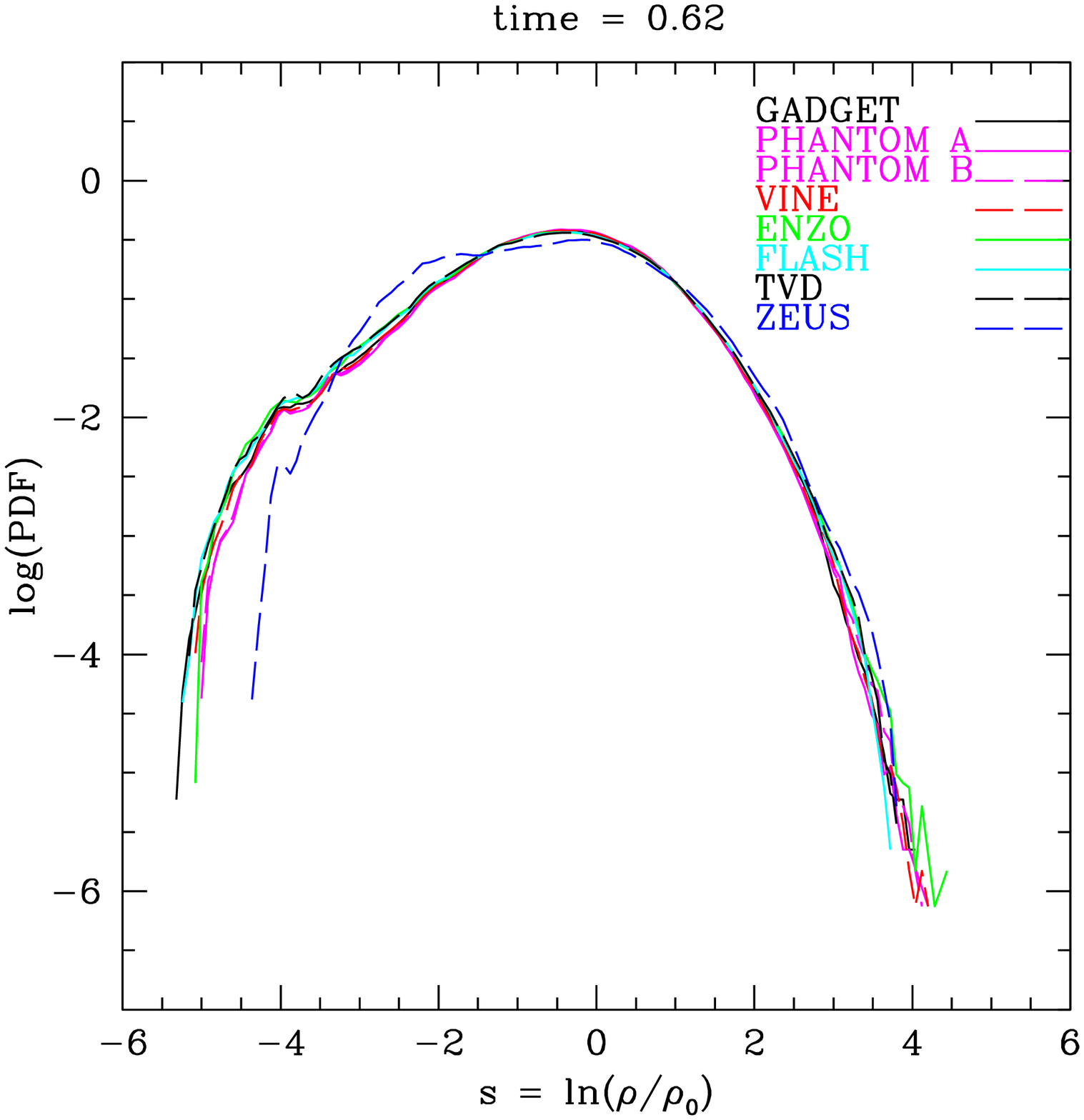} &
\includegraphics[width=0.29\linewidth]{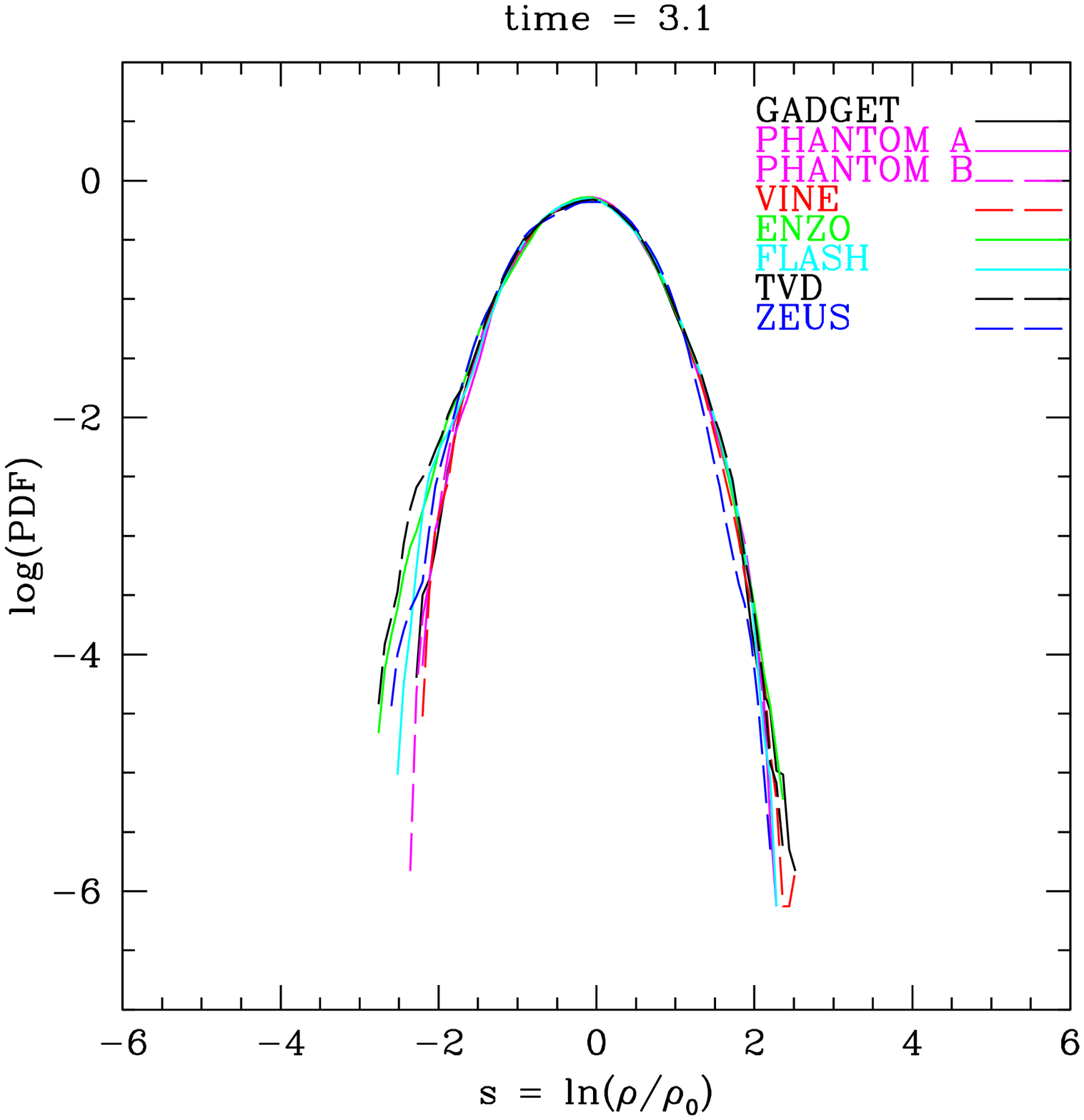} &
\includegraphics[width=0.29\linewidth]{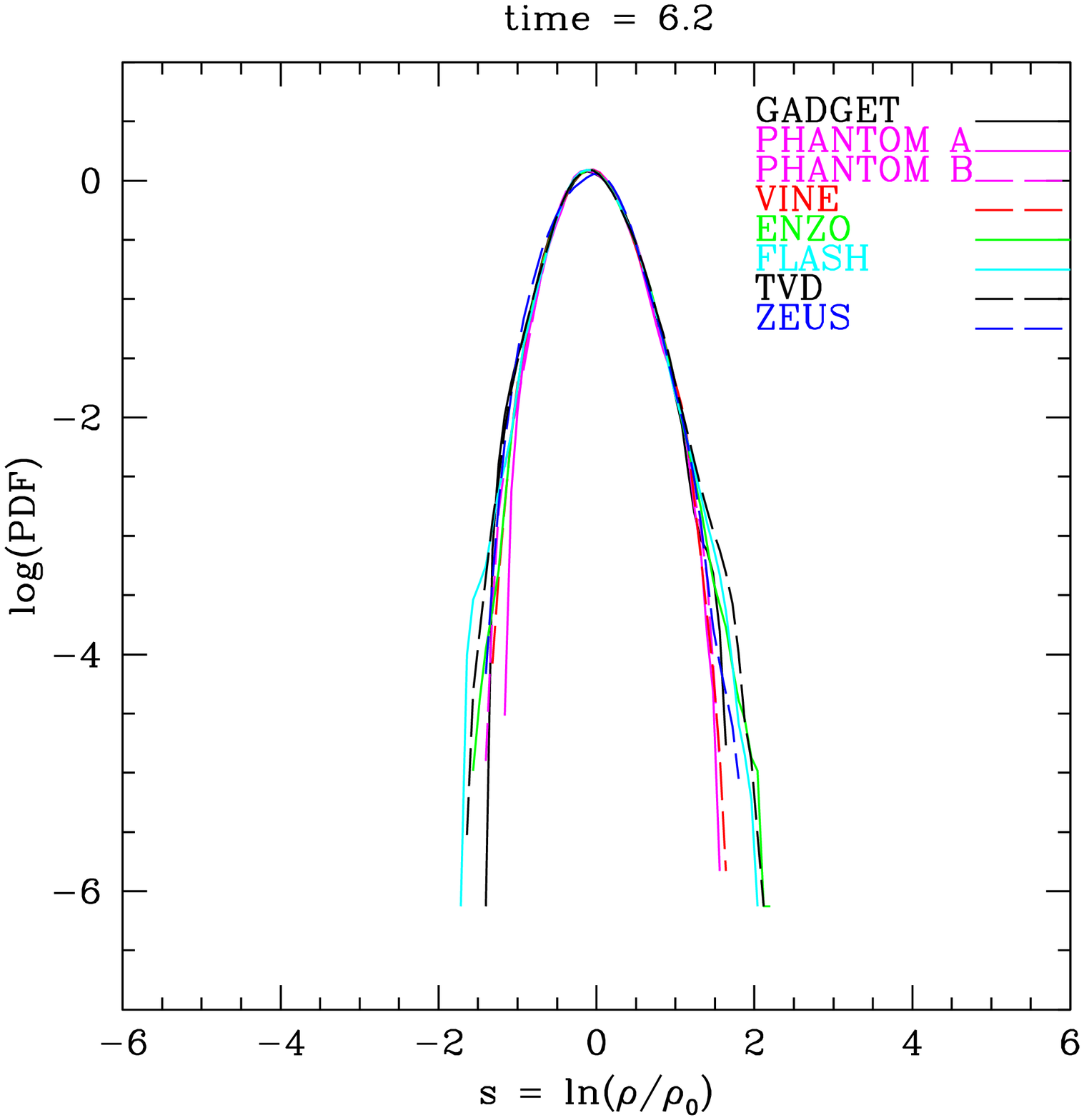}
\end{tabular}
\end{center}
\caption{Comparison of the volume-weighted density PDFs of all codes at different times along the decay: $t\,=\,0.0$, $0.06$, $0.31$, $0.62$, $3.1$, and $6.2\;\tcross$.}
\label{fig:densityPDFs}
\end{figure*}

\begin{figure*}
\begin{center}
\begin{tabular}{ccc}
\includegraphics[width=0.29\linewidth]{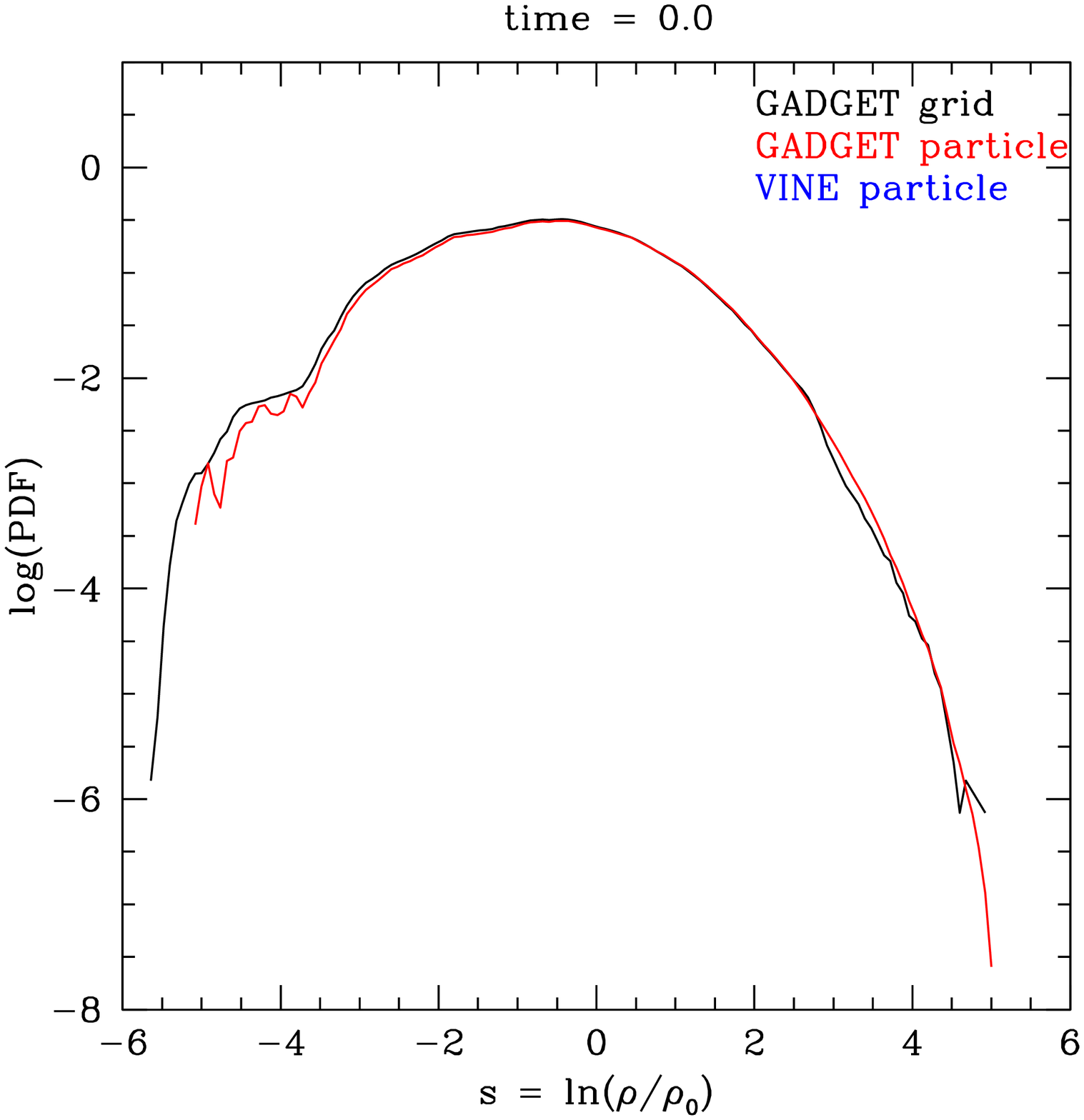} &
\includegraphics[width=0.29\linewidth]{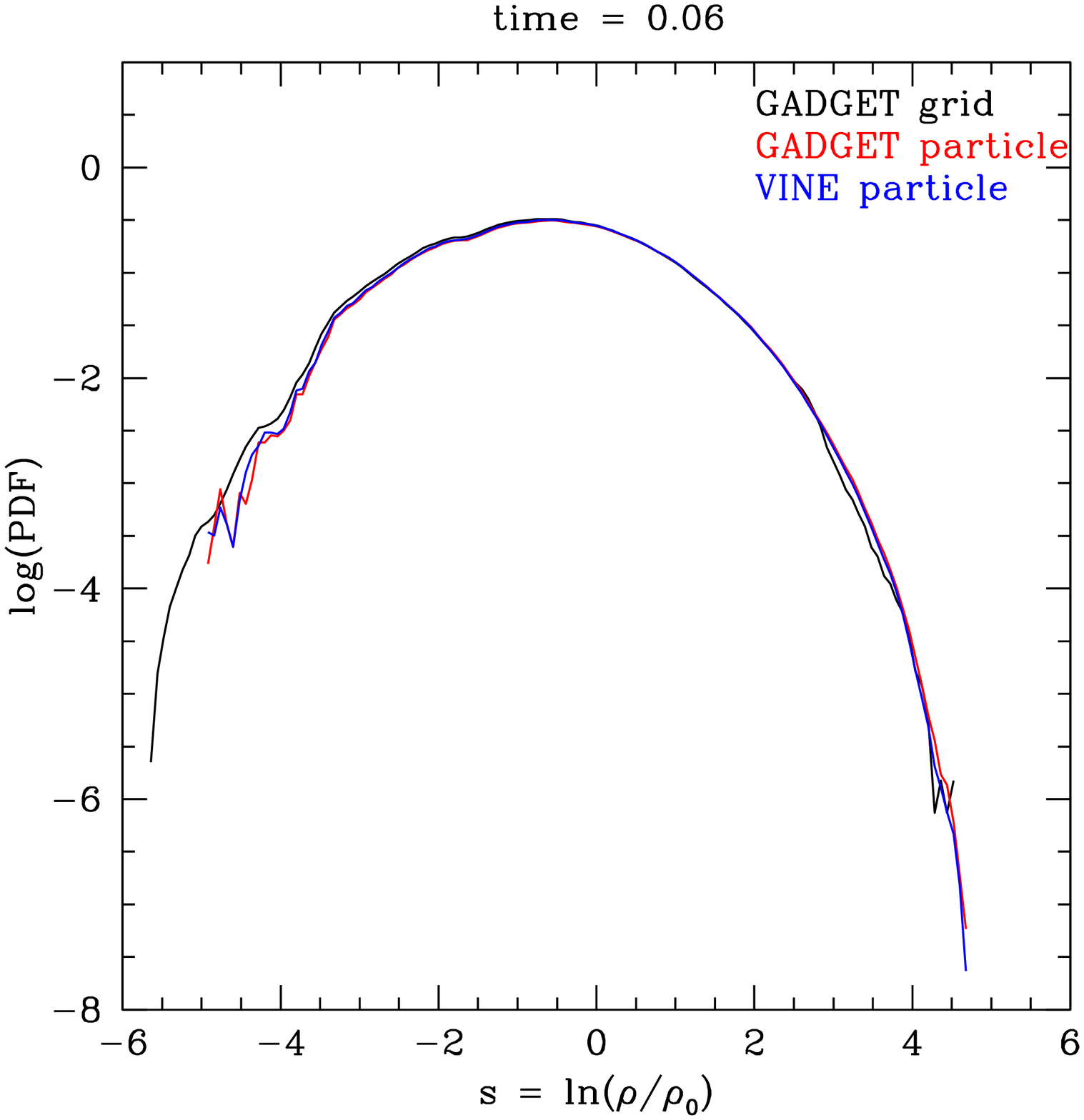} &
\includegraphics[width=0.29\linewidth]{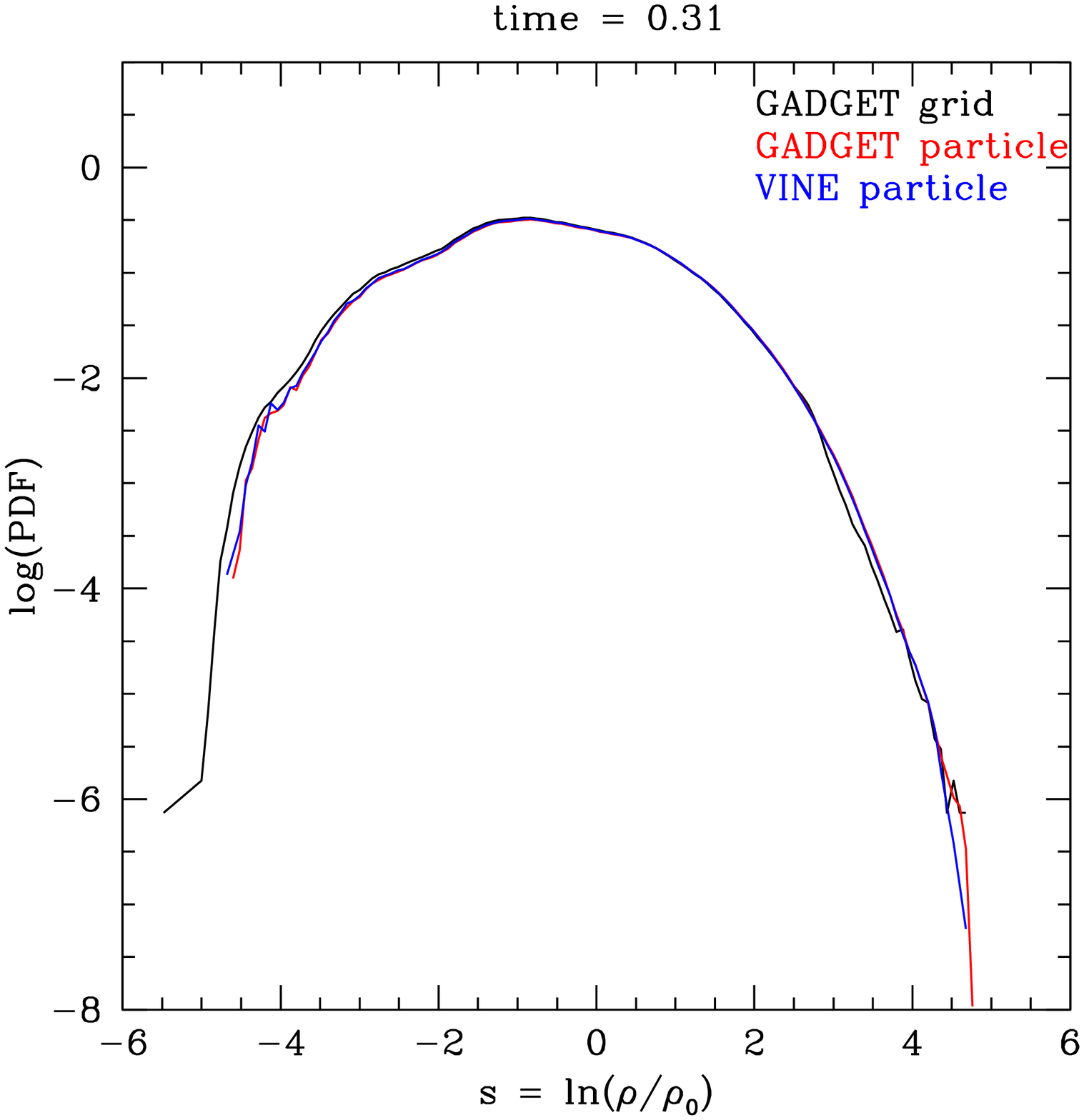} \\
\includegraphics[width=0.29\linewidth]{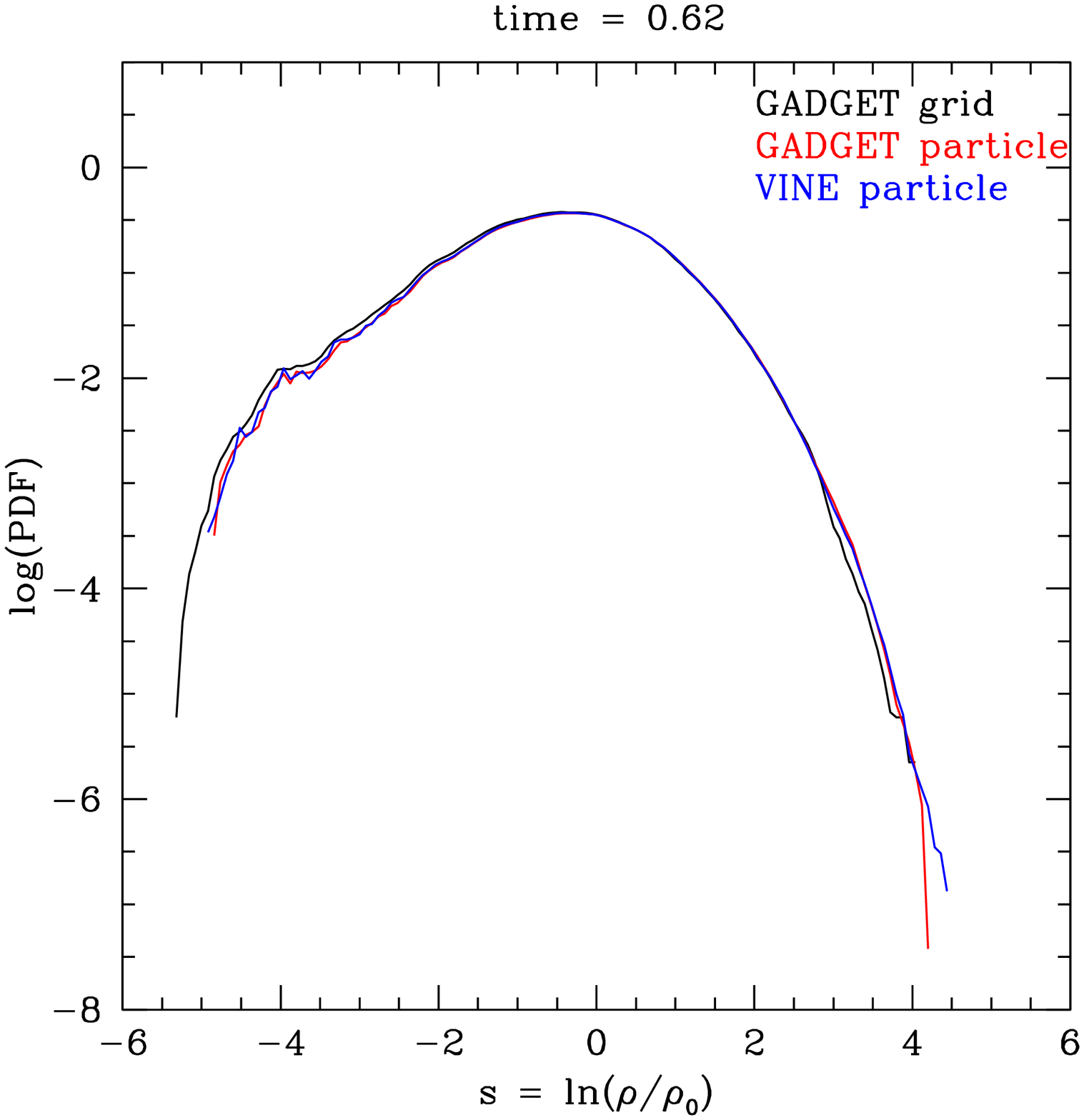} &
\includegraphics[width=0.29\linewidth]{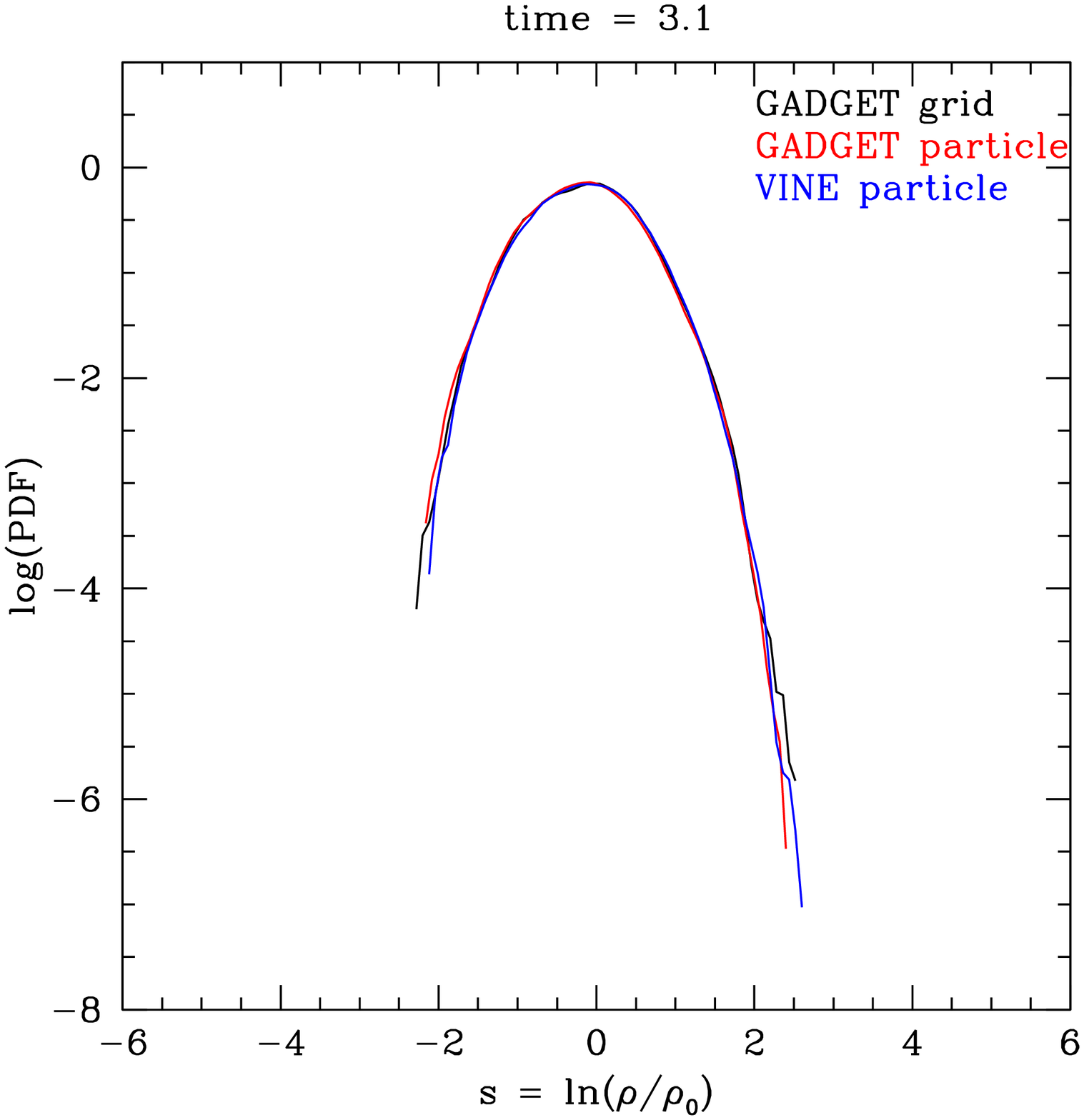} &
\includegraphics[width=0.29\linewidth]{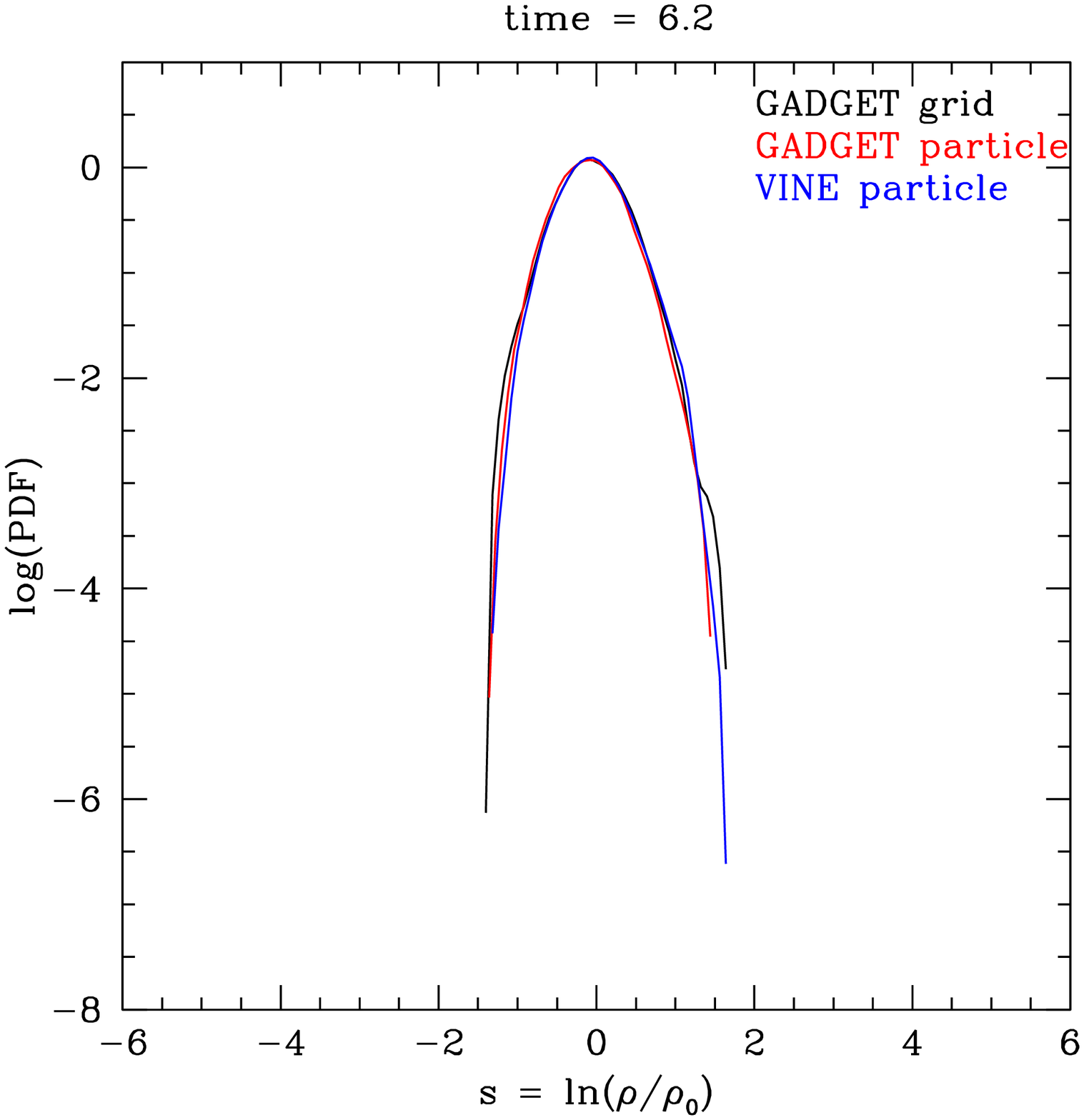}
\end{tabular}
\end{center}
\caption{Volume-weighted density PDFs at $t\,=\,0.0$, $0.06$, $0.31$, $0.62$, $3.1$, and $6.2\;\tcross$ based on all cells of the interpolated grid (black line), and based on all SPH particle densities for \texttt{GADGET} (red line), and all SPH particle densities for \texttt{VINE} (blue line). The SPH particle density PDFs were computed on the SPH particles directly. Thus, they do not involve any interpolation to a grid, but they needed to be converted to a volume-weighted density PDF by using $p_v=p_m/\rho$, in order to allow for a direct comparison to the grid-interpolated PDF (black line). Both the density PDF computed from the grid and the SPH densities agree very well, with the SPH density PDF extending to slightly higher densities, and exhibiting slightly less scatter at the high-density tail.}
\label{fig:densityPDFsInterpolation}
\end{figure*}

\begin{figure*}
\begin{center}
\begin{tabular}{ccc}
\includegraphics[width=0.29\linewidth]{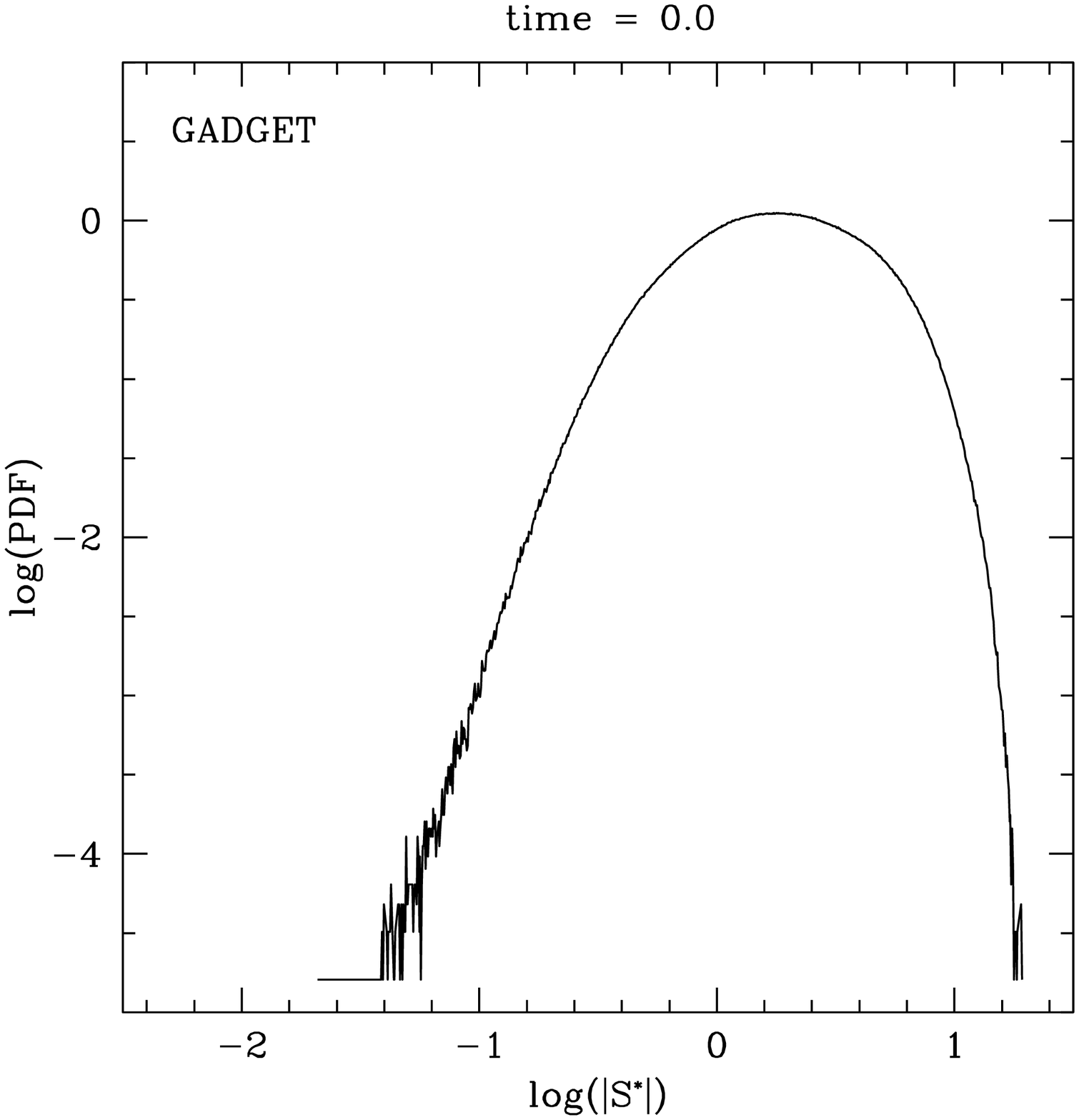} &
\includegraphics[width=0.29\linewidth]{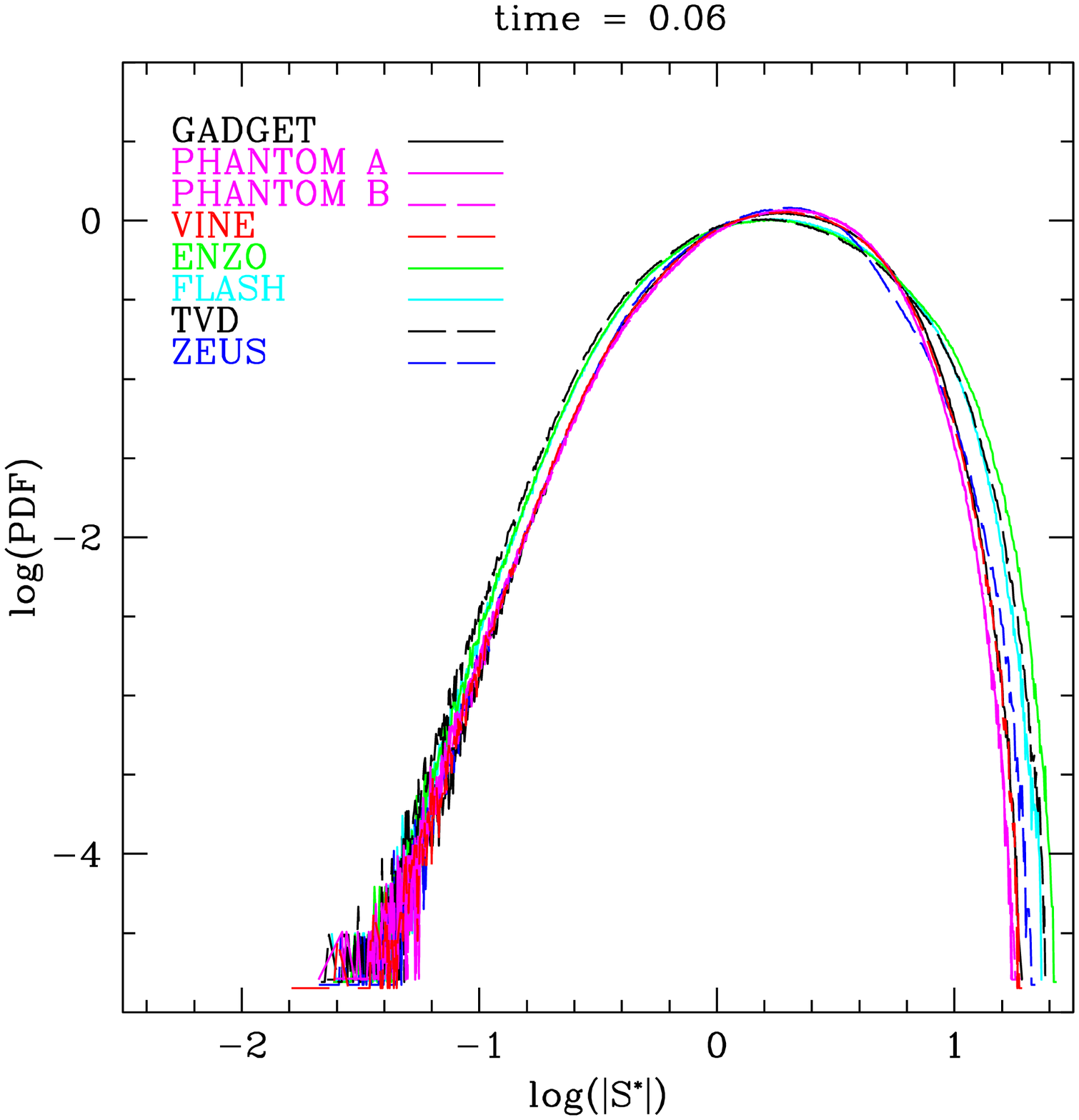} &
\includegraphics[width=0.29\linewidth]{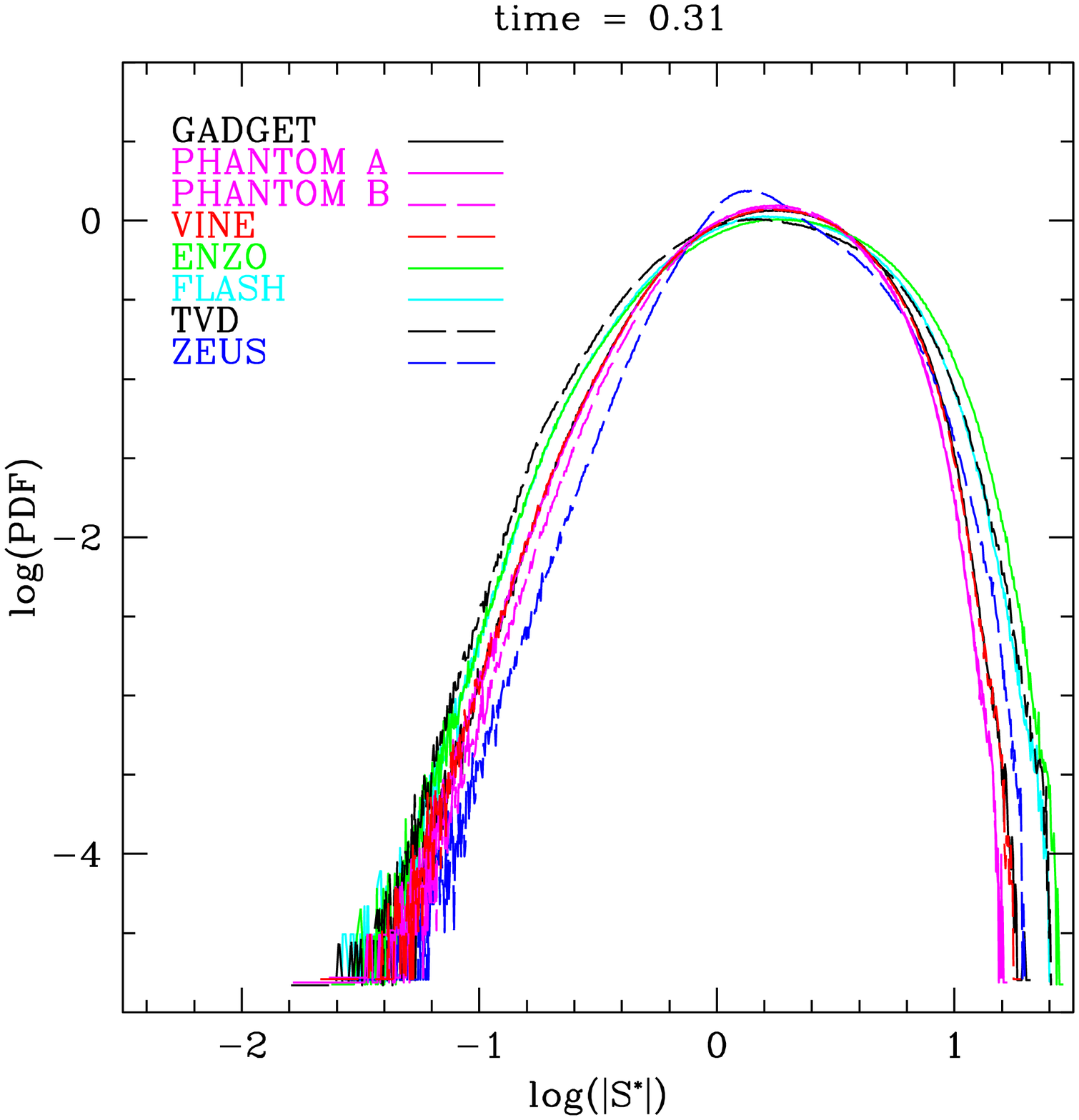} \\
\includegraphics[width=0.29\linewidth]{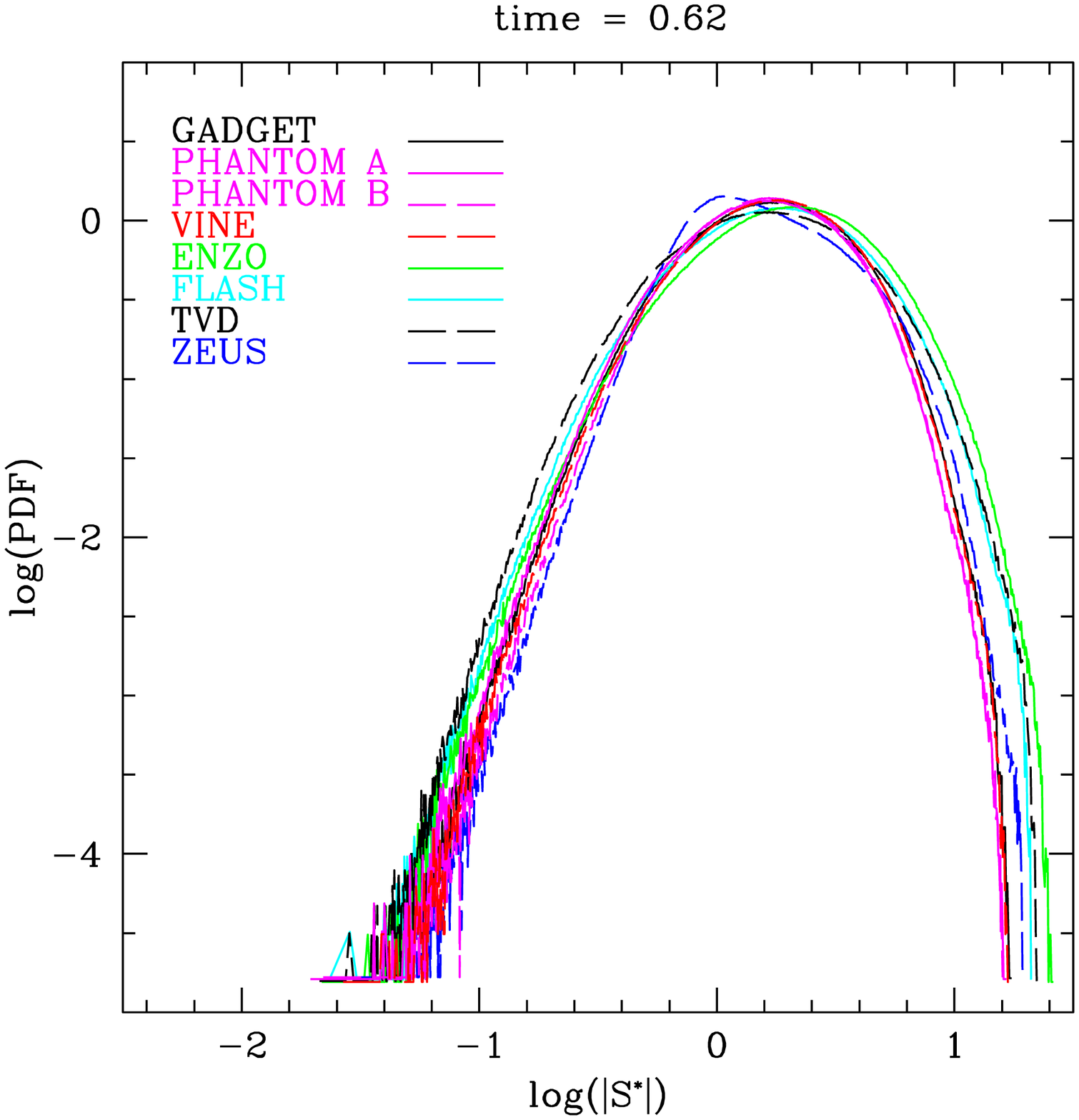} &
\includegraphics[width=0.29\linewidth]{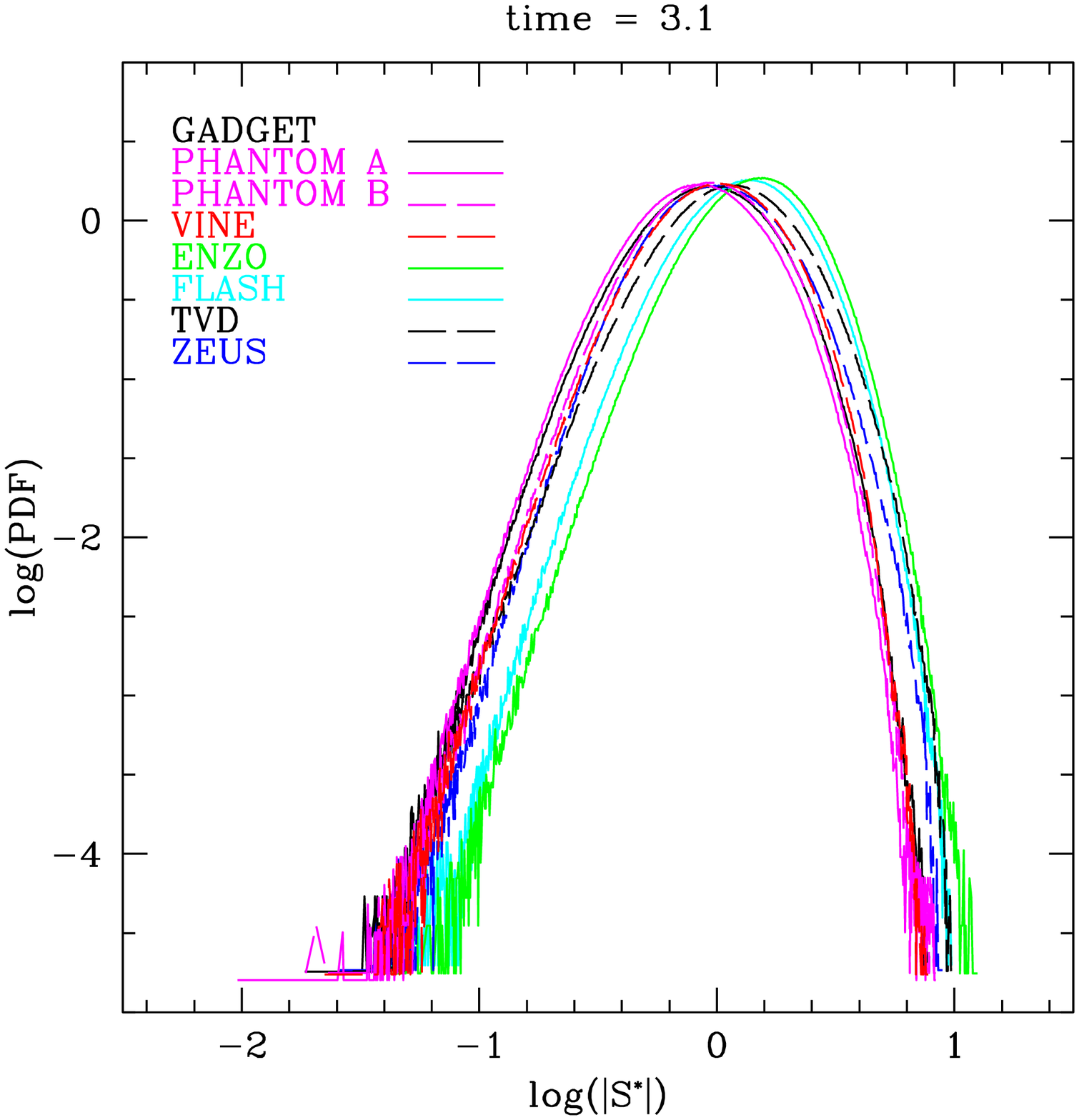} &
\includegraphics[width=0.29\linewidth]{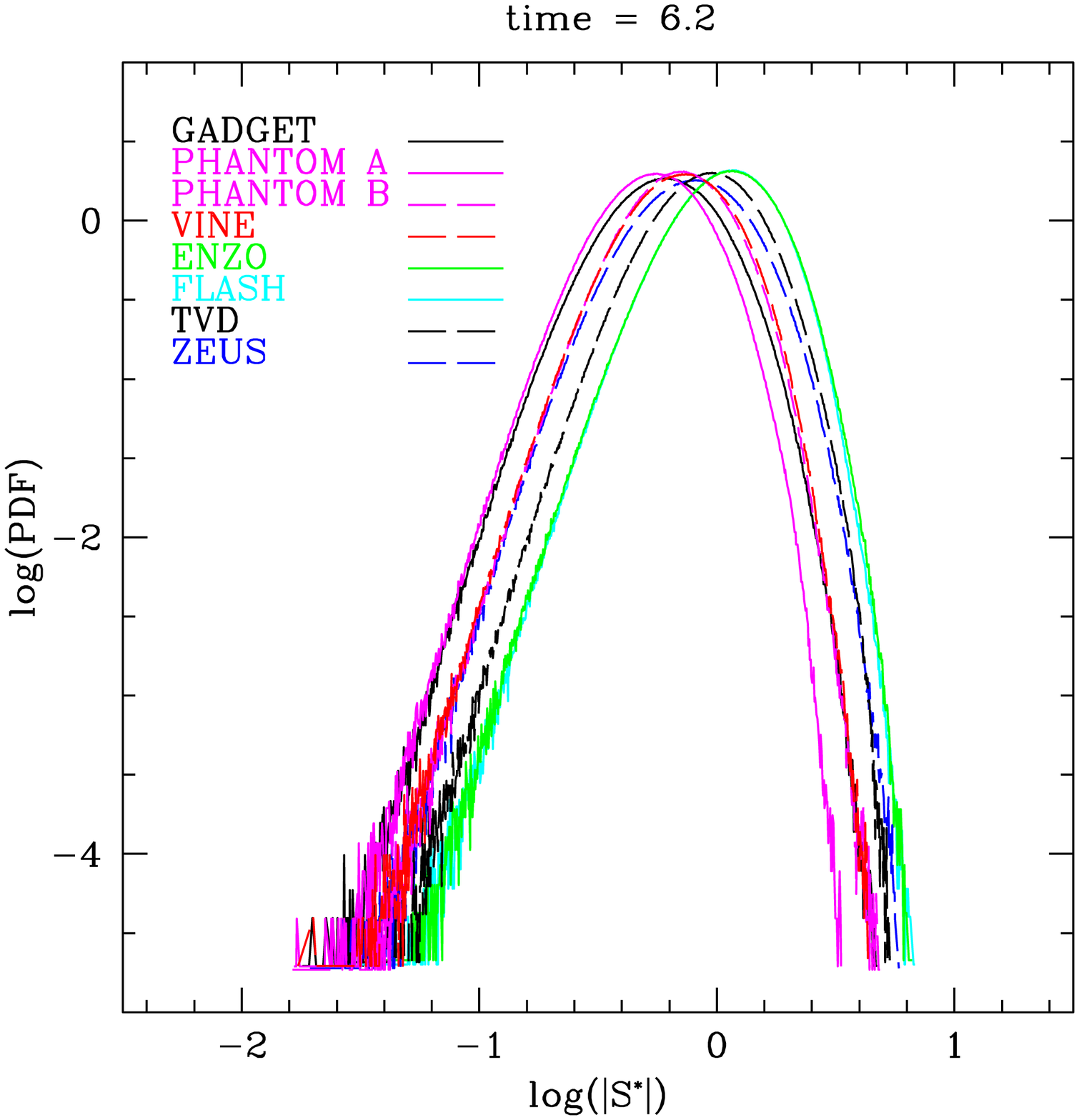}
\end{tabular}
\end{center}
\caption{Same as Figure~\ref{fig:densityPDFs}, but the PDFs of the trace-free rate of strain are shown, as defined in eq.~(\ref{eq:strain}).}
\label{fig:strainPDFs}
\end{figure*}

\begin{figure*}
\begin{center}
\begin{tabular}{ccc}
\includegraphics[width=0.29\linewidth]{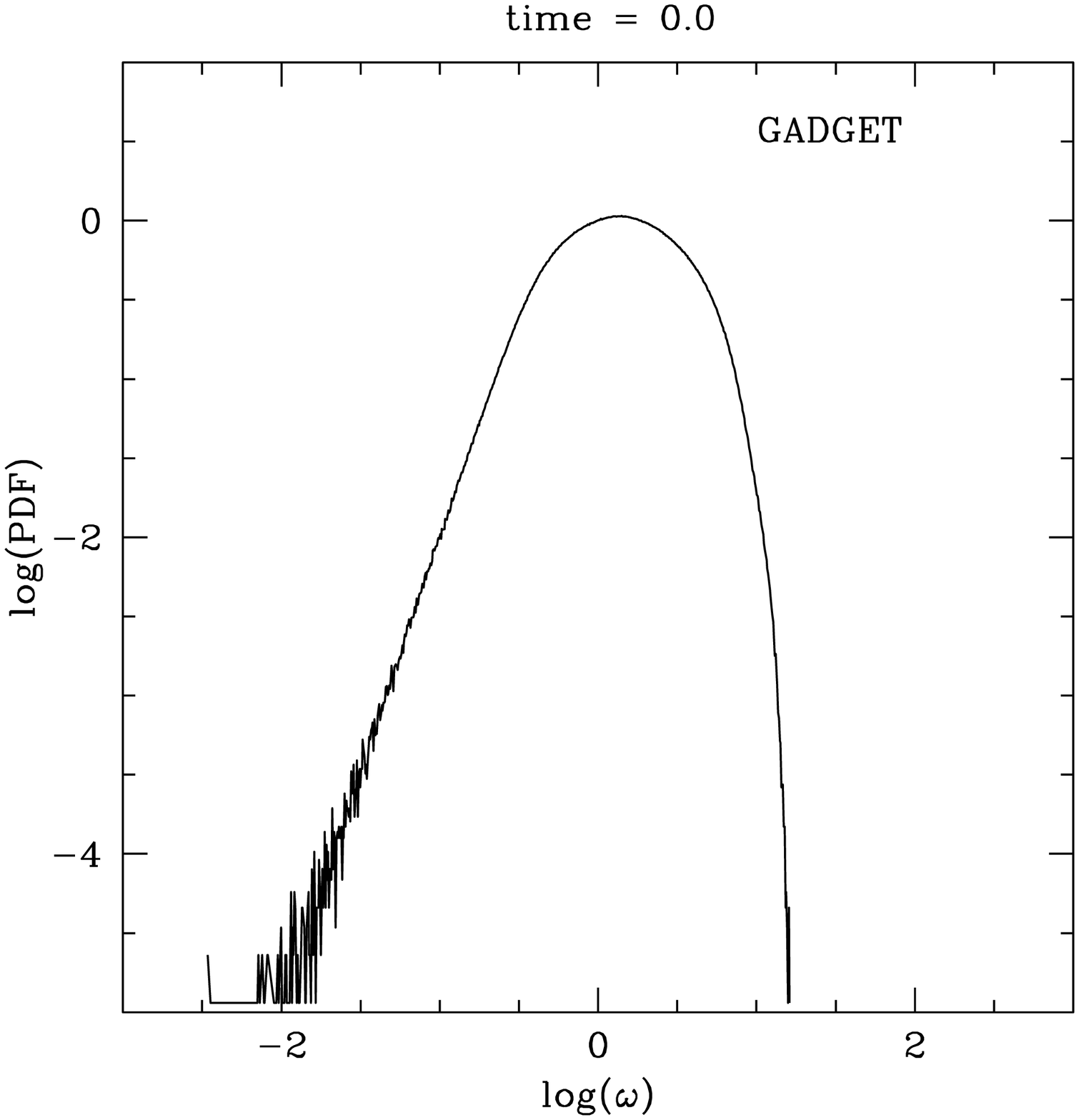} &
\includegraphics[width=0.29\linewidth]{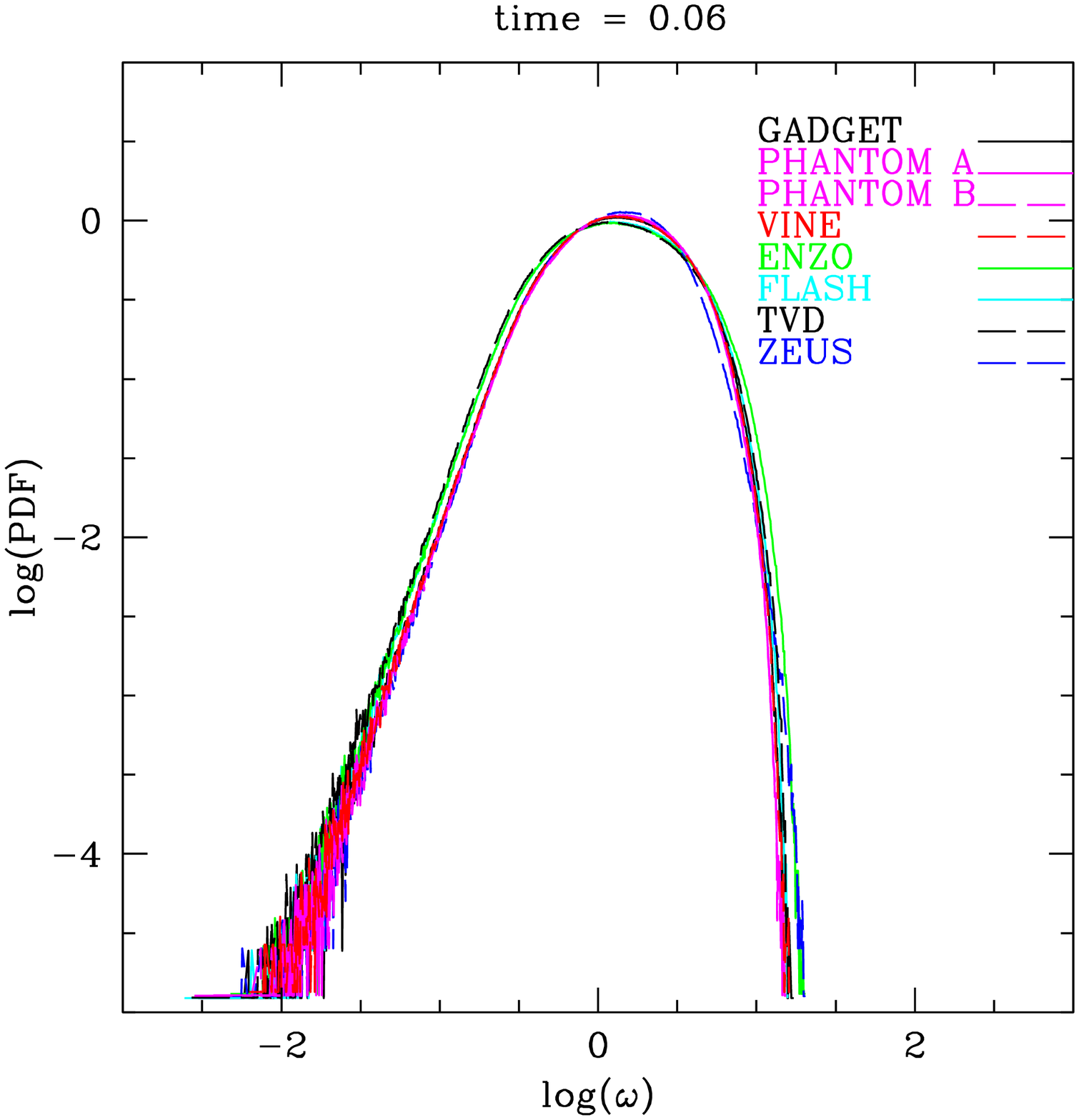} &
\includegraphics[width=0.29\linewidth]{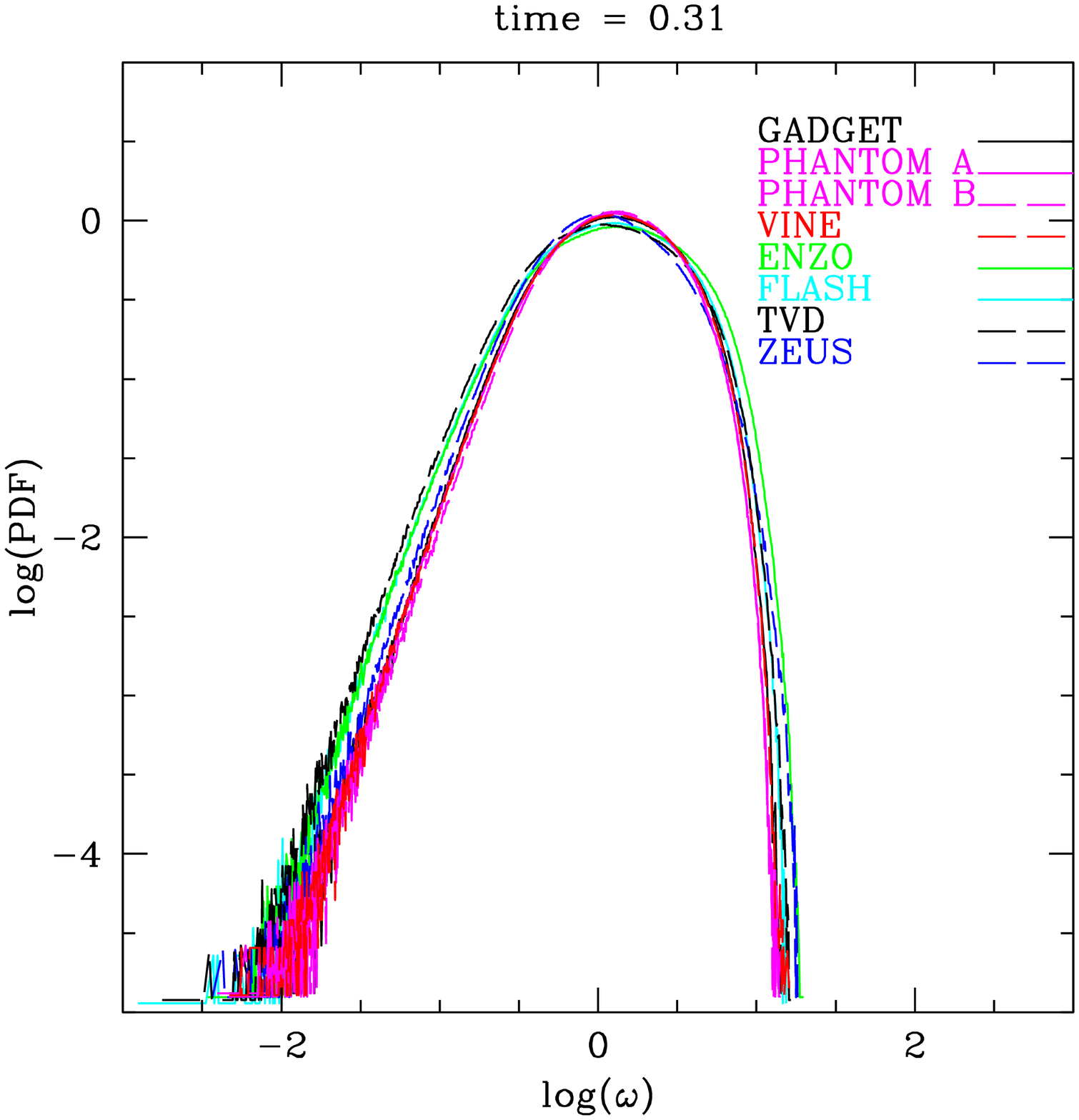} \\
\includegraphics[width=0.29\linewidth]{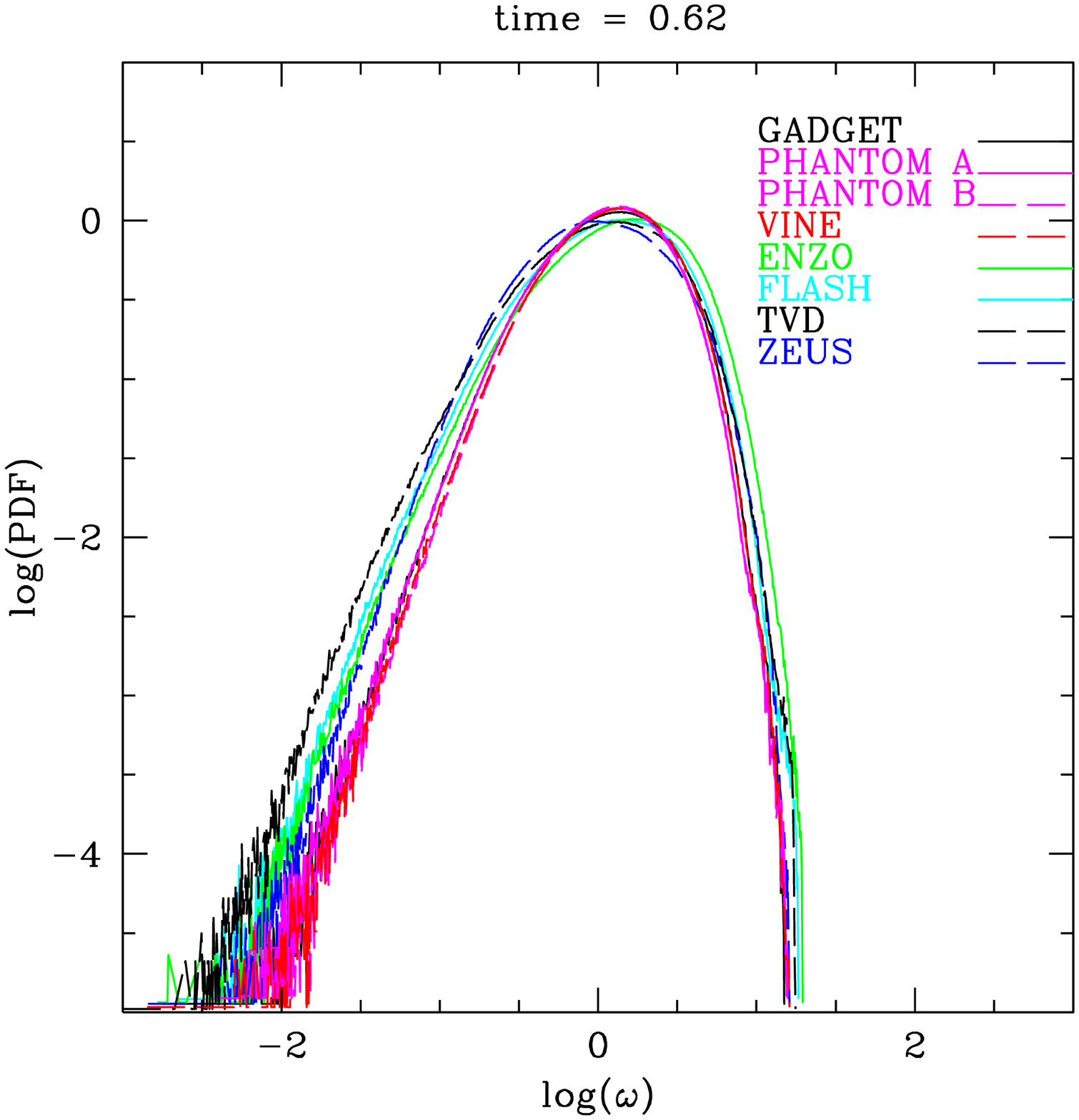} &
\includegraphics[width=0.29\linewidth]{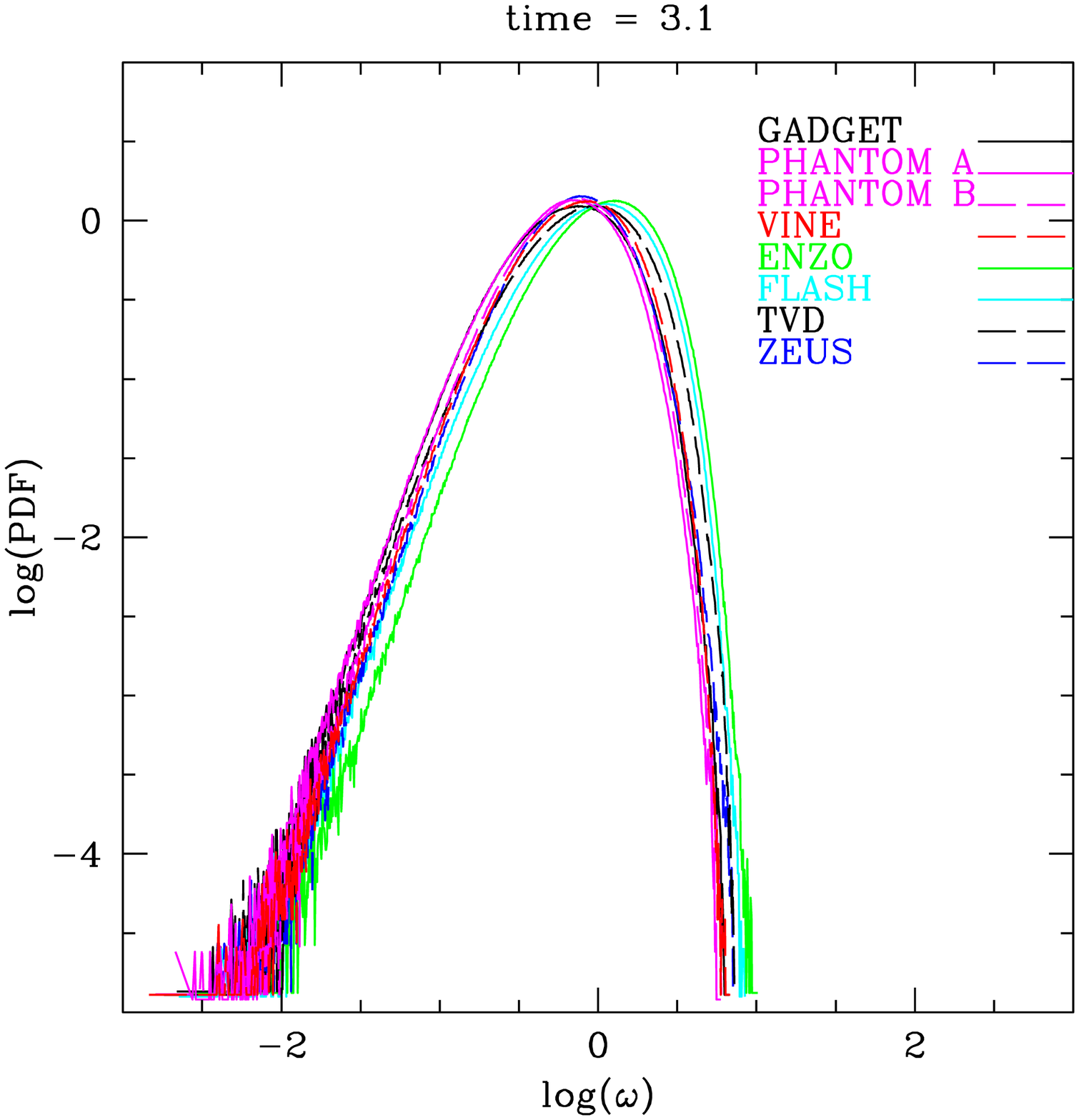} &
\includegraphics[width=0.29\linewidth]{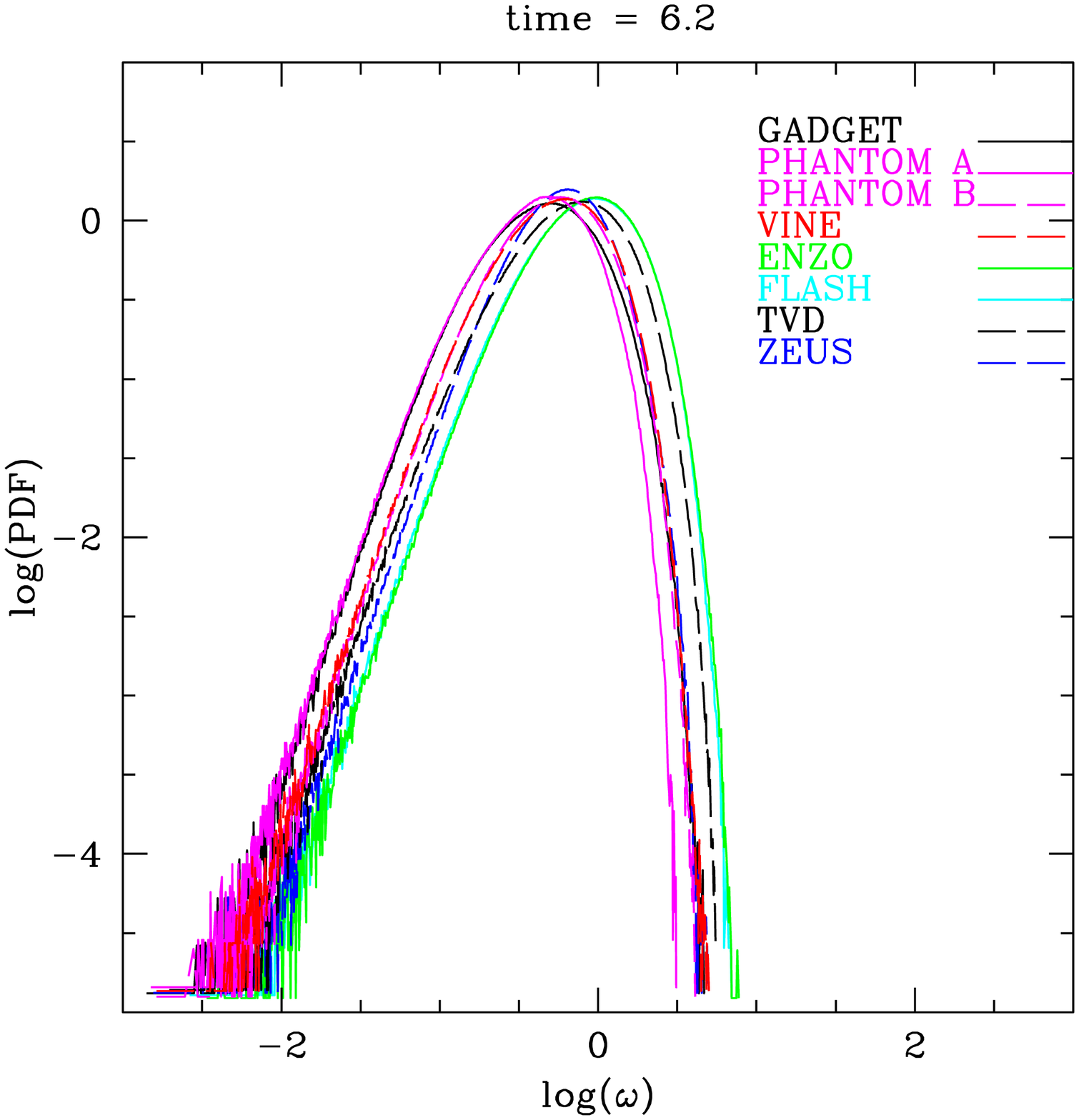}
\end{tabular}
\end{center}
\caption{Same as Figure~\ref{fig:densityPDFs}, but the PDFs of the vorticity are shown.}
\label{fig:vorticityPDFs}
\end{figure*}

\begin{figure*}
\begin{center}
\begin{tabular}{ccc}
\includegraphics[width=0.29\linewidth]{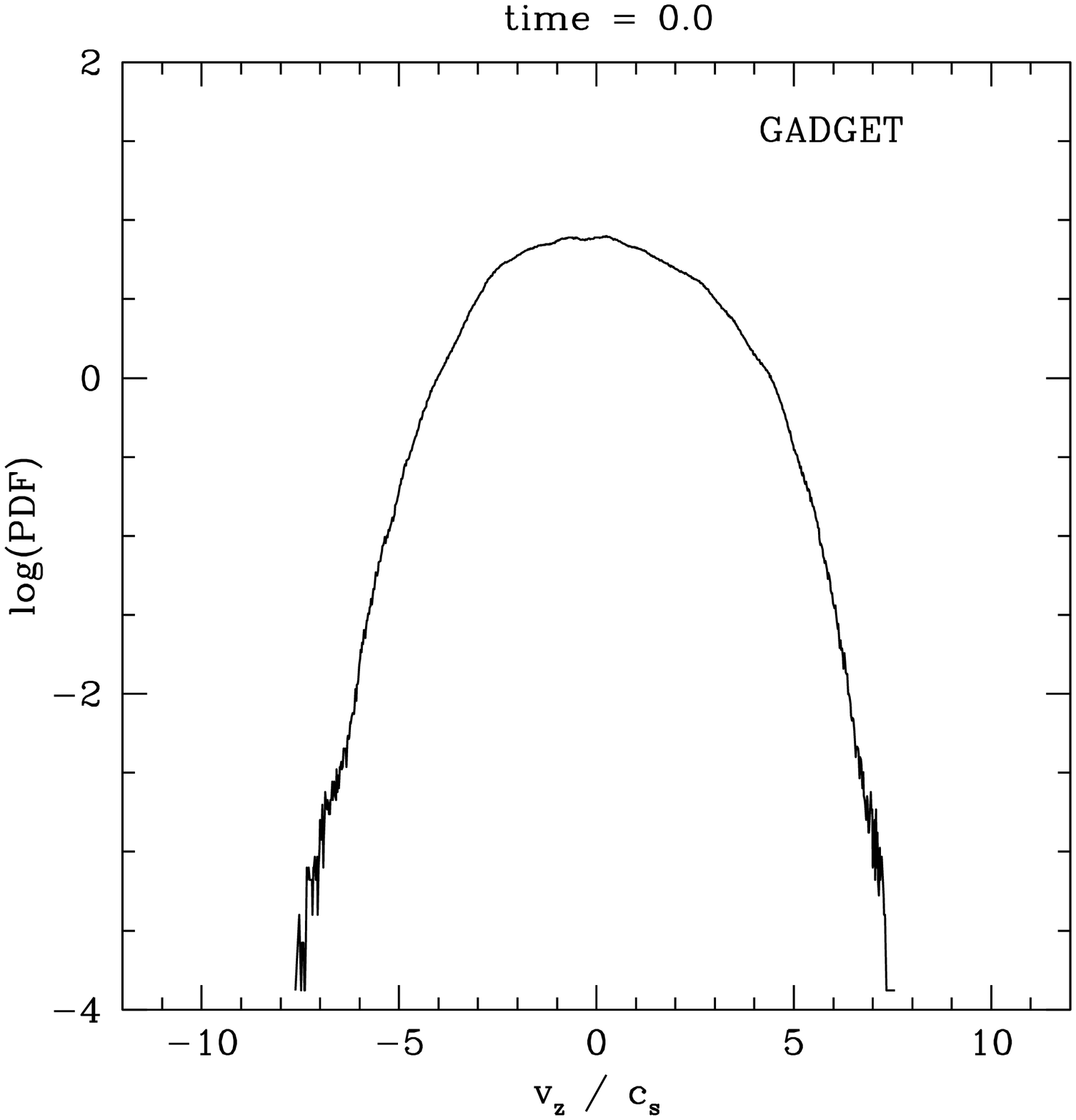} &
\includegraphics[width=0.29\linewidth]{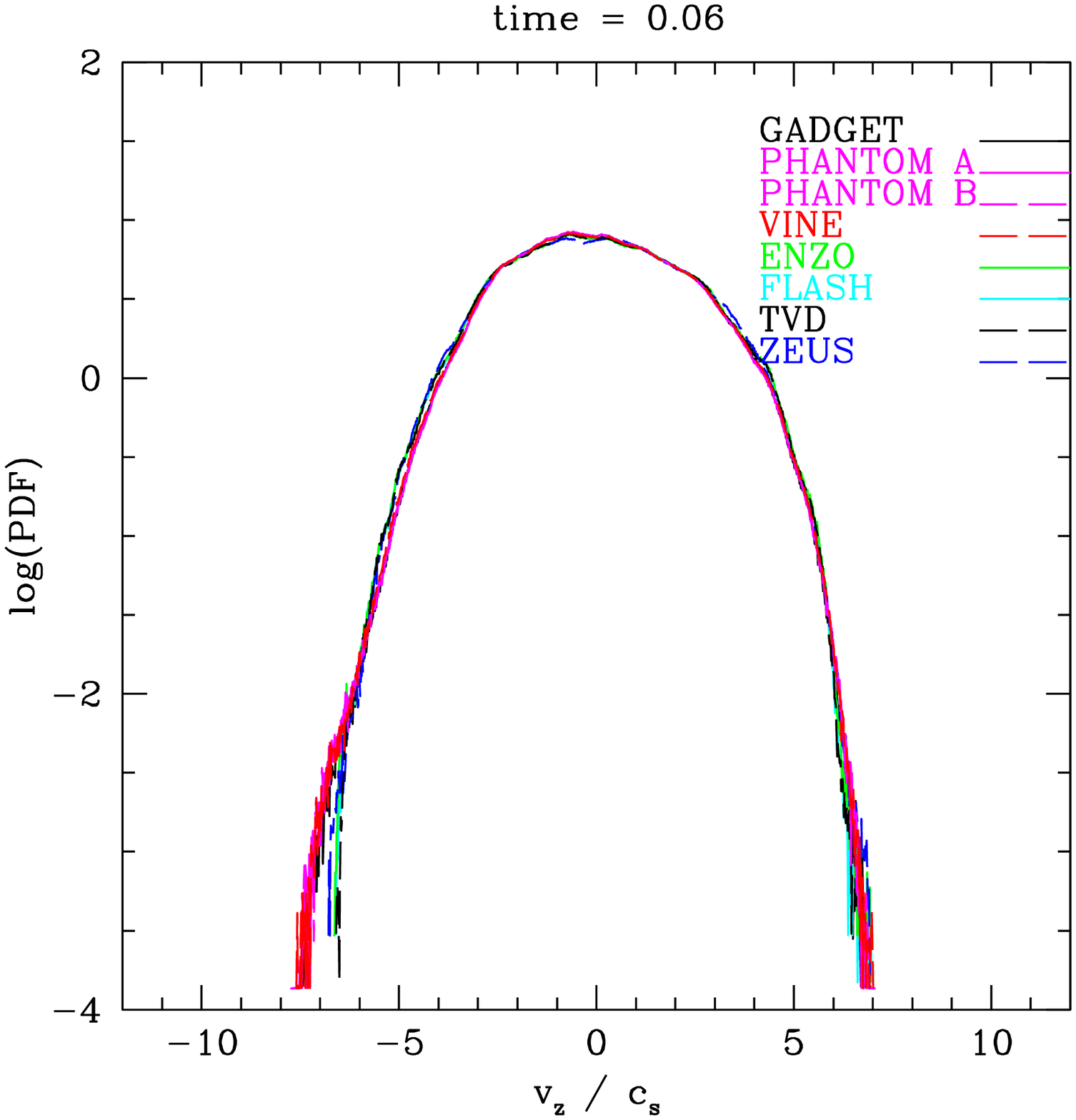} &
\includegraphics[width=0.29\linewidth]{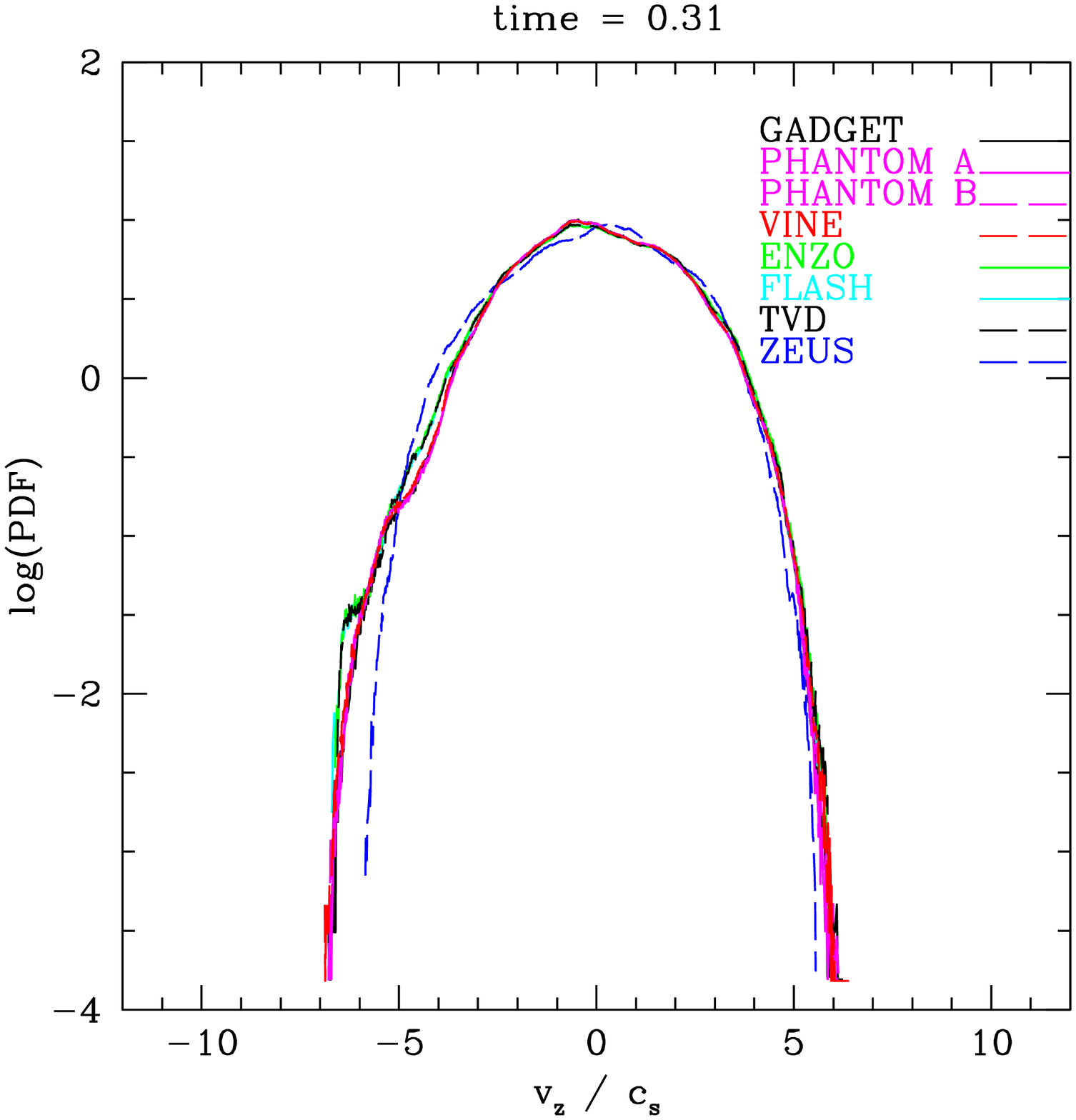} \\
\includegraphics[width=0.29\linewidth]{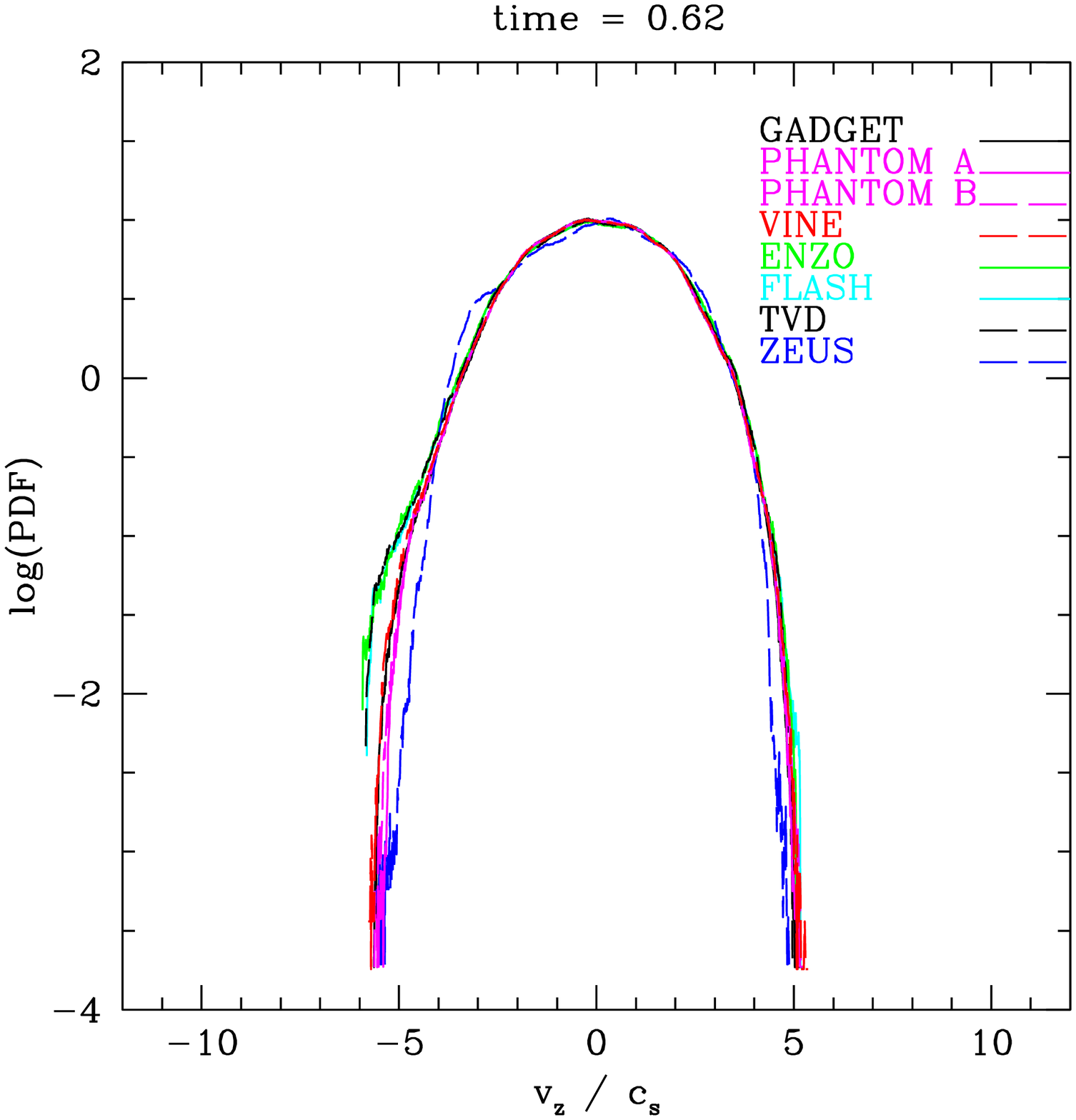} &
\includegraphics[width=0.29\linewidth]{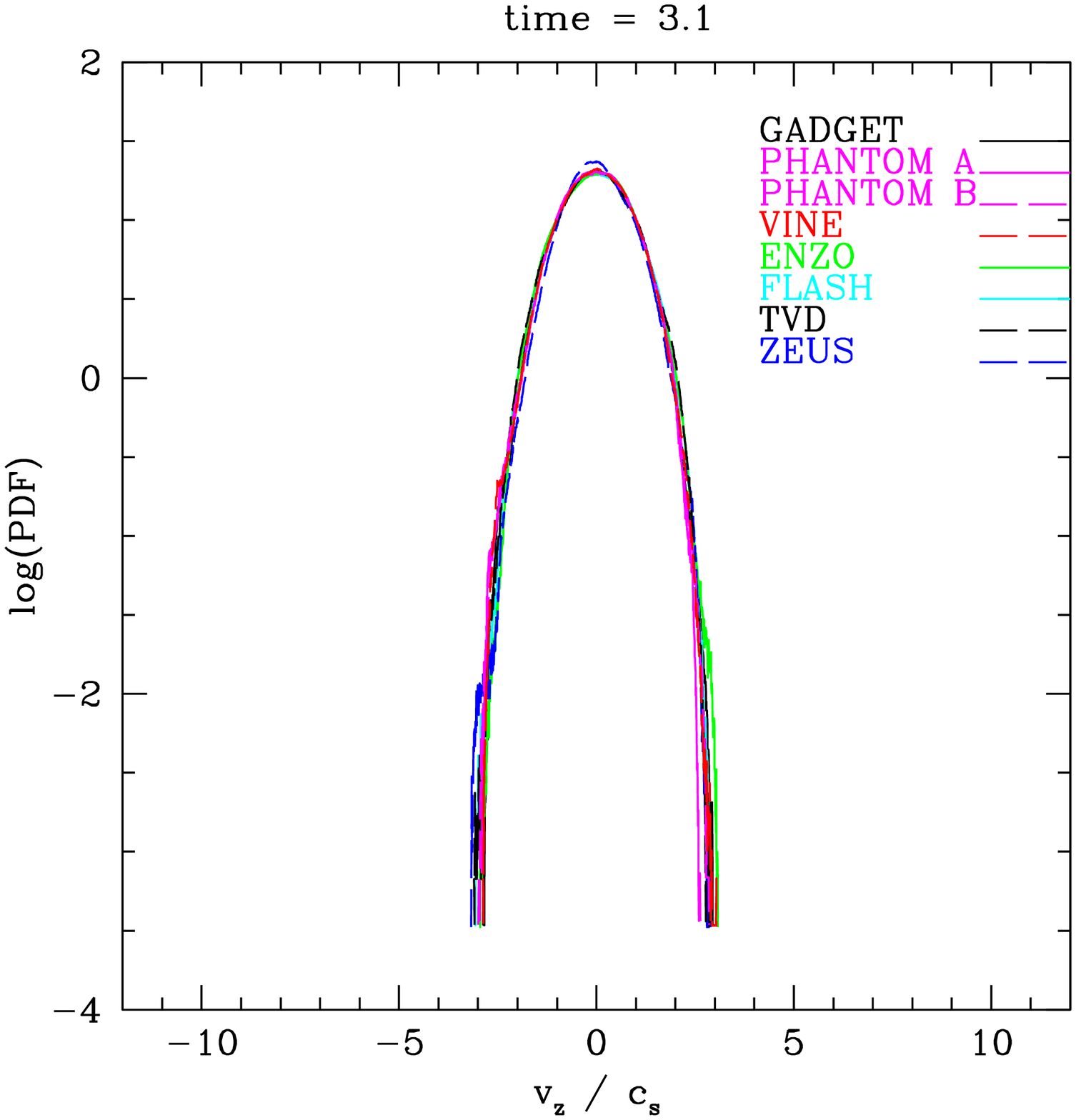} &
\includegraphics[width=0.29\linewidth]{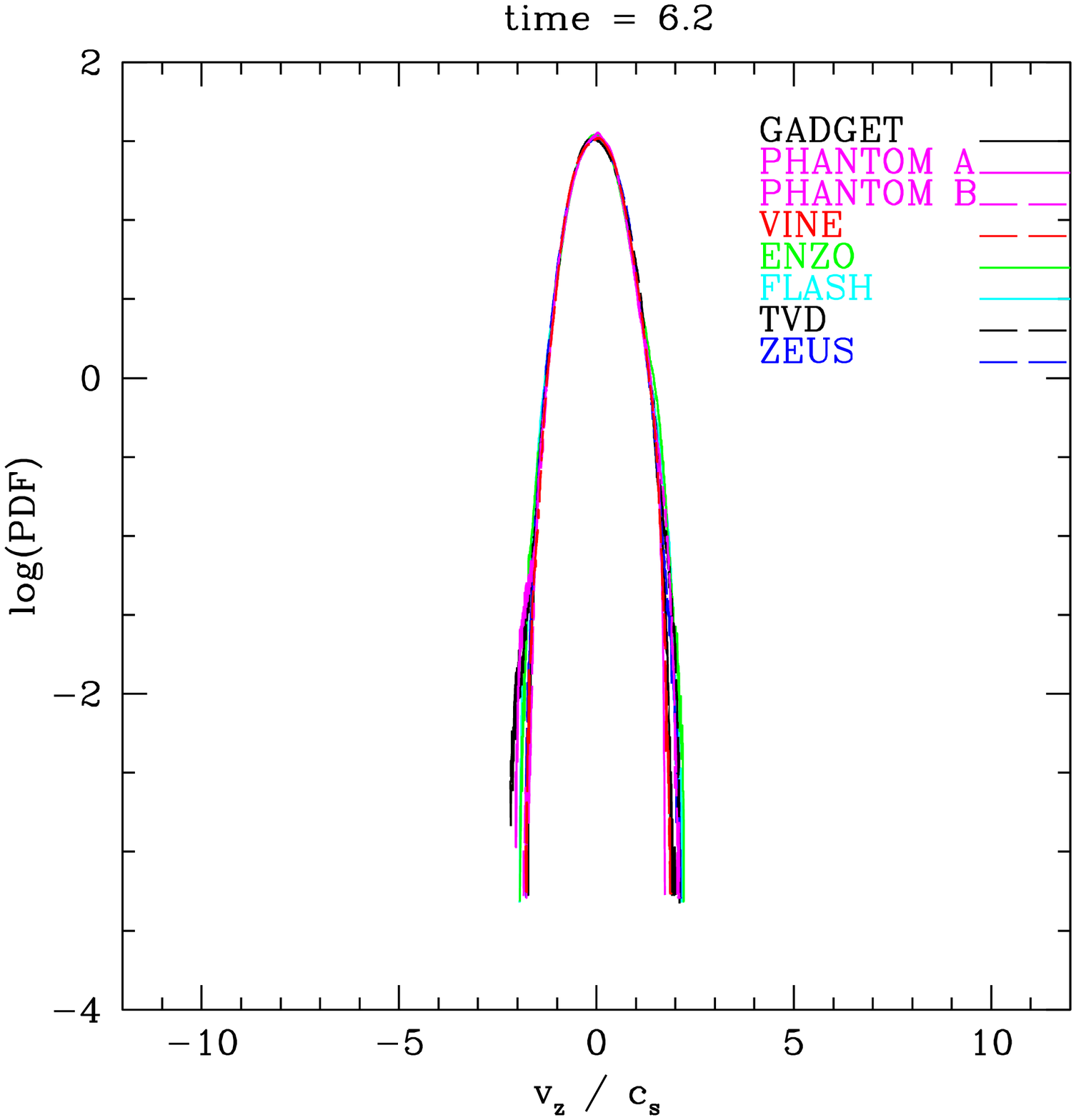}
\end{tabular}
\end{center}
\caption{Same as Figure~\ref{fig:densityPDFs}, but the PDFs of the $z$-component of the velocity ($v_{\rm z}$) in units of the sound speed $c_\mathrm{s}$ are shown.}
\label{fig:velocityPDFs}
\end{figure*}

Our results demonstrate that different codes have different dissipation mechanisms affecting spatial scales $k\gtrsim N/32$. However, our code comparison also shows that SPH and grid codes give similar results for an equivalent number of resolution elements $N$ for each direction in space on scales $k\lesssim N/32$, though with the SPH runs being about ten times more computationally expensive than the grid runs on average. Careful choice of numerical algorithm can extend this scaling range slightly, indicating that grid codes tend to show a slightly longer scaling range than SPH codes (cf.~Figures~\ref{fig:spect_codes_vol} and~\ref{fig:spect_codes_rho}). However, at the numerical resolutions employed in the present study, all slopes inferred from the volume- and density-weighted spectra are too steep compared with higher resolution data from the literature. It is thus rather a question of resolution than a question of the specific properties of the hydrodynamical codes used in this study that determines their performance in reproducing turbulence scaling relations. This must be tested in a future comparison employing at least one order of magnitude more resolution elements for both grid and SPH codes.

\acknowledgements
The authors would like to thank Jens Niemeyer for many interesting and stimulating discussions, especially during the setup phase of this project. We thank the anonymous referee for a balanced and detailed report. SK kindly acknowledges support by an EU Commission ''Marie Curie Intra-European (Individual) Fellowship'' of the 6th Framework Programme with contract number MEIF-CT-2004-011226. SK also acknowledges financial assistance by the EU Commission Research Training Network ''Constellation'' of the 6th Framework Programme. CF acknowledges financial support from the International Max Planck Research School for Astronomy and Cosmic Physics (IMPRS-A) and the Heidelberg Graduate School of Fundamental Physics (HGSFP). The HGSFP is funded by the Excellence Initiative of the German Research Foundation DFG GSC 129/1. The \texttt{ENZO} and \texttt{GADGET} simulations used computational resources from the HLRBII project grant h0972 and pr32lo at the Leibniz Rechenzentrum Garching and from a project grand by CASPUR, Italy. CF and RSK are grateful for subsidies from the DFG SFB 439 Galaxies in the Early Universe. RSK and CF acknowkledge financial support from the German Bundesministerium f\"{u}r Bildung und Forschung via the ASTRONET project STAR FORMAT (grant 05A09VHA) and from the Deutsche Forschungsgemeinschaft (DFG) under grants no.~KL 1358/1, KL 1358/4, KL 1359/5. RSK furthermore thank for subsidies from a Frontier grant of Heidelberg University sponsored by the German Excellence Initiative and for support from the Landesstiftung Baden-W\"{u}rttemberg via their programme International Collaboration II. DJP is supported by a Royal Society University Research Fellowship (UK). The PDF figures were partly produced using {\sc splash} \citep{splashpaper}. MG and SW acknowledge support by the DFG cluster of excellence ''Origin and Structure of the Universe''. All \texttt{VINE} and part of the \texttt{FLASH} calculations were performed on an SGI Altix 3700 Bx2 supercomputer that was funded by the DFG cluster of excellence "Origin and Structure of the Universe". The simulations performed by JK utilized a high performance cluster that was built with funding from the Korea Astronomy and Space Science Institute (KASI) and the Korea Science and Engineering Foundation through the Astrophysical Research Center for the Structure and Evolution of Cosmos (ARCSEC). The work of JK was supported by ARCSEC. AKJ acknowledges support by the Human Resources and Mobility Programme of the European Community under the contract MEIF-CT-2006-039569. MMML acknowledges partial support for his work from NASA Origins of Solar Systems grant NNX07AI74G. The software used in this work was in part developed by the DOE-supported ASC / Alliance Center for Astrophysical Thermonuclear Flashes at the University of Chicago.

\bibliographystyle{aa}

\begin{thebibliography}{112}
\expandafter\ifx\csname natexlab\endcsname\relax\def\natexlab#1{#1}\fi

\bibitem[{{Attwood} {et~al.}(2007){Attwood}, {Goodwin}, \&
  {Whitworth}}]{AttwoodGoodwinWhitworth2007}
{Attwood}, R.~E., {Goodwin}, S.~P., \& {Whitworth}, A.~P. 2007, \aap, 464, 447

\bibitem[{{Ballesteros-Paredes} {et~al.}(2006){Ballesteros-Paredes}, {Gazol},
  {Kim}, {Klessen}, {Jappsen}, \& {Tejero}}]{BallesterosEtAl2006}
{Ballesteros-Paredes}, J., {Gazol}, A., {Kim}, J., {et~al.} 2006, \apj, 637,
  384

\bibitem[{{Ballesteros-Paredes} {et~al.}(2007){Ballesteros-Paredes}, {Klessen},
  {Mac Low}, \& {Vazquez-Semadeni}}]{BallesterosEtAl2007}
{Ballesteros-Paredes}, J., {Klessen}, R.~S., {Mac Low}, M.-M., \&
  {Vazquez-Semadeni}, E. 2007, in Protostars and Planets V, ed. B.~{Reipurth},
  D.~{Jewitt}, \& K.~{Keil}, 63--80

\bibitem[{{Beetz} {et~al.}(2008){Beetz}, {Schwarz}, {Dreher}, \&
  {Grauer}}]{BeetzEtAl2008}
{Beetz}, C., {Schwarz}, C., {Dreher}, J., \& {Grauer}, R. 2008, Physics Letters
  A, 372, 3037

\bibitem[{{Benzi} {et~al.}(2008){Benzi}, {Biferale}, {Fisher}, {Kadanoff},
  {Lamb}, \& {Toschi}}]{BenziEtAl2008}
{Benzi}, R., {Biferale}, L., {Fisher}, R.~T., {et~al.} 2008, Physical Review
  Letters, 100, 234503

\bibitem[{{Berger} \& {Colella}(1989)}]{BergerColella1989}
{Berger}, M.~J. \& {Colella}, P. 1989, Journal of Computational Physics, 82, 64

\bibitem[{{Blitz} {et~al.}(2007){Blitz}, {Fukui}, {Kawamura}, {Leroy},
  {Mizuno}, \& {Rosolowsky}}]{BlitzEtAl2007}
{Blitz}, L., {Fukui}, Y., {Kawamura}, A., {et~al.} 2007, in Protostars and
  Planets V, ed. B.~{Reipurth}, D.~{Jewitt}, \& K.~{Keil}, 81--96

\bibitem[{{Boldyrev}(2002)}]{Boldyrev2002}
{Boldyrev}, S. 2002, \apj, 569, 841

\bibitem[{{Boldyrev} {et~al.}(2002{\natexlab{a}}){Boldyrev}, {Nordlund}, \&
  {Padoan}}]{BoldyrevNordlundPadoan2002}
{Boldyrev}, S., {Nordlund}, {\AA}., \& {Padoan}, P. 2002{\natexlab{a}}, \apj,
  573, 678

\bibitem[{{Boldyrev} {et~al.}(2002{\natexlab{b}}){Boldyrev}, {Nordlund}, \&
  {Padoan}}]{BoldyrevNordlundPadoan2002b}
{Boldyrev}, S., {Nordlund}, {\AA}., \& {Padoan}, P. 2002{\natexlab{b}},
  Physical Review Letters, 89, 031102

\bibitem[{{Bonazzola} {et~al.}(1992){Bonazzola}, {Perault}, {Puget},
  {Heyvaerts}, {Falgarone}, \& {Panis}}]{BonazzolaEtAl1992}
{Bonazzola}, S., {Perault}, M., {Puget}, J.~L., {et~al.} 1992, Journal of Fluid
  Mechanics, 245, 1

\bibitem[{{Calder} {et~al.}(2002){Calder}, {Fryxell}, {Plewa}, {Rosner},
  {Dursi}, {Weirs}, {Dupont}, {Robey}, {Kane}, {Remington}, {Drake}, {Dimonte},
  {Zingale}, {Timmes}, {Olson}, {Ricker}, {MacNeice}, \&
  {Tufo}}]{CalderEtAl2002}
{Calder}, A.~C., {Fryxell}, B., {Plewa}, T., {et~al.} 2002, \apjs, 143, 201

\bibitem[{{Chandrasekhar}(1949)}]{Chandrasekhar1949}
{Chandrasekhar}, S. 1949, \apj, 110, 329

\bibitem[{{Chandrasekhar}(1951{\natexlab{a}})}]{Chandrasekhar1951a}
{Chandrasekhar}, S. 1951{\natexlab{a}}, Royal Society of London Proceedings
  Series A, 210, 18

\bibitem[{{Chandrasekhar}(1951{\natexlab{b}})}]{Chandrasekhar1951b}
{Chandrasekhar}, S. 1951{\natexlab{b}}, Royal Society of London Proceedings
  Series A, 210, 26

\bibitem[{{Colella} \& {Woodward}(1984)}]{ColellaWoodward1984}
{Colella}, P. \& {Woodward}, P.~R. 1984, Journal of Computational Physics, 54,
  174

\bibitem[{{Commer{\c c}on} {et~al.}(2008){Commer{\c c}on}, {Hennebelle},
  {Audit}, {Chabrier}, \& {Teyssier}}]{CommerconEtAl2008}
{Commer{\c c}on}, B., {Hennebelle}, P., {Audit}, E., {Chabrier}, G., \&
  {Teyssier}, R. 2008, \aap, 482, 371

\bibitem[{{de Avillez} \& {Mac Low}(2002)}]{AvillezMacLow2002}
{de Avillez}, M.~A. \& {Mac Low}, M.-M. 2002, \apj, 581, 1047

\bibitem[{{Dimonte} {et~al.}(2004){Dimonte}, {Youngs}, {Dimits}, {Weber},
  {Marinak}, {Wunsch}, {Garasi}, {Robinson}, {Andrews}, {Ramaprabhu}, {Calder},
  {Fryxell}, {Biello}, {Dursi}, {MacNeice}, {Olson}, {Ricker}, {Rosner},
  {Timmes}, {Tufo}, {Young}, \& {Zingale}}]{DimonteEtAl2004}
{Dimonte}, G., {Youngs}, D.~L., {Dimits}, A., {et~al.} 2004, Physics of Fluids,
  16, 1668

\bibitem[{Dobler {et~al.}(2003)Dobler, Haugen, Yousef, \&
  Brandenburg}]{DoblerEtAl2003}
Dobler, W., Haugen, N. E.~L., Yousef, T.~A., \& Brandenburg, A. 2003, Phys.
  Rev. E, 68, 026304

\bibitem[{{Dubey} {et~al.}(2008){Dubey}, {Fisher}, {Graziani}, {Jordan},
  {Lamb}, {Reid}, {Rich}, {Sheeler}, {Townsley}, \& {Weide}}]{DubeyEtAl2008}
{Dubey}, A., {Fisher}, R., {Graziani}, C., {et~al.} 2008, in Astronomical
  Society of the Pacific Conference Series, Vol. 385, Numerical Modeling of
  Space Plasma Flows, ed. N.~V. {Pogorelov}, E.~{Audit}, \& G.~P. {Zank}, 145

\bibitem[{{Elmegreen}(2002)}]{Elmegreen2002}
{Elmegreen}, B.~G. 2002, \apj, 577, 206

\bibitem[{{Elmegreen}(2008)}]{Elmegreen2008}
{Elmegreen}, B.~G. 2008, \apj, 672, 1006

\bibitem[{{Elmegreen} \& {Scalo}(2004)}]{ElmegreenScalo2004}
{Elmegreen}, B.~G. \& {Scalo}, J. 2004, \araa, 42, 211

\bibitem[{{Federrath} {et~al.}(2009{\natexlab{a}}){Federrath}, {Duval},
  {Klessen}, {Schmidt}, \& {Mac Low}}]{FederrathDuvalKlessenSchmidtMacLow2009}
{Federrath}, C., {Duval}, J., {Klessen}, R.~S., {Schmidt}, W., \& {Mac Low},
  M.-M. 2009{\natexlab{a}}, \aap, submitted (arXiv:0905.1060)

\bibitem[{{Federrath} {et~al.}(2008{\natexlab{a}}){Federrath}, {Glover},
  {Klessen}, \& {Schmidt}}]{FederrathGloverKlessenSchmidt2008}
{Federrath}, C., {Glover}, S.~C.~O., {Klessen}, R.~S., \& {Schmidt}, W.
  2008{\natexlab{a}}, Phys. Scr. T, 132, 014025

\bibitem[{{Federrath} {et~al.}(2008{\natexlab{b}}){Federrath}, {Klessen}, \&
  {Schmidt}}]{FederrathKlessenSchmidt2008}
{Federrath}, C., {Klessen}, R.~S., \& {Schmidt}, W. 2008{\natexlab{b}}, \apjl,
  688, L79

\bibitem[{{Federrath} {et~al.}(2009{\natexlab{b}}){Federrath}, {Klessen}, \&
  {Schmidt}}]{FederrathKlessenSchmidt2009}
{Federrath}, C., {Klessen}, R.~S., \& {Schmidt}, W. 2009{\natexlab{b}}, \apj,
  692, 364

\bibitem[{Frisch(1995)}]{Frisch1995}
Frisch, U. 1995, Turbulence (Cambridge Univ. Press)

\bibitem[{{Fryxell} {et~al.}(2000){Fryxell}, {Olson}, {Ricker}, {Timmes},
  {Zingale}, {Lamb}, {MacNeice}, {Rosner}, {Truran}, \&
  {Tufo}}]{FryxellEtAl2000}
{Fryxell}, B., {Olson}, K., {Ricker}, P., {et~al.} 2000, \apjs, 131, 273

\bibitem[{{Gingold} \& {Monaghan}(1977)}]{GingoldMonaghan1977}
{Gingold}, R.~A. \& {Monaghan}, J.~J. 1977, \mnras, 181, 375

\bibitem[{{Glover} {et~al.}(2009){Glover}, {Federrath}, {Mac Low}, \&
  {Klessen}}]{GloverFederrathMacLowKlessen2009}
{Glover}, S.~C.~O., {Federrath}, C., {Mac Low}, M.-M., \& {Klessen}, R.~S.
  2009, \mnras, accepted (arXiv:0907.4081)

\bibitem[{{Glover} \& {Mac Low}(2007)}]{GloverMacLow2007b}
{Glover}, S.~C.~O. \& {Mac Low}, M.-M. 2007, \apj, 659, 1317

\bibitem[{{Gritschneder} {et~al.}(2009){Gritschneder}, {Naab}, {Walch},
  {Burkert}, \& {Heitsch}}]{GritschnederEtAl2009}
{Gritschneder}, M., {Naab}, T., {Walch}, S., {Burkert}, A., \& {Heitsch}, F.
  2009, \apjl, 694, L26

\bibitem[{{Haugen} \& {Brandenburg}(2004)}]{HaugenBrandenburg2004}
{Haugen}, N.~E. \& {Brandenburg}, A. 2004, \pre, 70, 026405

\bibitem[{{Hayes} {et~al.}(2006){Hayes}, {Norman}, {Fiedler}, {Bordner}, {Li},
  {Clark}, {ud-Doula}, \& {Mac Low}}]{HayesEtAl2006}
{Hayes}, J.~C., {Norman}, M.~L., {Fiedler}, R.~A., {et~al.} 2006, \apjs, 165,
  188

\bibitem[{{Heitmann} {et~al.}(2005){Heitmann}, {Ricker}, {Warren}, \&
  {Habib}}]{HeitmannEtAl2005}
{Heitmann}, K., {Ricker}, P.~M., {Warren}, M.~S., \& {Habib}, S. 2005, \apjs,
  160, 28

\bibitem[{{Heitsch} {et~al.}(2001){Heitsch}, {Mac Low}, \&
  {Klessen}}]{HeitschMacLowKlessen2001}
{Heitsch}, F., {Mac Low}, M.-M., \& {Klessen}, R.~S. 2001, \apj, 547, 280

\bibitem[{{Hennebelle} \& {Chabrier}(2008)}]{HennebelleChabrier2008}
{Hennebelle}, P. \& {Chabrier}, G. 2008, \apj, 684, 395

\bibitem[{{Hennebelle} \& {Chabrier}(2009)}]{HennebelleChabrier2009}
{Hennebelle}, P. \& {Chabrier}, G. 2009, \apj, 702, 1428

\bibitem[{{Isichenko}(1992)}]{Isichenko1992}
{Isichenko}, M.~B. 1992, Reviews of Modern Physics, 64, 961

\bibitem[{{Jappsen} {et~al.}(2005){Jappsen}, {Klessen}, {Larson}, {Li}, \& {Mac
  Low}}]{JappsenEtAl2005}
{Jappsen}, A.-K., {Klessen}, R.~S., {Larson}, R.~B., {Li}, Y., \& {Mac Low},
  M.-M. 2005, \aap, 435, 611

\bibitem[{{Kim} {et~al.}(1999){Kim}, {Ryu}, {Jones}, \& {Hong}}]{KimEtAl1999}
{Kim}, J., {Ryu}, D., {Jones}, T.~W., \& {Hong}, S.~S. 1999, \apj, 514, 506

\bibitem[{{Kitsionas} \& {Whitworth}(2002)}]{KitsionasWhitworth2002}
{Kitsionas}, S. \& {Whitworth}, A.~P. 2002, \mnras, 330, 129

\bibitem[{{Klein} {et~al.}(2007){Klein}, {Inutsuka}, {Padoan}, \&
  {Tomisaka}}]{KleinEtAl2007}
{Klein}, R.~I., {Inutsuka}, S.-I., {Padoan}, P., \& {Tomisaka}, K. 2007, in
  Protostars and Planets V, ed. B.~{Reipurth}, D.~{Jewitt}, \& K.~{Keil},
  99--116

\bibitem[{{Klessen}(2000)}]{Klessen2000}
{Klessen}, R.~S. 2000, \apj, 535, 869

\bibitem[{{Klessen} {et~al.}(2000){Klessen}, {Heitsch}, \& {Mac
  Low}}]{KlessenHeitschMacLow2000}
{Klessen}, R.~S., {Heitsch}, F., \& {Mac Low}, M.-M. 2000, \apj, 535, 887

\bibitem[{{Klessen} \& {Lin}(2003)}]{KlessenLin2003}
{Klessen}, R.~S. \& {Lin}, D.~N. 2003, \pre, 67, 046311

\bibitem[{{Kolmogorov}(1941)}]{Kolmogorov1941c}
{Kolmogorov}, A.~N. 1941, Dokl. Akad. Nauk SSSR, 32, 16

\bibitem[{{Kowal} \& {Lazarian}(2007)}]{KowalLazarian2007}
{Kowal}, G. \& {Lazarian}, A. 2007, \apjl, 666, L69

\bibitem[{{Kravtsov}(2003)}]{Kravtsov2003}
{Kravtsov}, A.~V. 2003, \apjl, 590, L1

\bibitem[{{Kritsuk} {et~al.}(2006){Kritsuk}, {Norman}, \&
  {Padoan}}]{KritsukNormanPadoan2006}
{Kritsuk}, A.~G., {Norman}, M.~L., \& {Padoan}, P. 2006, \apjl, 638, L25

\bibitem[{{Kritsuk} {et~al.}(2007){Kritsuk}, {Norman}, {Padoan}, \&
  {Wagner}}]{KritsukEtAl2007}
{Kritsuk}, A.~G., {Norman}, M.~L., {Padoan}, P., \& {Wagner}, R. 2007, \apj,
  665, 416

\bibitem[{{Krumholz} \& {McKee}(2005)}]{KrumholzMcKee2005}
{Krumholz}, M.~R. \& {McKee}, C.~F. 2005, \apj, 630, 250

\bibitem[{{Lemaster} \& {Stone}(2008)}]{LemasterStone2008}
{Lemaster}, M.~N. \& {Stone}, J.~M. 2008, \apjl, 682, L97

\bibitem[{{Lemaster} \& {Stone}(2009)}]{LemasterStone2009}
{Lemaster}, M.~N. \& {Stone}, J.~M. 2009, \apj, 691, 1092

\bibitem[{Lesieur(1997)}]{Lesieur1997}
Lesieur, M. 1997, Turbulence in Fluids (Kluwer)

\bibitem[{{Li} {et~al.}(2003){Li}, {Klessen}, \& {Mac
  Low}}]{LiKlessenMacLow2003}
{Li}, Y., {Klessen}, R.~S., \& {Mac Low}, M.-M. 2003, \apj, 592, 975

\bibitem[{{Lucy}(1977)}]{Lucy1977}
{Lucy}, L.~B. 1977, \aj, 82, 1013

\bibitem[{{Mac Low}(1999)}]{MacLow1999}
{Mac Low}, M.-M. 1999, \apj, 524, 169

\bibitem[{{Mac Low} \& {Klessen}(2004)}]{MacLowKlessen2004}
{Mac Low}, M.-M. \& {Klessen}, R.~S. 2004, Reviews of Modern Physics, 76, 125

\bibitem[{{Mac Low} {et~al.}(1998){Mac Low}, {Klessen}, {Burkert}, \&
  {Smith}}]{MacLowEtAl1998}
{Mac Low}, M.-M., {Klessen}, R.~S., {Burkert}, A., \& {Smith}, M.~D. 1998,
  Physical Review Letters, 80, 2754

\bibitem[{{McKee} \& {Ostriker}(2007)}]{McKeeOstriker2007}
{McKee}, C.~F. \& {Ostriker}, E.~C. 2007, \araa, 45, 565

\bibitem[{{Metzler} \& {Klafter}(2000)}]{MetzlerKlafter2000}
{Metzler}, R. \& {Klafter}, J. 2000, \physrep, 339, 1

\bibitem[{{Monaghan}(1988)}]{Monaghan1988}
{Monaghan}, J.~J. 1988, Computer Physics Communications, 48, 89

\bibitem[{{Monaghan}(1992)}]{Monaghan1992}
{Monaghan}, J.~J. 1992, \araa, 30, 543

\bibitem[{{Monaghan}(1997)}]{Monaghan1997}
{Monaghan}, J.~J. 1997, Journal of Computational Physics, 136, 298

\bibitem[{{Monaghan}(2005)}]{Monaghan2005}
{Monaghan}, J.~J. 2005, Reports on Progress in Physics, 68, 1703

\bibitem[{{Monaghan} \& {Gingold}(1983)}]{MonaghanGingold1983}
{Monaghan}, J.~J. \& {Gingold}, R.~A. 1983, Journal of Computational Physics,
  52, 374

\bibitem[{{Monaghan} \& {Lattanzio}(1985)}]{MonaghanLattanzio1985}
{Monaghan}, J.~J. \& {Lattanzio}, J.~C. 1985, \aap, 149, 135

\bibitem[{{Morris} \& {Monaghan}(1997)}]{MorrisMonaghan1997}
{Morris}, J. \& {Monaghan}, J. 1997, Journal of Computational Physics, 136, 41

\bibitem[{{Nelson} {et~al.}(2008){Nelson}, {Wetzstein}, \&
  {Naab}}]{NelsonWetzsteinNaab2008}
{Nelson}, A.~F., {Wetzstein}, M., \& {Naab}, T. 2008, arXiv:0802.4253

\bibitem[{{Nordlund} \& {Padoan}(1999)}]{NordlundPadoan1999}
{Nordlund}, {\AA}. \& {Padoan}, P. 1999, in Interstellar Turbulence, ed.
  J.~{Franco} \& A.~{Carraminana}, 218

\bibitem[{{Norman} \& {Bryan}(1999)}]{NormanBryan1999}
{Norman}, M.~L. \& {Bryan}, G.~L. 1999, in Astrophysics and Space Science
  Library, Vol. 240, Numerical Astrophysics, ed. S.~M. {Miyama}, K.~{Tomisaka},
  \& T.~{Hanawa}, 19

\bibitem[{{O'Shea} {et~al.}(2004){O'Shea}, {Bryan}, {Bordner}, {Norman},
  {Abel}, {Harkness}, \& {Kritsuk}}]{OSheaEtAl2004}
{O'Shea}, B.~W., {Bryan}, G., {Bordner}, J., {et~al.} 2004, arXiv:0403044

\bibitem[{{Ostriker} {et~al.}(1999){Ostriker}, {Gammie}, \&
  {Stone}}]{OstrikerGammieStone1999}
{Ostriker}, E.~C., {Gammie}, C.~F., \& {Stone}, J.~M. 1999, \apj, 513, 259

\bibitem[{{Ostriker} {et~al.}(2001){Ostriker}, {Stone}, \&
  {Gammie}}]{OstrikerStoneGammie2001}
{Ostriker}, E.~C., {Stone}, J.~M., \& {Gammie}, C.~F. 2001, \apj, 546, 980

\bibitem[{{Padoan} {et~al.}(2004){Padoan}, {Jimenez}, {Nordlund}, \&
  {Boldyrev}}]{PadoanJimenezNordlundBoldyrev2004}
{Padoan}, P., {Jimenez}, R., {Nordlund}, {\AA}., \& {Boldyrev}, S. 2004,
  Physical Review Letters, 92, 191102

\bibitem[{{Padoan} \& {Nordlund}(2002)}]{PadoanNordlund2002}
{Padoan}, P. \& {Nordlund}, {\AA}. 2002, \apj, 576, 870

\bibitem[{{Padoan} \& {Nordlund}(2009)}]{PadoanNordlund2009}
{Padoan}, P. \& {Nordlund}, A. 2009, \apj, submitted (arXiv:0907.0248)

\bibitem[{{Padoan} {et~al.}(1997){Padoan}, {Nordlund}, \&
  {Jones}}]{PadoanNordlundJones1997}
{Padoan}, P., {Nordlund}, {\AA}., \& {Jones}, B.~J.~T. 1997, \mnras, 288, 145

\bibitem[{{Padoan} {et~al.}(2007){Padoan}, {Nordlund}, {Kritsuk}, {Norman}, \&
  {Li}}]{PadoanEtAl2007}
{Padoan}, P., {Nordlund}, {\AA}., {Kritsuk}, A.~G., {Norman}, M.~L., \& {Li},
  P.~S. 2007, \apj, 661, 972

\bibitem[{{Panis} \& {P{\'e}rault}(1998)}]{PanisPerault1998}
{Panis}, J.-F. \& {P{\'e}rault}, M. 1998, Physics of Fluids, 10, 3111

\bibitem[{{Porter} \& {Woodward}(1994)}]{PorterWoodward1994}
{Porter}, D.~H. \& {Woodward}, P.~R. 1994, \apjs, 93, 309

\bibitem[{{Price}(2004)}]{Price2004}
{Price}, D.~J. 2004, PhD Thesis, University of Cambridge

\bibitem[{{Price}(2007)}]{splashpaper}
{Price}, D.~J. 2007, Publ. Astron. Soc. Aust., 24, 159

\bibitem[{{Price} \& {Federrath}(2009)}]{PriceFederrath2009}
{Price}, D.~J. \& {Federrath}, C. 2009, \mnras, submitted

\bibitem[{{Price} \& {Monaghan}(2004)}]{PriceMonaghan2004}
{Price}, D.~J. \& {Monaghan}, J.~J. 2004, \mnras, 348, 139

\bibitem[{{Price} \& {Monaghan}(2007)}]{PriceMonaghan2007}
{Price}, D.~J. \& {Monaghan}, J.~J. 2007, \mnras, 374, 1347

\bibitem[{{Sasao}(1973)}]{Sasao1973}
{Sasao}, T. 1973, \pasj, 25, 1

\bibitem[{{Scalo} \& {Elmegreen}(2004)}]{ScaloElmegreen2004}
{Scalo}, J. \& {Elmegreen}, B.~G. 2004, \araa, 42, 275

\bibitem[{{Schmidt} {et~al.}(2009){Schmidt}, {Federrath}, {Hupp}, {Kern}, \&
  {Niemeyer}}]{SchmidtEtAl2009}
{Schmidt}, W., {Federrath}, C., {Hupp}, M., {Kern}, S., \& {Niemeyer}, J.~C.
  2009, \aap, 494, 127

\bibitem[{{Schmidt} {et~al.}(2008){Schmidt}, {Federrath}, \&
  {Klessen}}]{SchmidtFederrathKlessen2008}
{Schmidt}, W., {Federrath}, C., \& {Klessen}, R. 2008, Phys. Rev. Lett., 101,
  194505

\bibitem[{{Schmidt} {et~al.}(2006{\natexlab{a}}){Schmidt}, {Hillebrandt}, \&
  {Niemeyer}}]{SchmidtHillebrandtNiemeyer2006}
{Schmidt}, W., {Hillebrandt}, W., \& {Niemeyer}, J.~C. 2006{\natexlab{a}},
  Computers and Fluids, 35, 353

\bibitem[{{Schmidt} {et~al.}(2006{\natexlab{b}}){Schmidt}, {Niemeyer}, \&
  {Hillebrandt}}]{SchmidtNiemeyerHillebrandt2006a}
{Schmidt}, W., {Niemeyer}, J.~C., \& {Hillebrandt}, W. 2006{\natexlab{b}},
  \aap, 450, 265

\bibitem[{{Schmidt} {et~al.}(2006{\natexlab{c}}){Schmidt}, {Niemeyer},
  {Hillebrandt}, \& {R{\"o}pke}}]{SchmidtEtAl2006b}
{Schmidt}, W., {Niemeyer}, J.~C., {Hillebrandt}, W., \& {R{\"o}pke}, F.~K.
  2006{\natexlab{c}}, \aap, 450, 283

\bibitem[{{She} \& {Leveque}(1994)}]{SheLeveque1994}
{She}, Z.-S. \& {Leveque}, E. 1994, Physical Review Letters, 72, 336

\bibitem[{{Springel} {et~al.}(2001){Springel}, {Yoshida}, \&
  {White}}]{SpringelYoshidaWhite2001}
{Springel}, V., {Yoshida}, N., \& {White}, S.~D.~M. 2001, New Astronomy, 6, 79

\bibitem[{{Stone} \& {Norman}(1992{\natexlab{a}})}]{StoneNorman1992a}
{Stone}, J.~M. \& {Norman}, M.~L. 1992{\natexlab{a}}, \apjs, 80, 753

\bibitem[{{Stone} \& {Norman}(1992{\natexlab{b}})}]{StoneNorman1992b}
{Stone}, J.~M. \& {Norman}, M.~L. 1992{\natexlab{b}}, \apjs, 80, 791

\bibitem[{{Stone} {et~al.}(1998){Stone}, {Ostriker}, \&
  {Gammie}}]{StoneOstrikerGammie1998}
{Stone}, J.~M., {Ostriker}, E.~C., \& {Gammie}, C.~F. 1998, \apjl, 508, L99

\bibitem[{{Strang}(1968)}]{Strang1968}
{Strang}, G. 1968, SIAM J. Numer. Anal., 5, 506

\bibitem[{{Tasker} {et~al.}(2008){Tasker}, {Brunino}, {Mitchell}, {Michielsen},
  {Hopton}, {Pearce}, {Bryan}, \& {Theuns}}]{TaskerEtAl2008}
{Tasker}, E.~J., {Brunino}, R., {Mitchell}, N.~L., {et~al.} 2008, \mnras, 390,
  1267

\bibitem[{{Tassis}(2007)}]{Tassis2007}
{Tassis}, K. 2007, \mnras, 382, 1317

\bibitem[{Toro(1997)}]{Toro1997}
Toro, E.~F. 1997, Riemann solvers and numerical methods for fluid dynamics
  (Springer)

\bibitem[{{van Leer}(1977)}]{VanLeer1977}
{van Leer}, B. 1977, Journal of Computational Physics, 23, 276

\bibitem[{{V{\'a}zquez-Semadeni}(1994)}]{Vazquez1994}
{V{\'a}zquez-Semadeni}, E. 1994, \apj, 423, 681

\bibitem[{{V{\'a}zquez-Semadeni} {et~al.}(2003){V{\'a}zquez-Semadeni},
  {Ballesteros-Paredes}, \& {Klessen}}]{VazquezBallesterosKlessen2003}
{V{\'a}zquez-Semadeni}, E., {Ballesteros-Paredes}, J., \& {Klessen}, R.~S.
  2003, \apjl, 585, L131

\bibitem[{{Vernaleo} \& {Reynolds}(2006)}]{VernaleoReynolds2006}
{Vernaleo}, J.~C. \& {Reynolds}, C.~S. 2006, \apj, 645, 83

\bibitem[{{von Weizs{\"a}cker}(1951)}]{Weizsaecker1951}
{von Weizs{\"a}cker}, C.~F. 1951, \apj, 114, 165

\bibitem[{{von Weizs{\"a}cker}(1943)}]{Weizsaecker1943}
{von Weizs{\"a}cker}, C.~F.~V. 1943, Zeitschrift fur Astrophysik, 22, 319

\bibitem[{{Wetzstein} {et~al.}(2008){Wetzstein}, {Nelson}, {Naab}, \&
  {Burkert}}]{WetzsteinEtAl2008}
{Wetzstein}, M., {Nelson}, A.~F., {Naab}, T., \& {Burkert}, A. 2008,
  arXiv:0802.4245

\end{thebibliography}

\clearpage

\begin{table*}
\caption{Time evolution of $E_\mathrm{tot}\,=\,\int E(k)\,\mathrm{d}k$ in units of $[c_\mathrm{s}^2]$, for all codes/runs.}
\label{tab:e_vw}
\centering
\def\arraystretch{1.1}
\begin{tabular}{l c c c c c c c c}
\hline\hline
 Time $t\;[\tcross]$ & \texttt{GADGET} & \texttt{PHANTOM~A} & \texttt{PHANTOM~B} & \texttt{VINE} & \texttt{ENZO} & \texttt{FLASH} & \texttt{TVD} & \texttt{ZEUS} \\
\hline
 0.0~\mydotfill  & 7.64  & 7.64  & 7.64  & 7.64  & 7.64  & 7.64  & 7.64  & 7.64  \\
 0.06~\mydotfill & 7.16  & 7.12  & 7.12  & 7.31  & 7.58  & 7.60  & 7.61  & 7.51  \\
 0.31~\mydotfill & 5.72  & 5.72  & 5.71  & 5.83  & 6.32  & 6.16  & 6.17  & 5.88  \\
 0.62~\mydotfill & 4.68  & 4.72  & 4.72  & 4.76  & 5.22  & 5.03  & 5.04  & 4.51  \\
 3.1~\mydotfill  & 1.72  & 1.70  & 1.74  & 1.71  & 1.75  & 1.74  & 1.75  & 1.64  \\
 6.2~\mydotfill  & 1.16  & 1.14  & 1.32  & 1.15  & 1.14  & 1.14  & 1.14  & 1.15  \\
\hline 
\end{tabular}
\end{table*}

\begin{table*}
\caption{Time evolution of $E_\mathrm{mw,tot}\,=\,\int E_\mathrm{mw}(k)\,\mathrm{d}k$ in units of $[\rho_0\,c_\mathrm{s}^2]$, for all codes/runs.}
\label{tab:e_mw}
\centering
\def\arraystretch{1.1}
\begin{tabular}{l c c c c c c c c}
\hline\hline
 Time $t\;[\tcross]$ & \texttt{GADGET} & \texttt{PHANTOM~A} & \texttt{PHANTOM~B} & \texttt{VINE} & \texttt{ENZO} & \texttt{FLASH} & \texttt{TVD} & \texttt{ZEUS} \\
\hline
 0.0~\mydotfill  & 7.22  & 7.22  & 7.22  & 7.22  & 7.22  & 7.22  & 7.22  & 7.22  \\
 0.06~\mydotfill & 6.84  & 6.93  & 6.95  & 7.00  & 7.00  & 6.82  & 6.83  & 6.82  \\
 0.31~\mydotfill & 5.42  & 5.49  & 5.50  & 5.54  & 5.70  & 5.41  & 5.43  & 5.39  \\
 0.62~\mydotfill & 4.30  & 4.36  & 4.38  & 4.38  & 4.55  & 4.31  & 4.32  & 4.17  \\
 3.1~\mydotfill  & 1.56  & 1.56  & 1.61  & 1.56  & 1.57  & 1.56  & 1.55  & 1.56  \\
 6.2~\mydotfill  & 1.12  & 1.11  & 1.29  & 1.12  & 1.11  & 1.10  & 1.10  & 1.13  \\
\hline 
\end{tabular}
\end{table*}

\begin{table*}
\caption{Time evolution of the sonic scale $k_{\rm s}$ as defined by equation~(\ref{eq:sonic_scale}) for all codes/runs.}
\label{tab:sonic_scale}
\centering
\def\arraystretch{1.1}
\begin{tabular}{l c c c c c c c c}
\hline\hline
Time $t\;[\tcross]$ & \texttt{GADGET} & \texttt{PHANTOM~A} & \texttt{PHANTOM~B} & \texttt{VINE} & \texttt{ENZO} & \texttt{FLASH} & \texttt{TVD} & \texttt{ZEUS} \\
\hline
 0.0~\mydotfill  & 10     & 10    & 10    & 10    & 10    & 10    & 10    & 10   \\
 0.06~\mydotfill & 9      & 9     & 9     & 9     & 10    & 10    & 10    & 9    \\
 0.31~\mydotfill & 9      & 8     & 8     & 8     & 10    & 9     & 9     & 8    \\
 0.62~\mydotfill & 8      & 7     & 7     & 8     & 10    & 9     & 9     & 7    \\
 3.1~\mydotfill  & 3      & 3     & 3     & 3     & 3     & 3     & 3     & 3    \\
 6.2~\mydotfill  & 2      & 2     & 2     & 2     & 2     & 2     & 2     & 2    \\
\hline 
\end{tabular}
\end{table*}

\begin{table*}
\caption{Time evolution of the mean $s_0$ and standard deviation $\sigma_s$ of the logarithmic density $s=\ln{(\rho/\rho_0)}$ for all codes/runs.}
\label{tab:moments}
\centering
\def\arraystretch{1.1}
\begin{tabular}{l rl rl rl rl rl rl rl rl}
\hline
\hline
      & \multicolumn{2}{c}{\texttt{GADGET}} & \multicolumn{2}{c}{\texttt{PHANTOM~A}} & \multicolumn{2}{c}{\texttt{PHANTOM~B}} & \multicolumn{2}{c}{\texttt{VINE}} & \multicolumn{2}{c}{\texttt{ENZO}} & \multicolumn{2}{c}{\texttt{FLASH}} & \multicolumn{2}{c}{\texttt{TVD}} & \multicolumn{2}{c}{\texttt{ZEUS}}  \\
 Time $t\;[\tcross]$ & $s_0$ & $\sigma_s$ & $s_0$ & $\sigma_s$ & $s_0$ & $\sigma_s$ & $s_0$ & $\sigma_s$ & $s_0$ & $\sigma_s$ & $s_0$ & $\sigma_s$ & $s_0$ & $\sigma_s$ & $s_0$ & $\sigma_s$ \\
\hline
 0.0~\mydotfill  & -0.73 & 1.23   & -0.73 & 1.23   & -0.73 & 1.23   & -0.73 & 1.23   & -0.73 & 1.23   & -0.73 & 1.23   & -0.73 & 1.23   & -0.73 & 1.23  \\
 0.06~\mydotfill & -0.72 & 1.22   & -0.71 & 1.21   & -0.71 & 1.21   & -0.72 & 1.22   & -0.73 & 1.23   & -0.73 & 1.23   & -0.73 & 1.23   & -0.73 & 1.23  \\
 0.31~\mydotfill & -0.72 & 1.22   & -0.70 & 1.20   & -0.70 & 1.21   & -0.72 & 1.22   & -0.73 & 1.23   & -0.73 & 1.23   & -0.73 & 1.23   & -0.73 & 1.23  \\
 0.62~\mydotfill & -0.54 & 1.10   & -0.52 & 1.08   & -0.53 & 1.08   & -0.54 & 1.09   & -0.58 & 1.14   & -0.58 & 1.13   & -0.59 & 1.14   & -0.68 & 1.19  \\
 3.1~\mydotfill  & -0.15 & 0.56   & -0.15 & 0.55   & -0.15 & 0.56   & -0.15 & 0.56   & -0.16 & 0.56   & -0.16 & 0.56   & -0.16 & 0.58   & -0.16 & 0.58  \\
 6.2~\mydotfill  & -0.06 & 0.34   & -0.05 & 0.32   & -0.06 & 0.33   & -0.06 & 0.34   & -0.06 & 0.34   & -0.06 & 0.34   & -0.06 & 0.35   & -0.06 & 0.36  \\
\hline 
\end{tabular}
\end{table*}

\begin{table*}
\begin{center}
\caption{Computational efficiency of all codes/runs.}
\label{tab:code_efficiency}
\def\arraystretch{1.1}
\begin{tabular}{l c c c c c c c c}
\hline\hline
      & \texttt{GADGET} & \texttt{PHANTOM~A} & \texttt{PHANTOM~B} & \texttt{VINE} & \texttt{ENZO} & \texttt{FLASH} & \texttt{TVD} & \texttt{ZEUS}  \\
\hline
architecture (CPU type)    & Itanium     & Clovertown  & Clovertown      & Itanium    & Itanium     & Xeon         & Xeon         & Xeon        \\
CPU clock rate             & $1.6\,$GHz  & $2.66\,$GHz & $2.66\,$GHz     & $1.3\,$GHz & $1.6\,$GHz  & $1.6\,$GHz   & $2.66\,$GHz  & $2.33\,$GHz \\
number of CPUs used        & 32          & 8           & 8               & 64         & 8 (32)      & 64           & 8            & 64          \\
parallelisation            & MPI         & OpenMP      & OpenMP          & OpenMP     & MPI         & MPI          & MPI          & MPI         \\
total CPU-h                & 6,490       & 248         & 248             & 15,000     & 165 (203)   & 256          & 10           & 315         \\
total CPU-h (norm.$^a$)    & 10,960      & 697         & 697             & 20,600     & 165 (203)   & 256          & 17           & 459         \\
\hline 
\end{tabular}
\end{center}
$^a$ CPU hours normalised to $256^3$ resolution elements, and normalised to a clock rate of $1.6\,$GHz on an Intel Itanium/Xeon chip. These are very rough estimates that should only be accurate to within factors of a few, because of the different parallelisation hardware used on the various supercomputing platforms. Note however that \texttt{GADGET} and \texttt{ENZO} were run on the same supercomputing platform, and with the best run time and parameter optimisations that we could achieve on that supercomputer for both codes. The total CPU time for these two runs can thus be compared directly. \texttt{ENZO} was also run on 32 CPUs (values given in brackets).
\end{table*}

\clearpage

\end{document}